\documentclass[apj]{emulateapj}
\usepackage{graphicx}
\usepackage{subfigure,captcont}
\usepackage{amsmath, amsfonts, amssymb}
\usepackage{natbib}

\newcommand{\loh}{L_{\rm OH}}
\newcommand{\lfir}{L_{\rm FIR}}
\newcommand{\lloh}{\log(\loh / L_\odot)}

\newcommand{\llfir}{\log(\lfir / L_\odot)}

\newcommand{\cmc}{\;{\rm cm}^{-3}}
\newcommand{\kms}{\;{\rm km\; s}^{-1}}

\newcommand{\nhtwo}{\bar{n}\left({\rm H_2}\right)}
\newcommand{\nsubhtwo}{n_{\rm H_2}}

\begin{document}

\title{An Arecibo Survey for Zeeman Splitting in OH Megamaser Galaxies}
\author{James McBride, Carl Heiles}
\affil{Department of Astronomy, University of California, Berkeley, CA 94720-3411; 
\\jmcbride@astro.berkeley.edu, heiles@astro.berkeley.edu
}

\begin{abstract}
We present the results of a comprehensive survey using the Arecibo Observatory for Zeeman splitting of OH lines in OH megamasers (OHMs). A total of seventy-seven sources were observed with the Arecibo telescope. Of these, maser emission could not be detected for eight sources, and two sources were only ambiguously detected. Another twenty-seven sources were detected at low signal-to-noise ratios or with interference that prevented placing any useful limits on the presence of magnetic fields. In twenty-six sources, it was possible to place upper limits on the magnitude of magnetic fields, typically between 10--30~mG. For fourteen sources, the Stokes $V$ spectra exhibit features consistent with Zeeman splitting. Eleven of these fourteen are new detections, and the remaining three are re-detections of Stokes $V$ detections in \citet{Robishaw2008}. Among confident new detections, we derive magnetic fields associated with maser regions with magnitudes ranging from 6.1--27.6~mG. The distribution of magnetic field strengths suggests the magnetic fields in OH masing clouds in OHMs are larger than those in Galactic OH masers. The results are consistent with magnetic fields playing a dynamically important role in OH masing clouds in OHMs.
\end{abstract}

\keywords{galaxies: magnetic fields --- ISM: magnetic fields --- magnetic fields --- masers --- polarization}

\section{Introduction} \label{sec:intro}
This work follows that of \citet{Robishaw2008} (hereafter R08), who detected Zeeman splitting in a sample of five OH megamasers, and is structured similarly for ease of reference to that work. OH megamasers (OHMs) are powerful extragalactic masers that operate in the 18-cm transitions of the hydroxyl molecule, with typical luminosities of order $\lloh = 3$---one hundred million ($10^8$) times greater than the luminosity of OH masers in the Milky Way. OHMs are found in luminous and ultra-luminous infrared galaxies (LIRGs and ULIRGs), classes of galaxies with $\llfir >$ 11 and $\llfir > 12$, respectively. Among these galaxies, OHMs have a preference for the most far infrared luminous \citep{Darling2002a}. 

OHM emission occurs primarily in the 1667~MHz main line transition, with the 1665~MHz main line either weaker or absent \citep{Darling2002}, and the satellite lines at 1612~MHz and/or 1720~MHz only detected in a few nearby OHMs \citet{Baan1987, Baan1992, Martin1989}. VLBI observations of Arp~220, IRAS~F17208--0014, and III~Zw~35, all with redshifts $z < 0.05$, find a combination of compact 1667~MHz emission on parsec sized scales and diffuse 1665~MHz and 1667~MHz emission on scales of hundreds of parsecs, and also see evidence for Keplerian motion in rotating maser rings \citep{Lonsdale1998, Diamond1999, Pihlstrom2001, Momjian2006}. Markarian~231, meanwhile, contained no compact emission in VLBI observations by \citet{Lonsdale2003} that had a physical resolution of roughly 80~pc, meaning emission is coming from regions larger than 80~pc. \citet{Pihlstrom2005} performed VLBI observations of IRAS~F12032+1707 and IRAS~F14070+0525, which are the third most distant and most distant OHMs, respectively, with $z > 0.20$. They found that the majority of emission from IRAS~F12032+1707 is ordered and compact on scales of less than 100~pc, while most of the emission in IRAS~F14070+0525 is resolved out. \citet{Parra2005a} modeled the emission of III~Zw~35 as arising from a rotating starburst ring with clumpy OH clouds, which \citet{Lo2005} suggests could explain other sources as well. In the context of the clumpy model of OHMs, \citet{Lockett2008} explained the observed line ratios of OHMs as being produced through 53~$\mu$m radiative pumping produced by dust with a minimum temperature of 45~K, and overlap of lines with $\sim20 \kms$ linewidths. 
 
[U]LIRGs that host OHMs are predominantly merging systems, in which molecular gas is funneled to a central merging nucleus. One or both of star formation and active galactic nucleus (AGN) activity produce far infrared and radio continuum emission \citep{Darling2002a, Darling2006}---evidence for AGN or star formation activity dominating in OHMs has been conflicting. \citet{Baan1998} used optical spectroscopy to look at the source of nuclear activity in OHMs, and found a preference for AGN among their sample of 42 galaxies. \citet{Darling2006} also studied optical spectra of OHMs, this time comparing to a sample of non-masing [U]LIRGs, and found no significant difference between the OHM hosts and non-masing galaxies. \citet{Vignali2005} observed seven OHMs with {\em Chandra}, detecting only one source weakly, a finding that is consistent with galaxies powered by star formation or low-luminosity AGN. \citet{Kandalian2007} looked at a larger sample of x-ray observations of 22 OHMs, using both {\em Chandra} and XMM-Newton data, and observed a weak relationship between the x-ray luminosity and OH luminosity that favored an AGN contribution in some sources, but noted that galaxies may host OHMs as a result of either AGN activity or rapid star formation. \citet{Willett2011a, Willett2011} compared OHM hosts and non-masing ULIRGs in the mid-infrared, finding that the fraction of AGN in OHMs is lower than that of non-masing ULIRGs, and also demonstrated support for a minimum dust temperature comparable to that used in the modeling in \citet{Lockett2008}. Altogether, the most recent work points to star formation powering OHMs, though AGN activity may still play an important role. In trying to distinguish OHM hosts from non-masing galaxies of similar luminosity, \citet{Darling2007} examined the conditions of the molecular gas, finding a clearer distinction between the two populations based on high mean molecular gas densities. Typical OHM hosts have $\nhtwo = 10^3$--$10^4\;\cmc$ and high dense gas fractions, with half of galaxies that have $L_{\rm HCN} / L_{\rm CO} > 0.07$ also hosting OHMs.

R08 noted that the conditions in which OHMs reside, particularly the high energy and molecular gas densities, suggest the presence of strong magnetic fields. Minimum energy arguments made using VLA observations of synchrotron emission argue for minimum volume averaged magnetic field strengths of 1~mG in the central 100~pc of ULIRGs \citep{Condon1991}. \citet{Thompson2006} argued that producing the FIR-radio correlation for starburst galaxies requires significant fine-tuning if magnetic fields are not significantly larger than the minimum energy estimates. They further show that for magnetic fields to be dynamically important relative to gravity, as in the Milky Way, the magnetic fields in Arp~220 or a similar galaxy would be of order 30~mG. These arguments for fields in [U]LIRGs with magnitudes between 1--30~mG provided a compelling case to search for Zeeman splitting in [U]LIRGs, and in particular, observe OHM lines. The hydroxyl molecule has a relatively large magnetic dipole moment, and OHMs provide bright, narrow lines in which Zeeman splitting may be detected. 

In an effort to directly Zeeman splitting and measure line-of-sight magnetic fields in [U]LIRGs, R08 used the Arecibo telescope and the Green Bank Telescope to perform Full-Stokes observations for eight OHMs. They detected magnetic fields with a median strength of $\simeq$~3~mG in five of the eight observed galaxies, and the strongest detected field had a magnitude of 17.9~mG. Their results confirmed that Full-Stokes observations of OHMs provide a viable way to directly measure magnetic fields in [U]LIRGs, even if magnetic fields are not of equal dynamical importance to gravity in the ISM of [U]LIRGs. Here, we go further, observing seventy-seven known extragalactic OH masers,\footnote{While most meet the definition of megamaser, a few would more accurately be called kilomasers or gigamasers. For simplicity, we subsequently call all of these sources OHMs.} including re-observations of six of the sources from R08. Eight OHMs are not detected at all, primarily as a result of increased radio-frequency interference (RFI) since their discovery. Three sources exhibit features in their Stokes $V$ spectra consistent with Zeeman splitting, but in which we are only marginally confident. Eleven OHMs have Stokes $V$ features that we confidently associate with Zeeman splitting, including re-detections of Zeeman splitting in three OHMs observed in R08. One source, IRAS F10173+0829, showed apparent Zeeman splitting in the observations of R08, but re-observation showed that the apparent features were interference. The remaining sources were detected in Stokes $I$, but did not show significant Zeeman splitting detections.

\section{Sources}
The seventy-seven OHMs make up the entire sample known at the time of observations that are observable with the Arecibo telescope\footnote{The Arecibo Observatory is part of the National Astronomy and Ionosphere Center, which is operated by Cornell University under a cooperative agreement with the National Science Foundation.} in Puerto Rico. All but two of the targets were taken from the OHM survey and compilation of past detections presented in \citet{Darling2000, Darling2001, Darling2002a} (hereafter DG00, DG01, DG02). The two exceptions are the only two Arecibo accessible sources that have been discovered in the decade since \citep{Willett2012}. While the choice to use only Arecibo excludes some viable targets, the results of R08 showed that detecting magnetic fields using Zeeman splitting of OHMs requires a level of sensitivity more easily achieved with Arecibo.

\section{Observations}
We used the $L$-band wide receiver on the 305-meter Arecibo telescope in full-Stokes mode to observe all known OH megamasers at declinations accessible to
Arecibo, over a period spanning December~2007--December~2009. The interim correlator was set up to observe four different bands: one 12.5 MHz band, centered halfway between the 1665 and 1667~MHz, and three 6.25 MHz bands, centered at 1612, 1667, and 1720 MHz. This set-up allowed simultaneous observation of all four ground state OH maser lines and all four Stokes parameters, $I$, $Q$, $U$, and $V$. This differs from the correlator setup used by R08, which focused only on the main lines, and had a range of bandwidths that depended upon the velocity extent of the source. The uniformity of the setup used here eased reduction and analysis. This work focuses on the 1665~MHz and 1667~MHz lines, as there were no Zeeman detections in the satellite lines at 1612~MHz or 1720~MHz. A future work will present the results of the observations of the satellite lines. 

We spent equal time on and off source, switching position every 4~minutes. 
The off source position had the same declination and a right 
ascension 4~minutes east of the source, which kept the hour angles 
of on and off source observations nearly equal. The integration time 
was 1~second, chosen to allow for elimination of RFI that appeared on 
short time scales. 

\section{Data Reduction} \label{sec:reduction}
Many of the details of the data
reduction are well described in R08. This includes calibration of spectra, 
RFI removal, bandpass and gain correction, and fitting 
Gaussian components to the Stokes $I$ profiles. The discussion here will focus
on describing new components of the data reduction, and mostly does not repeat
the description in R08 in cases where the procedure used in reducing 
the data in this paper was identical to that in R08. Section 4.1 of R08,
describing calibration, and 4.3, describing bandpass and gain correction, are
omitted entirely. 

We use the classical definition of Stokes $I$, which is the sum, rather
than the average, of two orthogonal polarizations. Our flux densities
are thus twice those of most previously published work, 
but are consistent with the choice of R08. We take the IAU
definition of Stokes $V$, with $V = RHCP - LHCP$. 
Following the IEEE definitions
$RHCP$ is clockwise rotation and $LHCP$ is counterclockwise rotation of the
electric field vector along the direction of propagation.

\subsection{RFI removal}
The data were examined for RFI in various stages. First, the on/off pairs of
1~second records were flagged for RFI and checked for significant rippling
caused by the Sun or Moon being within the beam sidelobes. At this point, the
data reduction forked: within each on/off pair of 4~minute records, the 1~
second records were averaged together to form a single 4~minute spectrum, and
also combined by taking the median to form a second 4~minute spectrum for
the same data. In cases where the RFI is limited, the average spectrum
tended to be less noisy, but forming a composite spectrum from the median of
the 1~second integrations was clearly better for sources with significant 
narrow-band RFI. 

Each source was typically observed for a few hours.
This resulted in $\sim$10--20 4~minute on and off spectra for each source. 
These too were combined in the same manner as the 1~second spectra, either by
taking the average or the median. Applying these choices to each set of mean
or median 4~minute spectra resulted in four different final spectra for each
source at a given frequency and polarization: mean/mean, mean/median, 
median/mean, median/median. Of these four options, the spectrum with the  
least noise was selected. 

\subsection{Confidence tests} \label{sec:conf_tests}
Many of the Zeeman measurements presented here represent marginal detections.
We describe here some of the primary obstacles to determining magnetic
field strengths, and tests applied to the data and fits that validate
the results. 

\subsubsection{Broadband rippling} \label{sec:broad_rippling}
Many of the Stokes $V$ spectra contain ripples of characteristic size $\sim 0.3$--1 MHz that can masquerade as Zeeman splitting of broad OHM components, and represent the most serious source of systematic error in estimating magnetic fields. The best fit fields to these ripples in the Stokes $V$ spectra are often in excess of 50 milligauss. These ripples are easily spotted by eye in most situations, but can confuse magnetic field fits in sources with particularly rich spectra. For this reason, magnetic field fits associated with spectral features wider than $0.3$ MHz are all considered to be untrustworthy. This choice is also physically motivated, in that particularly wide maser lines are likely produced by blending from many masing clouds, and are less likely to provide a detectable line-of-sight Zeeman splitting signal. 
\subsubsection{Spectral shifting}
\begin{figure}[h!]
\begin{center}
  \includegraphics[width=3.5in] {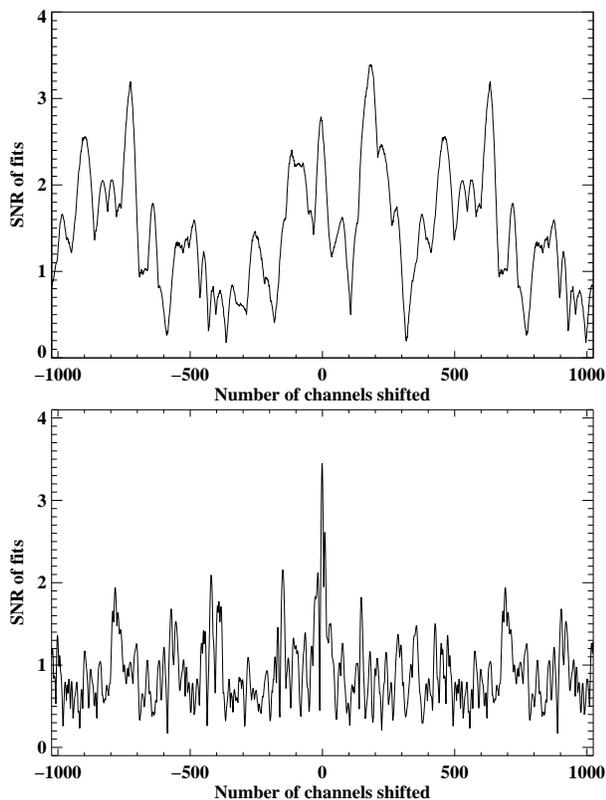}
\end{center}
\caption{The average signal-to-noise ratios of fitted fields 
to narrow components in the Stokes $I$ and $V$ spectra of 
IRAS F23028+0725 (top panel) and IRAS F09039+0503 (bottom panel) 
are plotted for fits when the 
Stokes $I$ and $V$ spectra are shifted relative to each other. The overall
fit quality at no shift for F23028+0725 is mediocre, 
and the fits are marginally better when the spectra are
shifted relative to each other, suggesting that the features being fit in the
Stokes $V$ spectrum are not Zeeman splitting. The fits for F09039+0503 are 
at their best for no shift, and no other shift produces fits of comparable
quality.} \label{fig:compare_shifts}
\end{figure}

Even after removing the broadest ripples, there are Stokes $V$ spectra with structured noise on smaller frequency scales that can confuse interpretation. In some cases, noisy structure in Stokes $V$ may be coincidentally lined up with features in the Stokes $I$ spectrum, and produce claimed magnetic field detections. To identify instances of dubious magnetic field claims, we ran our fitting procedure for Zeeman splitting with the Stokes $V$ spectrum shifted relative to the Stokes $I$ spectrum for all sources. With 2048 channels, this produces 2048 different fits to features in Stokes $V$. For each set of fits, we selected the features narrower than $0.3$~MHz, found the associated signal to noise ratio for the magnetic fields of each narrow component, and produced an overall quality of fit at each shift. 
For sources where non-zero shifts produced fits of comparable quality to the fits with no shift, the reported fields were regarded as highly dubious.
In Section \ref{sec:v_dets}, all sources for which we claim a magnetic field detection were best fit when no shift was applied between the Stokes $I$ and $V$ spectra. Other sources with claimed magnetic fields, but which were also well fit when the Stokes $I$ and $V$ were shifted relative to one another, were considered Stokes~$V$ non-detections or marginal detections, depending on other factors. An example of this process is shown in Figure \ref{fig:compare_shifts}. The source in the top panel is not fit particularly well at any one shift, whereas the source in the bottom panel is only well fit when no shift is applied, providing confidence that the features being fit are real. The full fits for these sources are shown in Sections \ref{sec:non_v_dets} and \ref{sec:v_dets}.

\subsubsection{Comparison of overlapping spectra}
For each source observed, there was a total of 25 MHz of spectral coverage,
with two 6.25 MHz bands covering the OH satellite lines at rest frequencies
of 1612 and 1720 MHz, and the broad 12.5
MHz band that encompassed both main lines at 1665 and 1667~MHz. 
With sources spread out over a redshift range of 0.007--0.265, the 
highest observed frequency was 1711~MHz, and the minimum was 1271~MHz. 
Just over 90\% of this range was covered by at least one source, including
the entire range between 1351.8~MHz--1663.7~MHz. Moreover, 
73\% of this band was covered by at least two sources, and 50\% was covered
by at least four sources.  

\begin{figure}[h!]
\begin{center}
  \includegraphics[width=3.5in] {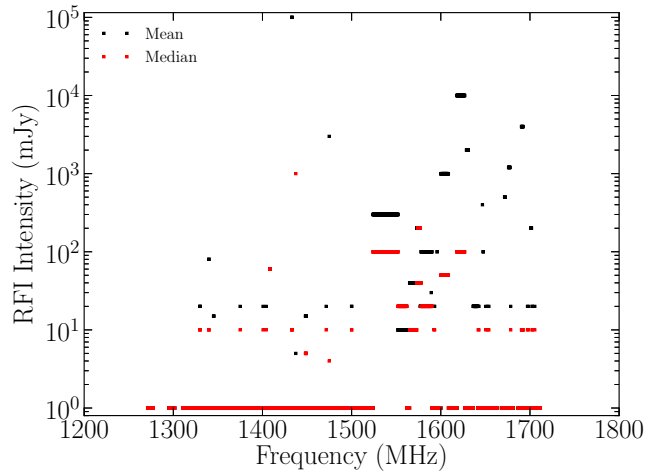}
\end{center}
\caption{The spectral coverage of the survey is shown here, along with the RFI environment at Arecibo during observations. Areas with observations and no observed RFI are plotted at a default level of 1~mJy, as that was the typical rms error for RFI free data. At frequencies where RFI was present in at least one data set, the approximate amplitude of the most severe RFI is shown. For a couple of wide regions with severe RFI, such as that between 1524--1552~MHz, the value plotted is typical of the large RFI spikes throughout the frequency range, but RFI of that amplitude was not necessarily observed in each channel.} \label{fig:rfi}
\end{figure}

This overlapping coverage was a valuable tool in RFI identification. Spectra
with features of unclear origin could be compared to spectra of other sources
in the same frequency range. If a similar feature appeared in both spectra,
then it strongly suggests interference. 
Comparing overlapping spectra of different sources was critical 
in discovering that the Stokes $V$ features of IRAS~F10173+0829, reported in
R08, were actually interference.
A rough characterization of RFI severity across the frequency range in 
the survey can be seen in Figure \ref{fig:rfi}.

\subsubsection{Field sign}
In a survey of Zeeman splitting in Galactic OH masers, \citet{Fish2006} tested for unknown systematics in their data by counting how many Zeeman pairs had greater flux in the LCP component, how many had greater flux in the RCP component, and comparing their result to the expectation that the each possibility was equally likely. We performed a similar test, counting the number of positive and negative magnetic fields derived, and comparing the results with the binomial distribution. 

For $N$ fields and an equal likelihood of positive or negative fields, the probability of a positive field is $p = 0.5$, the expected number of positive fields is $Np$, the expected number of negative fields is $N(1-p)$, and their variance is $Np(1-p)$. For all sources, we fit a total of 373 Gaussian components, and a magnetic field estimate was made for each component, giving an expected number of $186.5 \pm 9.7$ for positive and negative fields. Of derived fields, 172 had positive sign, and 201 were negative in sign, which are not statistically significant deviations. Likewise, for magnetic field components we considered a detection, there were a total of 35 derived fields. Compared to an expectation of $17.5 \pm 2.1$, 19 were positive in sign and 16 were negative, very close to the expected distribution.
\subsubsection{Bootstrap resampling}
An often used technique to distinguish a source of broadband interference from one of astrophysical interest is to split a dataset in half, and check that the feature is roughly the same strength throughout the observation. When looking for very low SNR features, this method proved to be only marginally helpful, as the fluctuations in amplitude for a low SNR astrophysical signal due to noise might be similar to those expected from broadband interference. As an extension of this, we bootstrap resampled our spectra, rather than splitting the data in to two halves. 

Bootstrapping is a resampling method that can be used for estimating confidence intervals associated with a statistic \citep{Efron1993}. For a statistic such as the median, bootstrap resampling makes it possible to characterize uncertainty, despite there being no formal way to do so. The procedure is relatively straightforward. Given a sample of $n$ observations, {\bf x} = ($x_1, x_2, ..., x_n$), the empirical probability of observing each of the $n$ values is $1/n$. A bootstrap sample of {\bf x} is made up of $n$ values drawn randomly from {\bf x}. One of the original observations may be drawn more than once, or not at all, in this bootstrap sample. To build a distribution of values for some statistic, such as the median, take the median of the original sample as well as the median of each of many bootstrap samples. This sampled distribution of the median measurements provides an estimate of the uncertainty in the measurement of the median. 

As the spectrum for each OHM was made by combining 4~minute spectra, either using the mean or median, we treated each of the 4~minute spectra as an individual observation. As discussed above, for some sources, taking the median of those 4~minute spectra produced a spectrum with less noise, usually because it better handles RFI, while for some sources the mean spectra were less noisy; the ``measurement'' in this case is whichever of the two was better. We then took 1000 bootstrap samples of that measurement. The resultant lower and upper limits are in a few cases useful in establishing whether a broad spectral feature is OH emission. We performed similar resampling in checking the derived errors of magnetic field estimates. For sources with claimed magnetic field detections, the Stokes $I$ and Stokes $V$ spectra were jointly resampled, and the magnetic field derivation procedure described below was applied. The error from bootstrap resampling was consistently comparable to those reported in the fitting procedure. 

\subsection{Fitting Gaussian Components to Stokes $I$ Profiles}
We followed the same general guidelines given in R08 for fitting Gaussians
to the overall Stokes $I$ profile. Fits were made by eye, so are inherently
subjective, but follow relatively straightforward guidelines that aim to
provide a balance between minimizing the number of Gaussian components while
also accounting for the majority of structure in the OHM profiles. 

\section{Results} \label{sec:results}

\subsection{Circular Polarization and Line-of-Sight Magnetic Fields}
\subsubsection{Magnetic Field Derivation}
As in the section on data reduction, much of the process for obtaining
magnetic field measurements is the same as R08. In particular, magnetic fields
for most sources are derived assuming that the Zeeman splitting produced by
the magnetic field is smaller than the intrinsic linewidth. In this situation, the Stokes $V$ spectrum is proportional to the derivative of the Stokes $I$ spectrum, with
\begin{equation}
	V = \left(\frac{\nu}{\nu_0}\right)\left(\frac{dI}{d\nu}\right)b B_\parallel + cI, \label{eq:narrow_split}
\end{equation}
where $B_\parallel$ is the line-of-sight magnetic field, $b$ is the splitting coefficient \citep{Heiles1993}, and the $cI$ term accounts for leakage of $I$ into $V$. The splitting coefficient is itself proportional to the Land\'e $g$-factor for the observed transition; for the 1667~MHz line in OH, it is 1.96~Hz~$\mu {\rm G}^{-1}$, and for the 1665~MHz OH line, it is 3.27~Hz~$\mu {\rm G}^{-1}$. In deriving line-of-sight magnetic fields, we perform least squares fits to Equation \ref{eq:narrow_split}. As explained and justified in R08, the fits to the Stokes $I$ profile serve as the independent variable in solving Equation \ref{eq:narrow_split}. For each Gaussian component used in the overall Stokes $I$ profile, a separate line-of-sight magnetic field is fit, along with estimated errors. The reported errors are purely statistical, and underestimate uncertainty in many spectra. For that reason, we do not rely on the reported errors alone when assigning confidence to a fitted magnetic field. 

This approach does not account for the blending of magnetic field measurements that may occur if multiple masing clouds contribute to the observed emission at a specific velocity, which likely introduces error for at least some fraction of observed OHMs. \citet{Parra2005a} explained the combined compact and diffuse emission in III~Zw~35 observed by \citet{Pihlstrom2001} as a result of clumpy OH masing clouds in a narrow ring. Bright, compact maser features are produced when more than one cloud is along the line-of-sight. R08 compared their single dish spectrum of III~Zw~35 with the interferometric observations of \citet{Pihlstrom2001}. They found that three of the Gaussian components that they fit to the total III~Zw~35 profile correspond to three compact features mapped by \citet{Pihlstrom2001}.

This adds a layer of complication to interpretation of Zeeman splitting results, as well as making detections more difficult. For instance, two masing clouds at the same velocity along the line-of-sight that have equal intensities and magnetic fields of equal magnitude but opposite sign will combine to produce a perfectly flat Stokes $V$ spectrum, meaning no magnetic field would be measured. More generally, the measured magnetic field is the average of the intensity weighted fields for each cloud. Thus the reported fields may reflect lower magnetic field magnitudes than are actually being probed, if blended lines are associated with fields oriented in opposite directions.

Descriptions of each source follow. Described properties of 
OH megamaser host galaxies are based upon the results compiled in 
\citet{Darling2000}, \citet{Darling2001}, and 
\citet{Darling2002a}. 
Lines are described as ``narrow'' or ``wide'' throughout the following 
sections. Wide lines are those fit with Gaussians that have a 
full width at half maximum (FWHM) greater than 0.3 MHz, and
narrow lines have FWHM less than 0.3 MHz. This cutoff is motivated by the
size of noisy rippling in Stokes $V$ spectra that can masquerade as Zeeman
splitting, making reported fields on wide lines highly dubious.  
For sources with features consistent with Zeeman splitting, and a small
sample of sources with no Zeeman splitting, we present
a table with the full fitting results.
Narrow lines with fitted fields that are $>3\sigma$ are highlighted in bold
text.

\subsubsection{Re-observed sources}
The sources with Full-Stokes observations made by
R08 in February~2006 were observed again in 2008--2009 as part of this survey.
This was done for two reasons: to look for evidence of variability in OHMs, 
and to confirm the magnetic field strengths derived by R08. 
Overall, our results are very similar to those presented in
R08, though our more recent observations are generally slightly noisier, and 
have higher reported errors associated with Gaussian components and fields. 
For this reason, we only reproduce the table with fits for one source, 
IRAS~F12032+1707, in order to discuss VLBI observations. 
However, we note that for all sources presented in R08, the tables reported an 
intensity for Gaussian~4 that is too small by a factor of 100. 

We find evidence of weak variability in the Stokes $I$ profiles of most of the sources, and no strong evidence of variability in the Stokes $V$ profiles.
For one source, IRAS F$10173+0829$, we discovered that apparent Zeeman
splitting seen in R08 was actually circularly polarized RFI.

\begin{figure*}[t!]
\begin{center}
    \subfigure[IRAS F01417+1651: The difference between the Stokes $I$ profiles was scaled by a factor of 2.]{
        \includegraphics[width=3.2in] {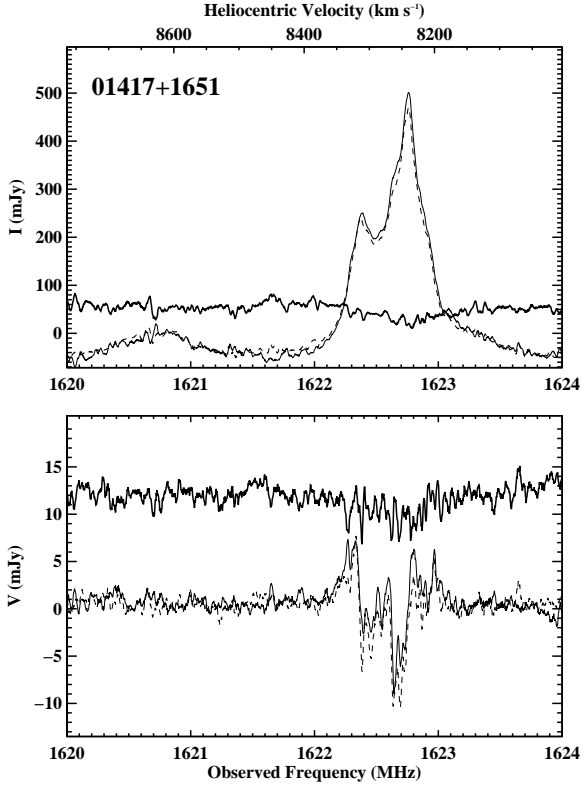}
        \label{fig:01417}
    } 
    \subfigure[IRAS F12032+1707: No scaling or offset was applied to the difference.]{
        \includegraphics[width=3.2in] {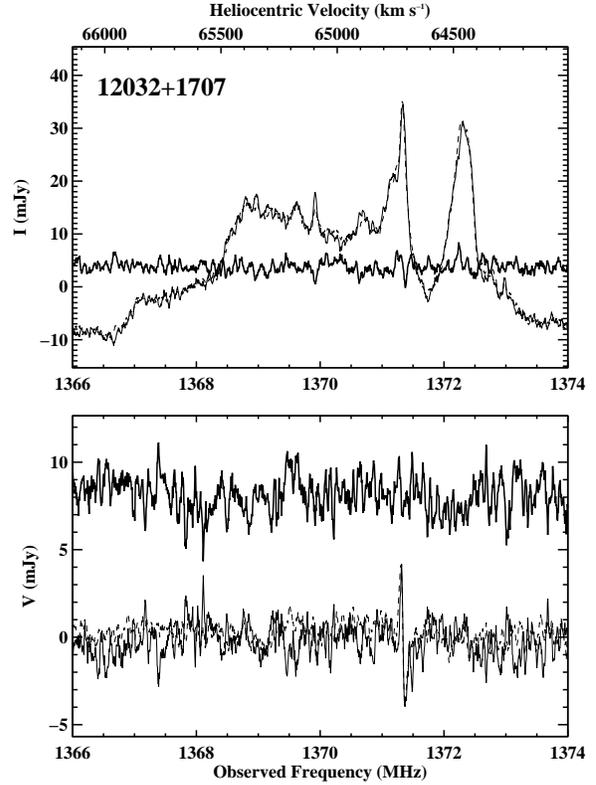}
        \label{fig:12032}
    } \\
    \subfigure[IRAS 12112+0305: The difference between the two spectra is 
        offset from a mean close to zero.]{
        \includegraphics[width=3.2in] {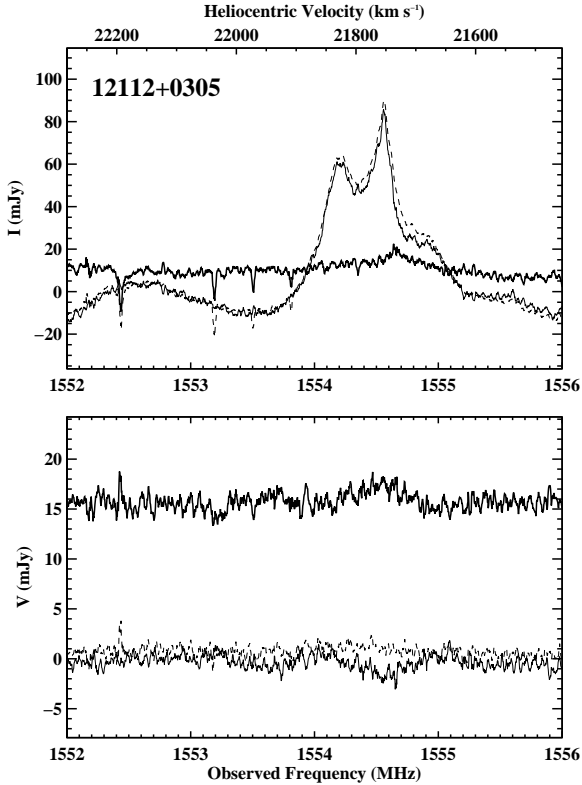}
        \label{fig:12112}
  }
    \subfigure[IRAS F15327+2340: The difference between the Stokes $I$ profiles was scaled by a factor of 4.]{
        \includegraphics[width=3.2in] {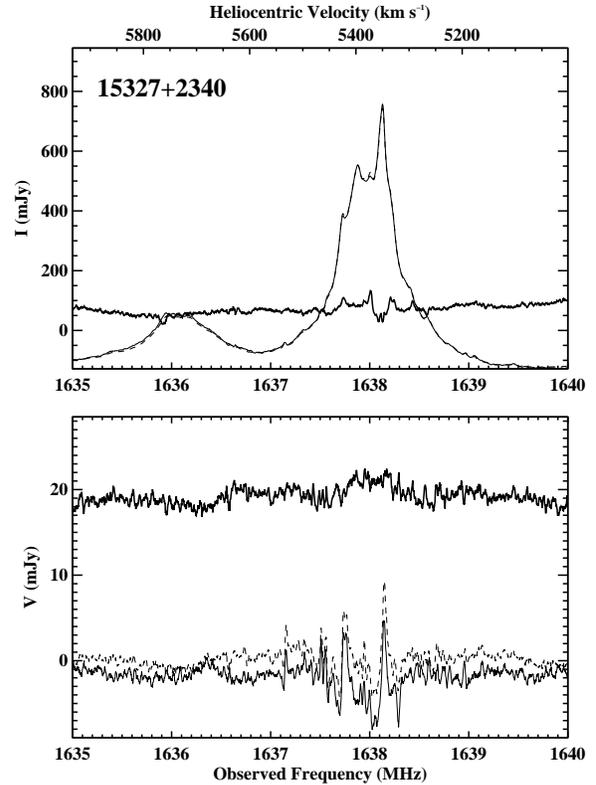}
        \label{fig:15327}
}
\caption{Stokes $I$ (top panel) and $V$ (bottom) profiles four sources observed
first in 2006 and then again in 2008--2009. The data taken in 2006 was first 
presented in R08, and shown here again with dashed lines, while the more 
recent data are shown with a thin solid line.
The difference between the two data sets (thick solid)
are in some cases offset or scaled to show variation more clearly.}\label{fig:variability}
\end{center}
\end{figure*}

{\bf IRAS F01417+1651} (III Zw 35): The shape of features in the Stokes~$I$
spectrum remain similar in the more recent observations, 
as shown in Figure \ref{fig:01417}, but the entire 1667~MHz feature
appears to be moderately brighter in 2008--2009 than it was in 2006.
The peak flux density feature, located at 1622.8~MHz, brightened by 
$\sim$40~mJy, a fractional change of $\sim$8\%. 
There are no clear changes between the old and more recent Stokes~$V$ spectra.
The individual Gaussian components used for fitting the new spectrum, 
and the magnetic field values associated with each, 
are in good agreement with those of R08. In addition, 
the same field reversal is observed. Despite brightening, the structure of 
the OHM in IRAS~F01417+1651 appears relatively unchanged. 

\begin{figure}[h!]
\begin{center}
  \includegraphics[width=3.5in] {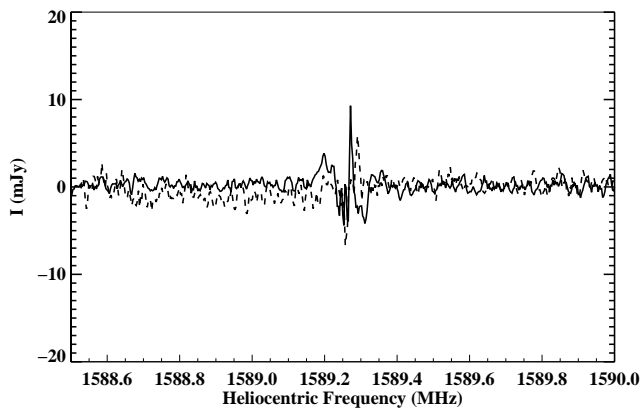}
\end{center}
\caption{The solid line shows the Stokes $V$
spectrum for IRAS~10173+0829, while the dashed line 
shows the Stokes $V$ spectrum for IRAS~15107+0724, which is of 
similar amplitude and shape, at the same frequency.
While the features in IRAS~10173+0829 nicely line up with
maser emission in its Stokes $I$ spectrum, 
the Stokes $I$ spectrum for IRAS~15107+0724 does not have any features
suggesting maser emission, indicating that the Stokes $V$ features in both
sources are interference.} 
\label{fig:10173}
\end{figure}

{\bf IRAS F10173+0829}: R08 reported features consistent with Zeeman splitting
in their observations of this OHM.  
Unfortunately, the Stokes $V$ spectrum composed by taking the
median of 1-second and 4-minute spectra found weaker features
in the Stokes $V$ spectra than were seen by R08. Further, 
the Stokes $V$ spectra of other
sources at the same frequency had Stokes $V$ features similar to those seen
in IRAS~F10173+0829. Figure \ref{fig:10173} shows the Stokes~$V$ spectrum 
of the 1612~MHz line of IRAS F15107+0724 compared to that of IRAS~F10173+0829. 
IRAS~F15107+0724 showed no evidence of
maser emission at frequencies between 1589.2--1589.4~MHz in its Stokes~$I$ 
spectrum, yet had a similar Stokes~$V$ signal to IRAS~F10173+0829.
This not only means that the apparently detected Zeeman splitting is not
real, but also prevents placing meaningful limits on magnetic fields in this
OHM.  

{\bf IRAS F12032+1707}: The results of the more recent observations are in
general agreement with those of R08, as shown in Figure \ref{fig:12032}. The 
rms error in the difference spectrum is dominated by that in the more recent
spectrum, which has $\sim$2~mJy channel-to-channel variation. There is weak
evidence for variation associated with the brightest, narrowest features in
the Stokes~$I$ spectrum, as the difference spectrum has small peaks
with amplitudes of $\sim$5~mJy associated with each bright emission feature,
but this variation is nowhere near as dramatic as the nearly factor of two
brightening observed in the brightest peak, located at 1371.4~MHz, between
the observations of DG01 and those of R08. The Stokes~$V$ features remain
consistent between observations, and the magnetic fields found by fits to 
the new spectra are entirely consistent with those published in R08. 

\begin{deluxetable*}{lccccc}
\tablecolumns{6}
\tablewidth{0pt}
\tablecaption{IRAS F12032$+$1707 Gaussian Fit Parameters \label{tab:12032}}
\tabletypesize{\footnotesize}
\tablehead{
& \colhead{$S$} & \colhead{$\nu$} & \colhead{$\Delta\nu$} & \colhead{$v_{\odot}$} & \colhead{$B_{\parallel}$} \\
\colhead{Gaussian} & \colhead{(mJy)} & \colhead{(MHz)} & \colhead{(MHz)} & \colhead{(km s$^{-1}$)} & \colhead{(mG)} \\
\colhead{(1)} & \colhead{(2)} & \colhead{(3)} & \colhead{(4)} & \colhead{(5)} & \colhead{(6)}
}
\startdata
     0 \dotfill & $\phantom{8}    4.31 \pm   0.54$ & $   1367.1002 \pm  0.0169$ & $   0.4082 \pm   0.0370$ & $  65844.0$ & $           \phantom{ }    -6.79 \pm   7.94\phantom{8}$ \\
     1 \dotfill & $\phantom{8}    3.95 \pm   0.96$ & $   1367.6739 \pm  0.0457$ & $   0.7915 \pm   0.1793$ & $  65690.6$ & $\phantom{-}               30.42 \pm  12.43           $ \\
     2 \dotfill & $\phantom{8}    8.42 \pm   0.99$ & $   1368.8016 \pm  0.0094$ & $   0.6264 \pm   0.0448$ & $  65389.5$ & $\phantom{-}\phantom{8}     6.91 \pm   4.86\phantom{8}$ \\
     3 \dotfill & $\phantom{8}    3.00 \pm   0.34$ & $   1369.5922 \pm  0.0071$ & $   0.1460 \pm   0.0207$ & $  65178.7$ & $           \phantom{ }    -1.05 \pm   6.48\phantom{8}$ \\
     4 \dotfill & $\phantom{8}    0.21 \pm   0.36$ & $   1369.6903 \pm  0.0452$ & $   2.6434 \pm   0.1574$ & $  65152.6$ & $           \phantom{ }    -3.25 \pm   3.98\phantom{8}$ \\
     5 \dotfill & $\phantom{8}    3.23 \pm   0.54$ & $   1369.9216 \pm  0.0036$ & $   0.0450 \pm   0.0089$ & $  65091.0$ & $\phantom{-}\phantom{8}     3.10 \pm   3.41\phantom{8}$ \\
     6 \dotfill & $\phantom{8}    6.67 \pm   0.34$ & $   1370.7194 \pm  0.0064$ & $   0.2954 \pm   0.0165$ & $  64878.6$ & $\phantom{-}               11.64 \pm   4.31\phantom{8}$ \\
     7 \dotfill & $              21.67 \pm   0.65$ & $   1371.2516 \pm  0.0030$ & $   0.4769 \pm   0.0115$ & $  64737.0$ & $\phantom{-}               {\bf 10.90 \pm   1.72}\phantom{8}$\footnote{Derived magnetic field strengths above $3\sigma$ that are considered believable are marked in bold.} \\
     8 \dotfill & $              16.99 \pm   0.46$ & $   1371.3316 \pm  0.0010$ & $   0.0849 \pm   0.0029$ & $  64715.8$ & $\phantom{-}               {\bf 17.92 \pm   0.89}\phantom{8}$ \\
     9 \dotfill & $              16.19 \pm   3.38$ & $   1372.1576 \pm  0.0568$ & $   0.4872 \pm   0.0531$ & $  64496.4$ & $\phantom{-}\phantom{8}     1.78 \pm   2.48\phantom{8}$ \\
    10 \dotfill & $              25.02 \pm   4.51$ & $   1372.3299 \pm  0.0027$ & $   0.2795 \pm   0.0165$ & $  64450.6$ & $           \phantom{ }    -1.45 \pm   1.12\phantom{8}$ \\
    11 \dotfill & $\phantom{8}    9.56 \pm   0.51$ & $   1372.7780 \pm  0.0412$ & $   0.7315 \pm   0.0588$ & $  64331.7$ & $                         -11.69 \pm   4.98\phantom{8}$ \\
    12 \dotfill & $\phantom{8}    2.02 \pm   0.21$ & $   1373.8371 \pm  0.0179$ & $   0.3807 \pm   0.0500$ & $  64051.0$ & $\phantom{-}               20.61 \pm  15.49           $
\enddata
\end{deluxetable*}
There are existing VLBI observations of this OHM in \citet{Pihlstrom2005} that
were not discussed in R08. For this reason, Table \ref{tab:12032} is reproduced
from that paper. The emission is compact on scales of less than
100~pc, and all of the single dish flux is recovered in this region, within
error. There is a clear north-south velocity gradient, with the bluest
peak located in the north. \citet{Pihlstrom2005} noted a peak centered at 
64,723~$\kms$ that appeared asymmetric in the R08 observations. 
R08 fit this peak with two Gaussian components, at velocities of 64,737~$\kms$ 
and 64,715~$\kms$. The Gaussian components, when used to fit the Stokes~$V$
spectrum, yielded fields of 10.9$\pm$1.7~mG and 17.9$\pm$0.9~mG, respectively.

{\bf IRAS F12112+0305}: In comparing the old and new spectra for this OHM,
shown in Figure \ref{fig:12112},
the most obvious difference is the appearance of weak 
interference spikes in the older spectra that were not a 
problem in the more recent observations.
There is also astrophysical change, as the brightest feature in the spectrum 
for IRAS~F12112+0305 has apparently dimmed by $\sim$10~mJy, relative to a flux
density of $\sim$100~mJy. At the same frequency as the change in the Stokes~$I$
spectrum, the Stokes~$V$ spectrum has also changed, as there is now a broad
dip associated with the peak of the maser emission. 

Despite this new dip, combined fits to the Stokes $I$ and $V$ profiles do not
produce any evidence of magnetic fields. 
At the frequencies corresponding to 1667~MHz emission, 
the more recent spectra have comparable noise, and errors
in derived Gaussian fits to the structure, as the original observations.
We find a comparable upper limit of $\sim$3 mG for magnetic
fields in the masing regions. 

{\bf IRAS F14070+0525}: 
R08 were only able to only place upper limits on magnetic fields in this OHM. 
Our Stokes $I$ and $V$ spectra are roughly a 
factor of two noisier than those presented in R08. 
Consequently, no Stokes $V$ features are detected, and the upper
limits we may place on magnetic fields are less stringent. Comparing the old
and new spectrum, there is no evidence for variations with amplitudes greater
than 3~mJy.

{\bf IRAS F15327+2340} (Arp 220): 
Variability has previously been observed in Arp~220 by \citet{Lonsdale2008}.
In Figure \ref{fig:15327}, our spectra are compared with those of R08, and 
show strong support for claims of variability.
In the Stokes~$I$ spectrum, narrow features between 1637.5--1638.3~MHz
vary by 10--15~mJy between the old and new spectra, some dimming and some
brightening, as well as a slight brightening of the ``shoulder'' at 1638.55~MHz.
Given the brightness of Arp~220, these difference represent only
$\sim$1\% variations in the flux density. 
To the eye, the most noticeable difference in the Stokes~$V$ spectrum 
is a $\sim$0.5~MHz 
hump that falls in the same frequency range as the changes in the Stokes~$I$
structure, but it is difficult to make out clear differences between the
narrower peaks. 

Overall, when applying the same fitting routine to the newer spectra
and comparing to the older spectra, the features are largely the same, 
with fitted components having center frequencies, widths, amplitudes
and associated magnetic fields that are the same within error of the fits. 
The one difference that does appear is in a Gaussian 
at 1638.3~MHz, which in the original spectrum
fit with a magnetic field with strength $2.03 \pm0.76$~mG, whereas the more
recent spectrum was fit with a field with strength $7.29 \pm 1.43$~mG. 
Upon careful inspection, the more recent Stokes~$V$ spectrum is
deeper at that frequency, and the difference has an amplitude larger than the
features in the difference spectrum at frequencies not associated with the 
Stokes~$I$ spectrum. Nonetheless, given other amplitude variations in the
Stokes~$V$ spectrum nearby, we do not consider this strong evidence of
variation in the magnetic field strength.

\subsubsection{Stokes $I$ non-detections}
Eight sources previously reported as OHM hosts were not detected in our
survey. In most cases, non-detections are due to increased RFI since the
original discovery of these sources. 

For non-detected sources, upper limits are 
calculated according to the prescription 
in DG02, with
\begin{equation}
	\loh^{\rm max} = \left(\frac{1}{2}\right)4 \pi D_L^2 1.5\sigma \left(\frac{\delta v}{c}\right) \left(\frac{\nu_o}{1 + z}\right). \label{eq:limit}
\end{equation}
This assumes a boxcar line profile of height 1.5 $\sigma$, which is the
RMS noise in the Stokes $I$ spectrum, and a rest frame
velocity line with of 150 $\kms$, which is the 
average FWHM of the 1667~MHz line of the known OHM sample. 
To calculate the luminosity distance, $D_L$, we use the results from WMAP5 and 
assume $H_0 = 70.5$ $\kms$ Mpc$^{-1}$, 
$\Omega_\Lambda$ = 0.726, and $\Omega_M$ = 0.274 \citep{Hinshaw2009}. The
factor of $\frac{1}{2}$ appears because we use the classical definition of 
Stokes $I$, equal to the sum of the orthogonal polarizations.

In some cases, comparison with previously published work is difficult.
\citet{Bottinelli1987}, \citet{Martin1989b}, and \citet{Bottinelli1990}
do not explicitly state the cosmology used to calculate
the isotropic luminosities they reported. For comparison with the values 
reported in those instances, we follow \citet{Martin1988}, 
who used $H_0 = 75 \kms$ Mpc$^{-1}$ and a deceleration parameter, $q_0 = 0$. \\

\begin{figure*}[t!]
    \begin{center}
    \subfigure[IRAS 08071+0509: Zeeman splitting non-detection.]{ 
        \includegraphics[width=3.4in] {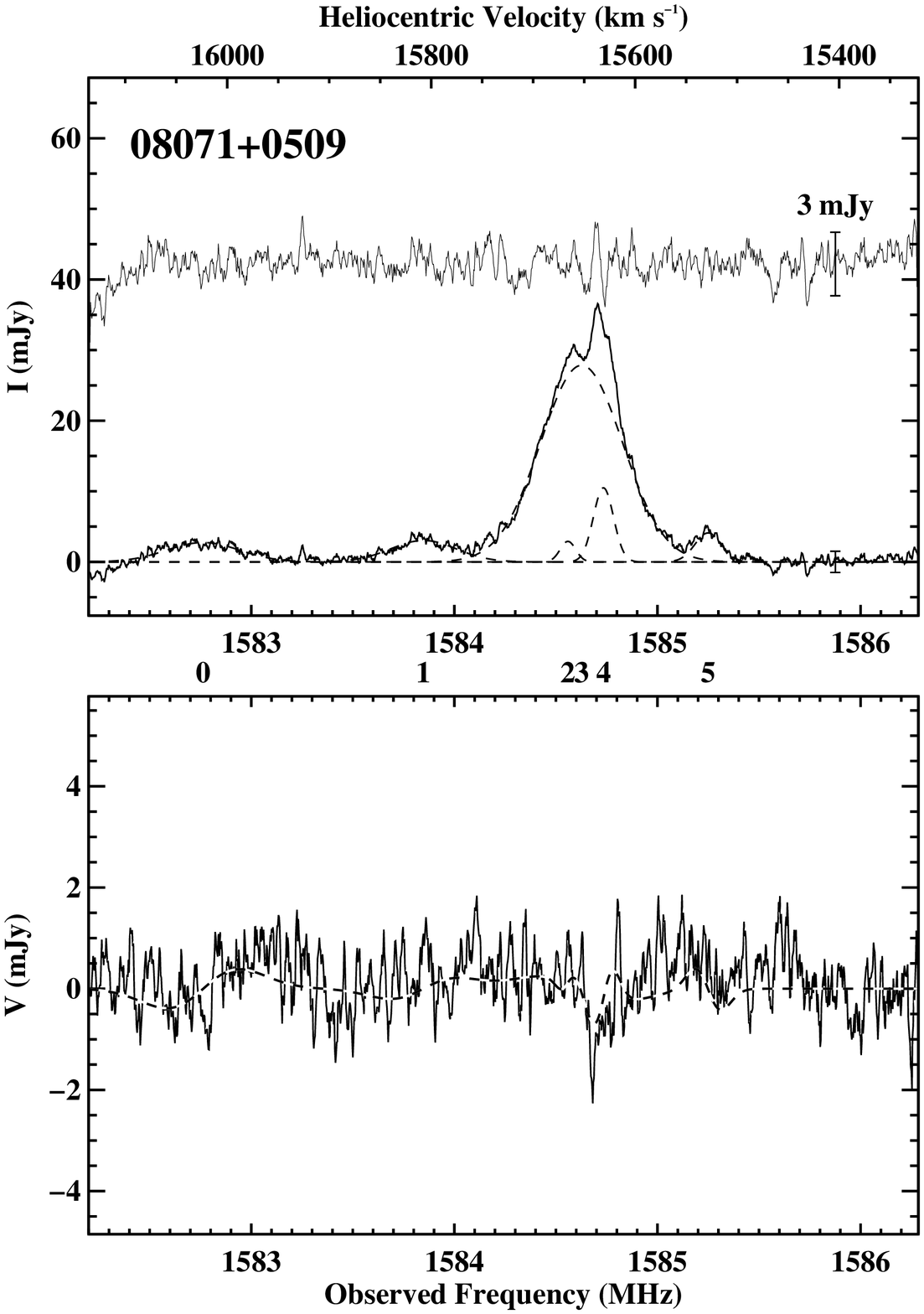}
        \label{fig:08071}
    }
    \subfigure[IRAS F10339+1548: Zeeman splitting non-detection.]{
        \includegraphics[width=3.4in] {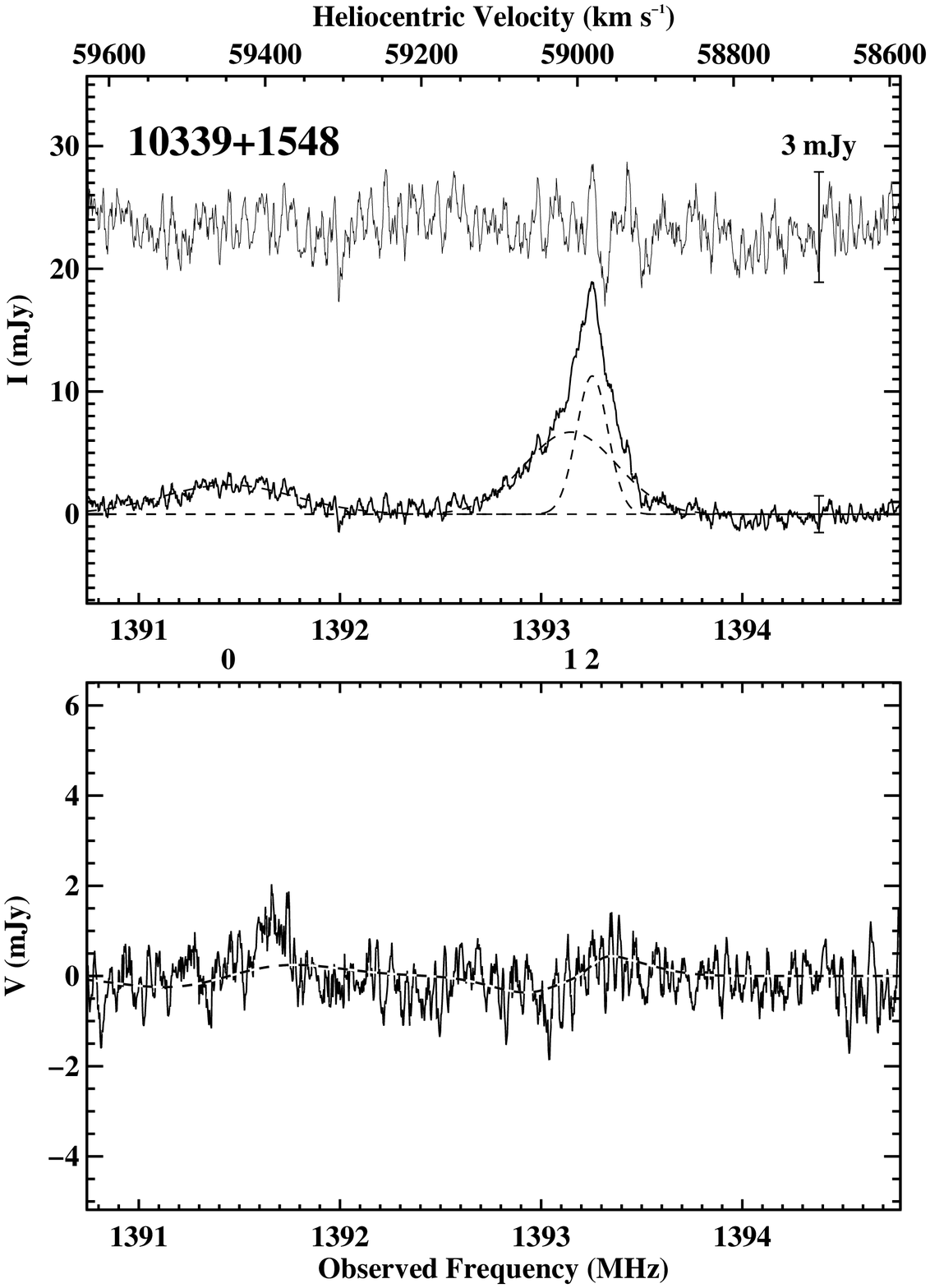}
        \label{fig:10339}
    }
\captcont{Each sub-figure shows Stokes $I$ and $V$ spectra in two panels.
    Sub-figures (a), (b), and (c) show non-detections of Zeeman splitting. 
    Sub-figures (d), (e), and (f) are considered marginal Zeeman splitting
    detections. The remaining sub-figures are new Stokes $V$ detections. 
    In each sub-figure, the {\em top panel} shows the Stokes $I$ spectrum for 
    each source. The flux density is defined as the sum of the orthogonal 
    polarization, $I = RHCP + LHCP$, rather than the average. The data are 
    shown by the solid line. The dashed lines are Gaussian components used to 
    fit the overall profile. The thin solid line shows the residuals of the 
    fits, expanded by a factor of three. 
    The {\em bottom panel} in each sub-figure shows the Stokes $V$ profile. 
    The data are indicated by the solid line. The fit, produced using the 
    Gaussians in the top panel as inputs, is shown with a dashed line.}
\label{fig:newiv1}
\end{center}
\end{figure*}

\begin{figure*}
\begin{center}
    \subfigure[IRAS F23028+0725: Zeeman splitting non-detection.]{
        \label{fig:23028}
        \includegraphics[width=3.4in] {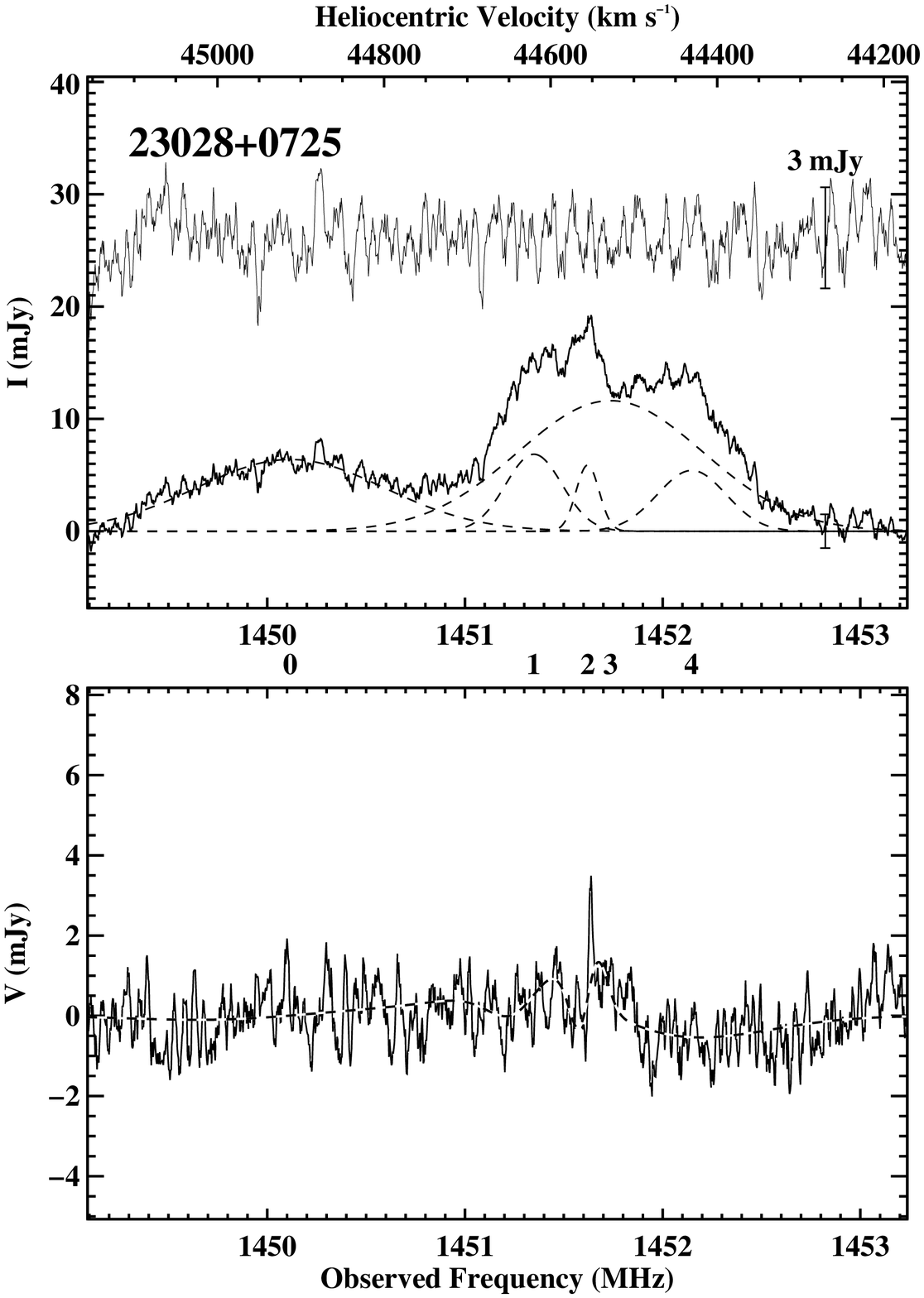}
    }
    \subfigure[IRAS F15224+1033: Marginal detection of Zeeman splitting.]{
        \label{fig:15224}
        \includegraphics[width=3.4in] {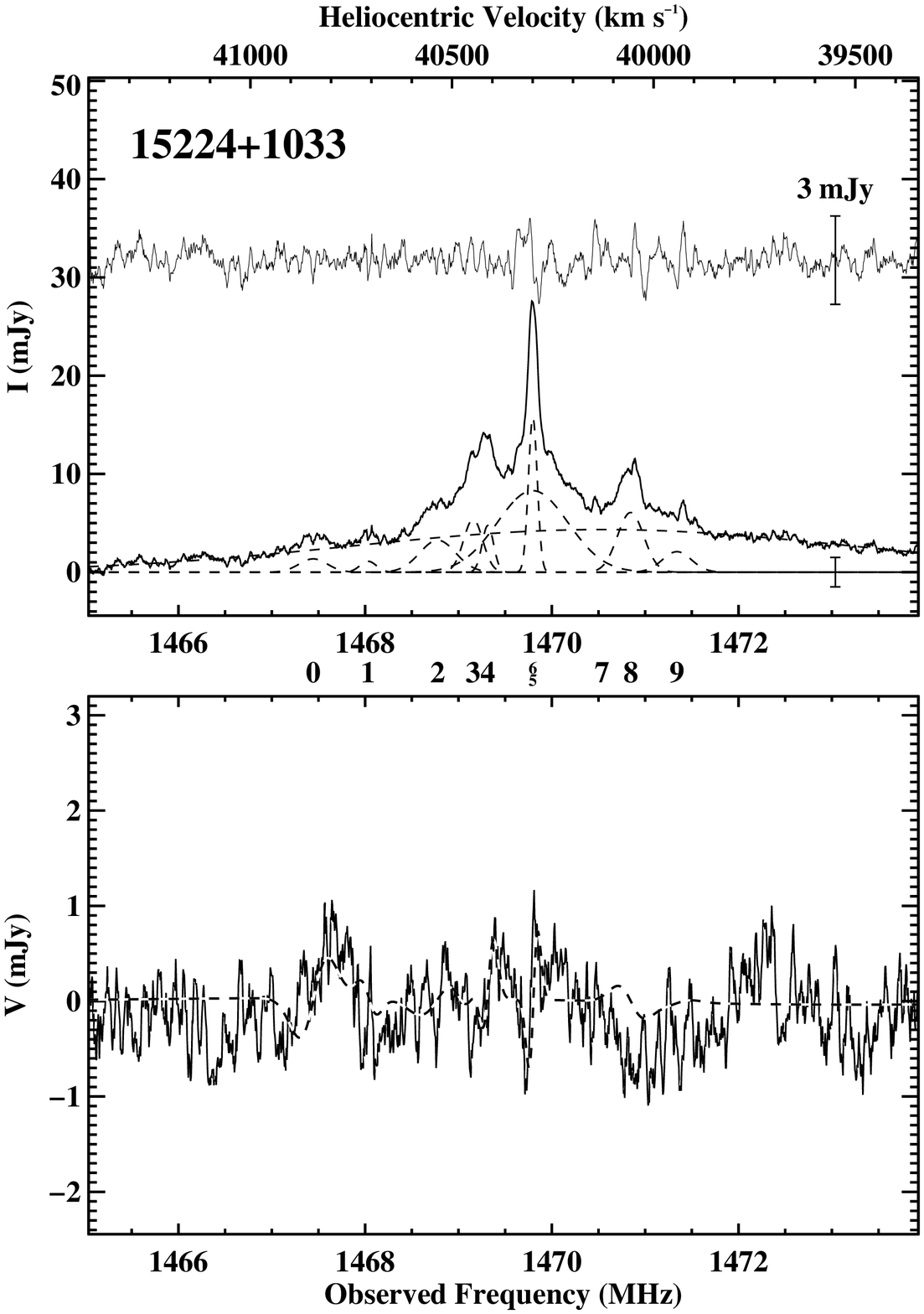}
    } \\
    \subfigure[IRAS F15587+1609: Marginal detection of Zeeman splitting.] {
        \label{fig:15587}
        \includegraphics[width=3.4in] {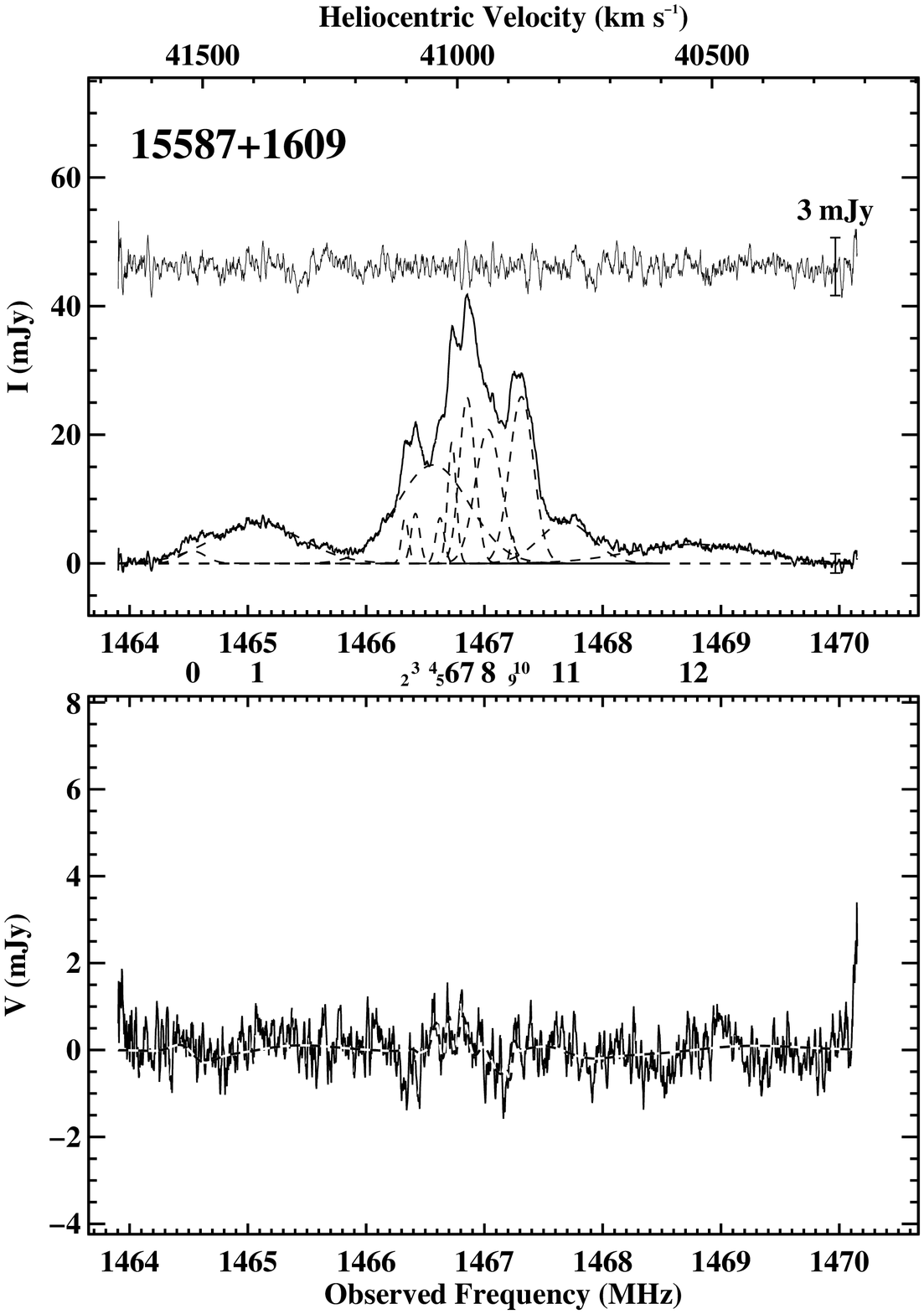}
    }
    \subfigure[IRAS F20550+1655: Marginal detection of Zeeman splitting.]{
        \label{fig:20550}
        \includegraphics[width=3.4in] {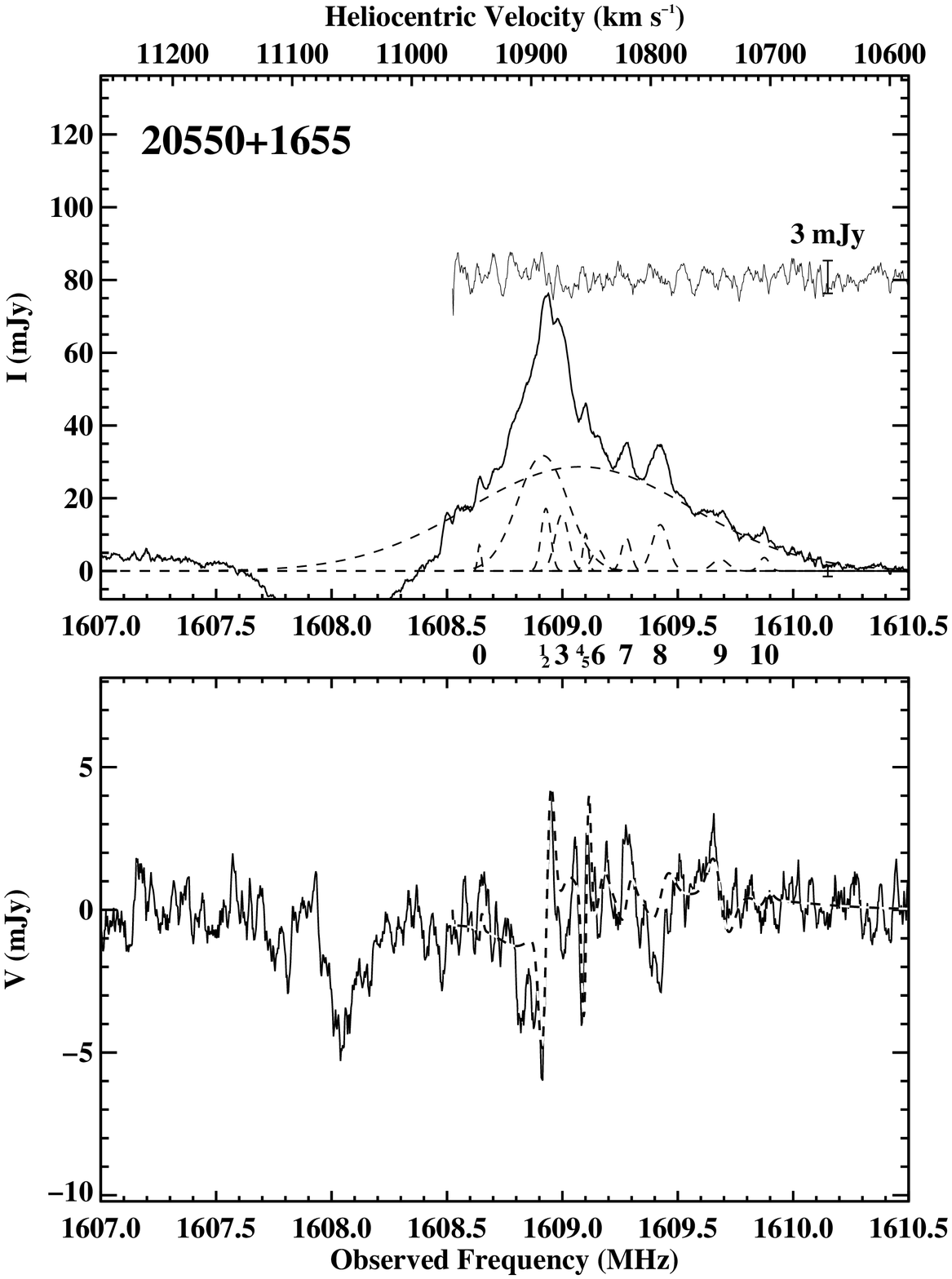}
    }
    \captcont{(continued)}
    \label{fig:newiv2}
\end{center}
\end{figure*}

\begin{figure*}
\begin{center}
    \subfigure[IRAS F02524+2046: Zeeman splitting detection.]{
        \label{fig:02524}
        \includegraphics[width=3.4in] {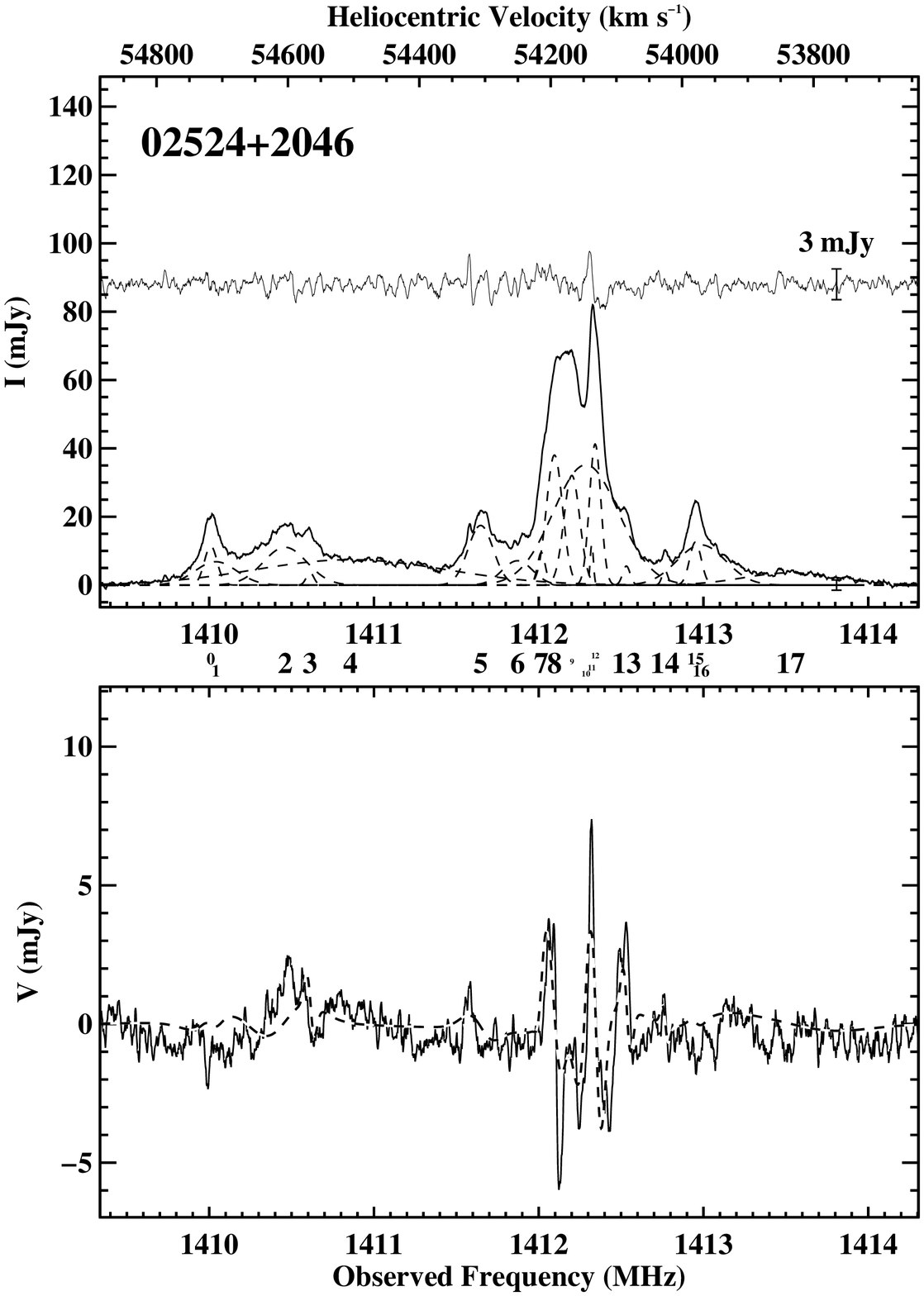}
    }
    \subfigure[IRAS F04332+0209: Zeeman splitting detection. Instead of showing fits to Stokes $I$, the $RHCP$ (dashed) and $LHCP$ (dash-dot) are plotted in the top panel, and $RHCP - LHCP$ for the fitted Gaussian is shown in the bottom panel.]{
        \label{fig:04332}
        \includegraphics[width=3.4in] {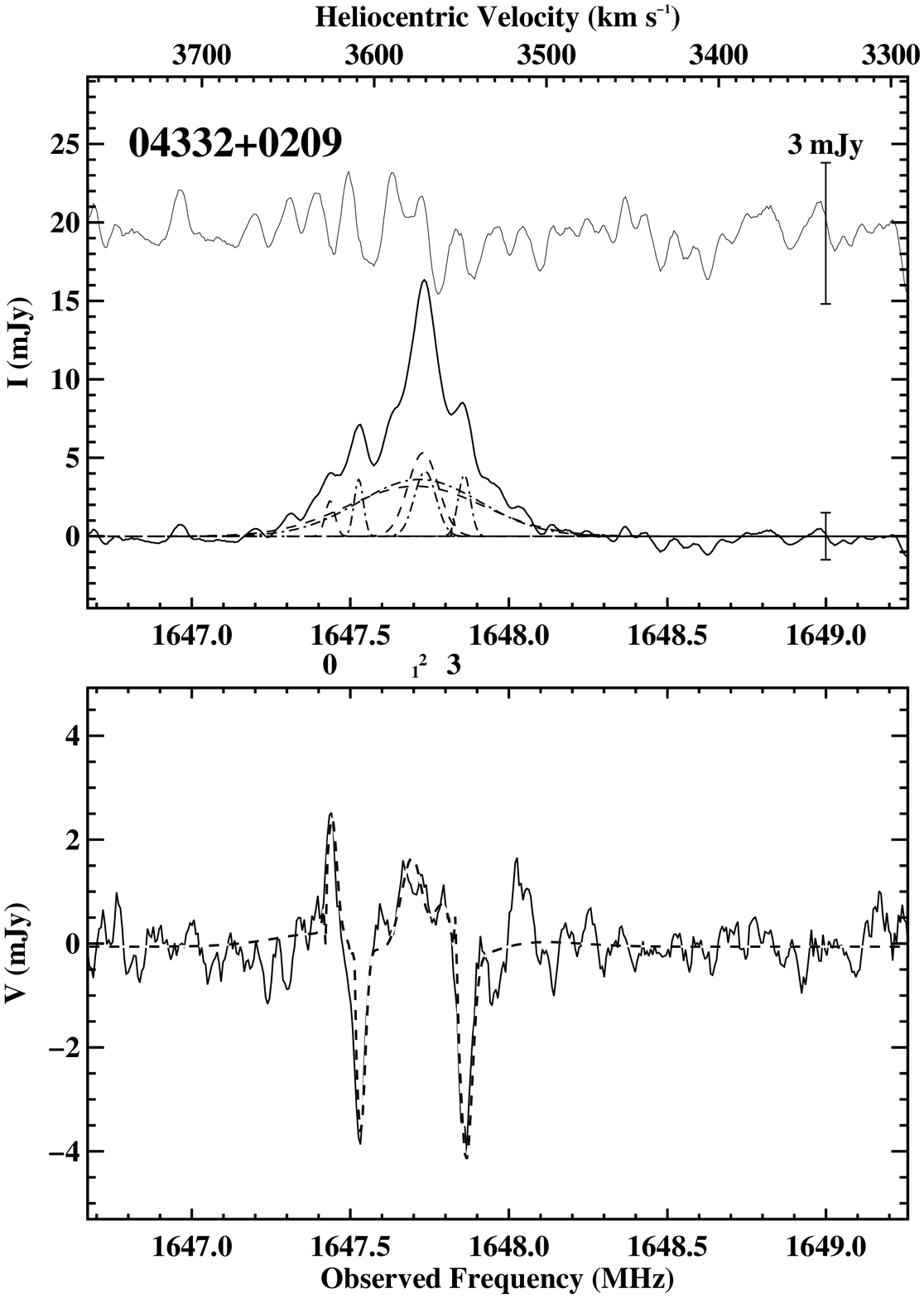}
    } \\
    \subfigure[IRAS F09039+0503: Zeeman splitting detection.]{
        \label{fig:09039}
        \includegraphics[width=3.4in] {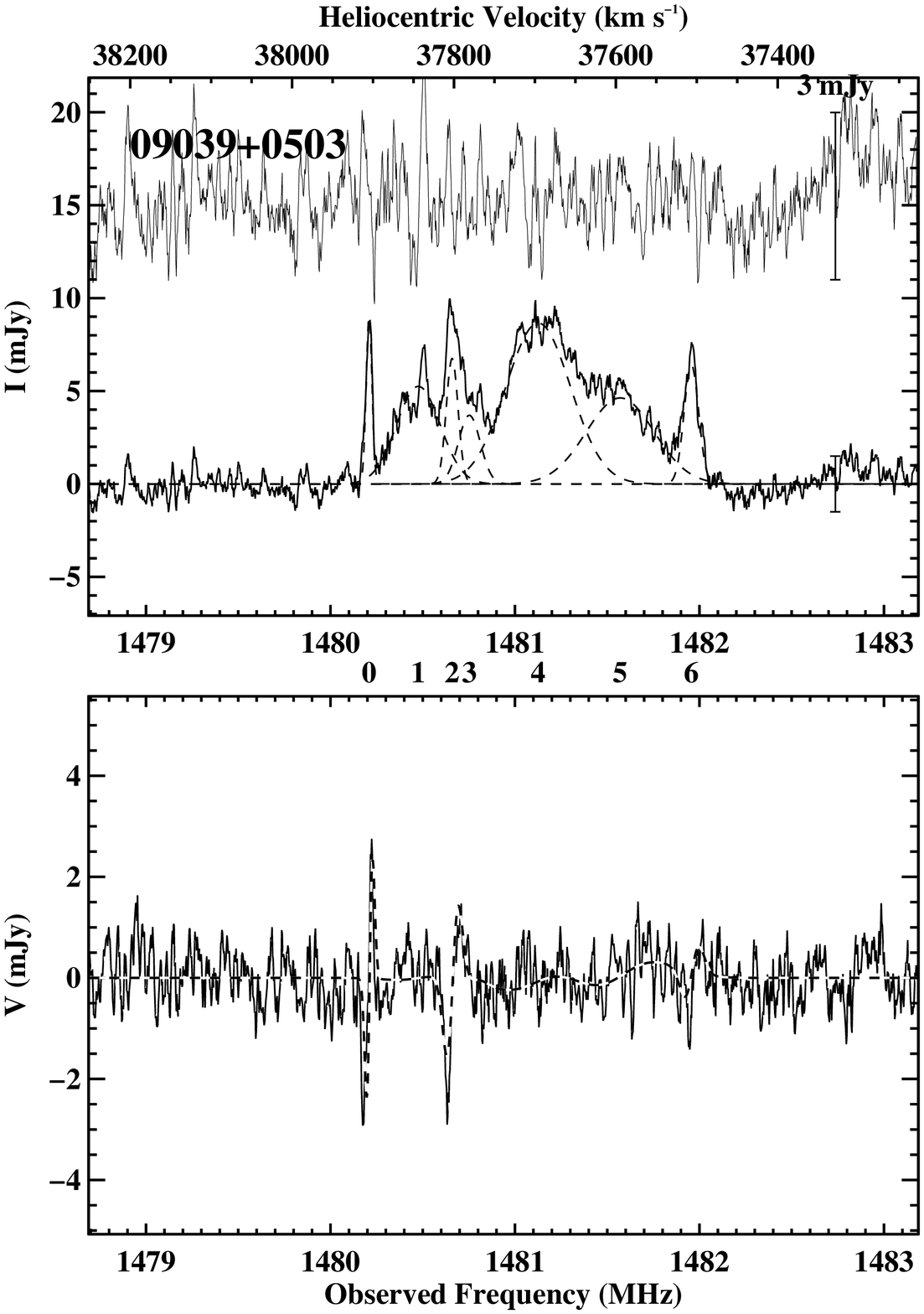}
    }
    \subfigure[IRAS F10378+1108: Zeeman splitting detection.]{
        \label{fig:10378}
        \includegraphics[width=3.4in] {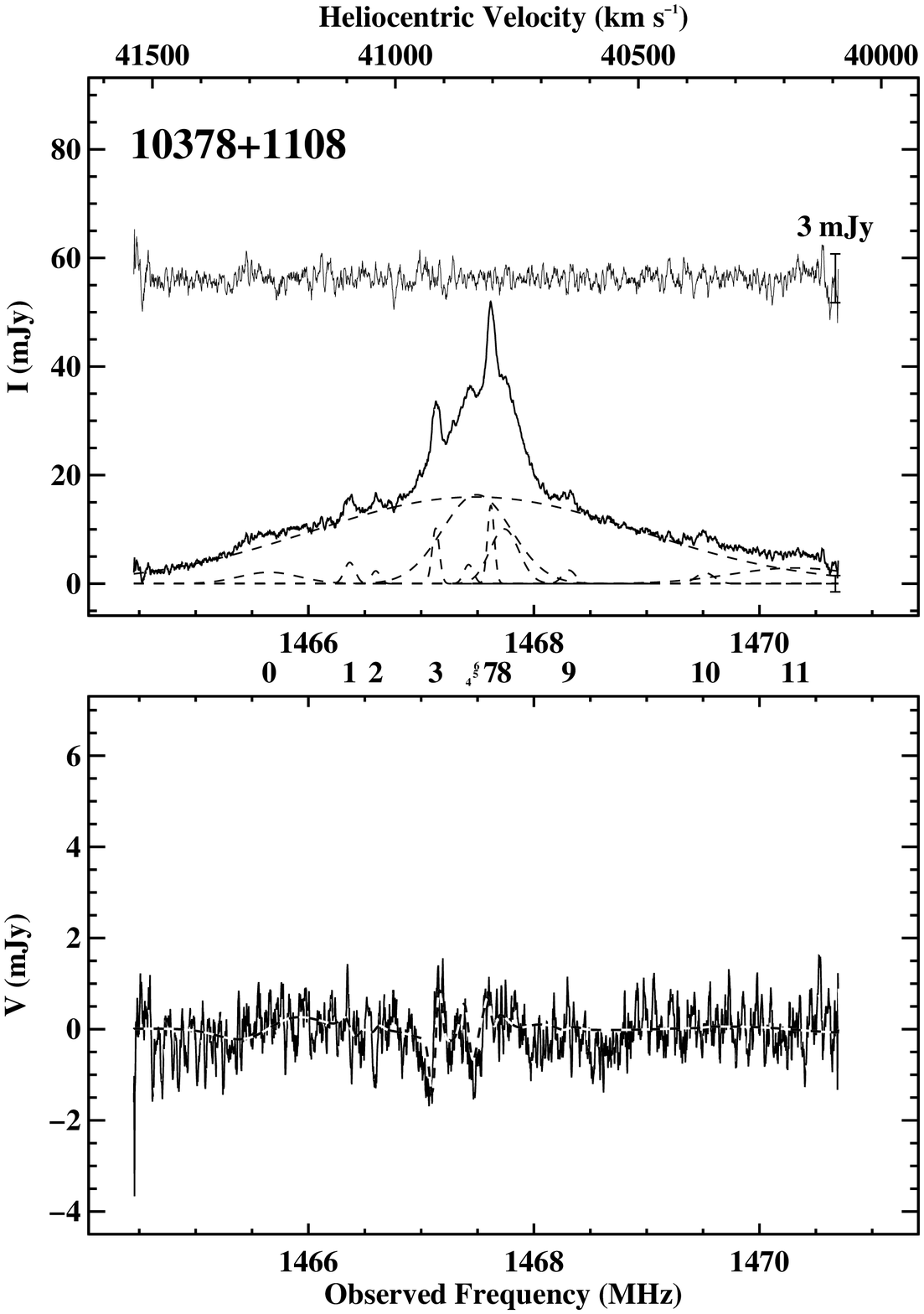}
    }
    \captcont{(continued)}
    \label{fig:newiv3}
\end{center}
\end{figure*}

\begin{figure*}
\begin{center}
    \subfigure[IRAS F16255+2801: Zeeman splitting detection.]{
        \label{fig:16255}
        \includegraphics[width=3.4in] {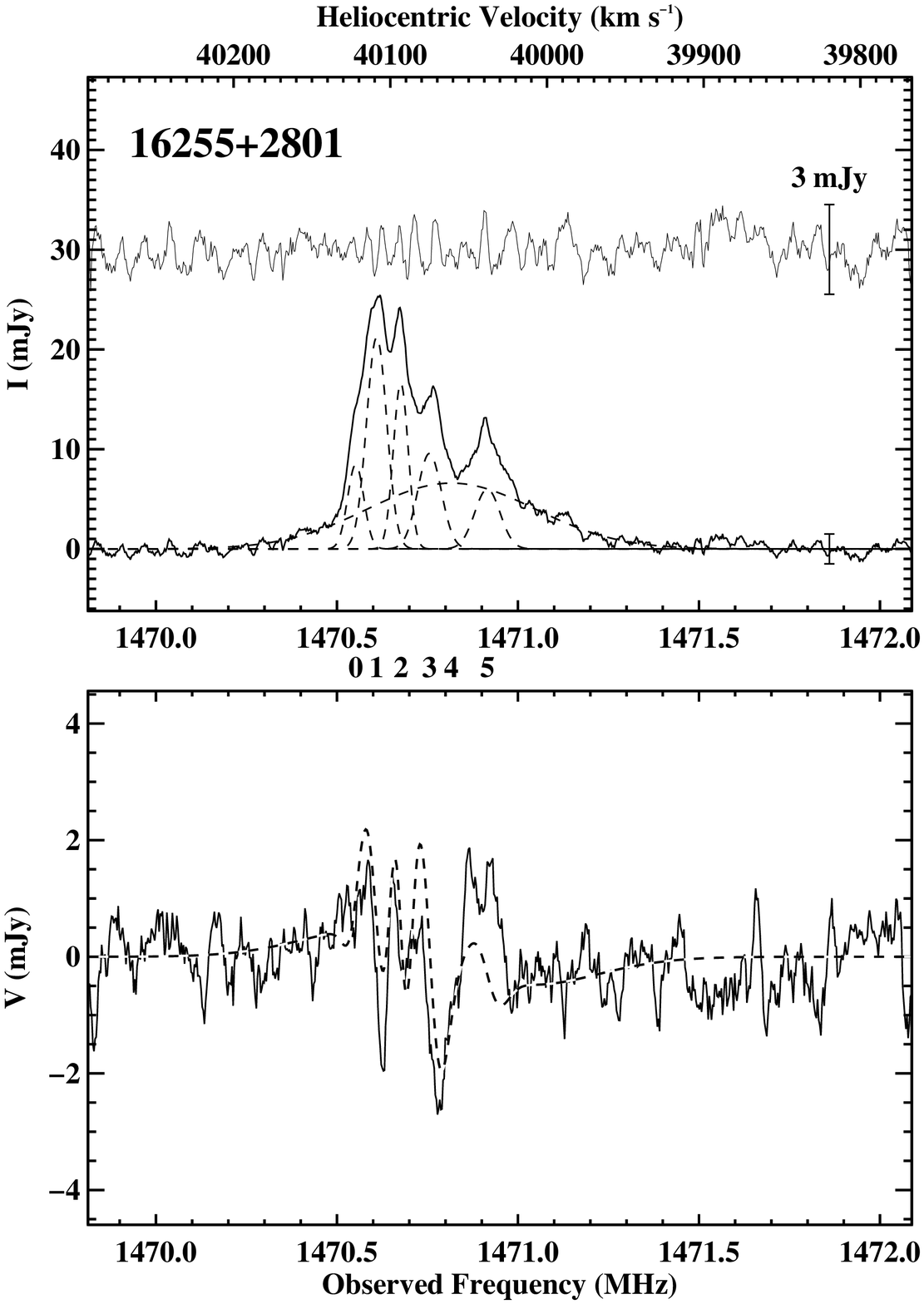}
    }
    \subfigure[IRAS F18368+3549: Zeeman splitting detection.]{
        \label{fig:18368}
        \includegraphics[width=3.4in] {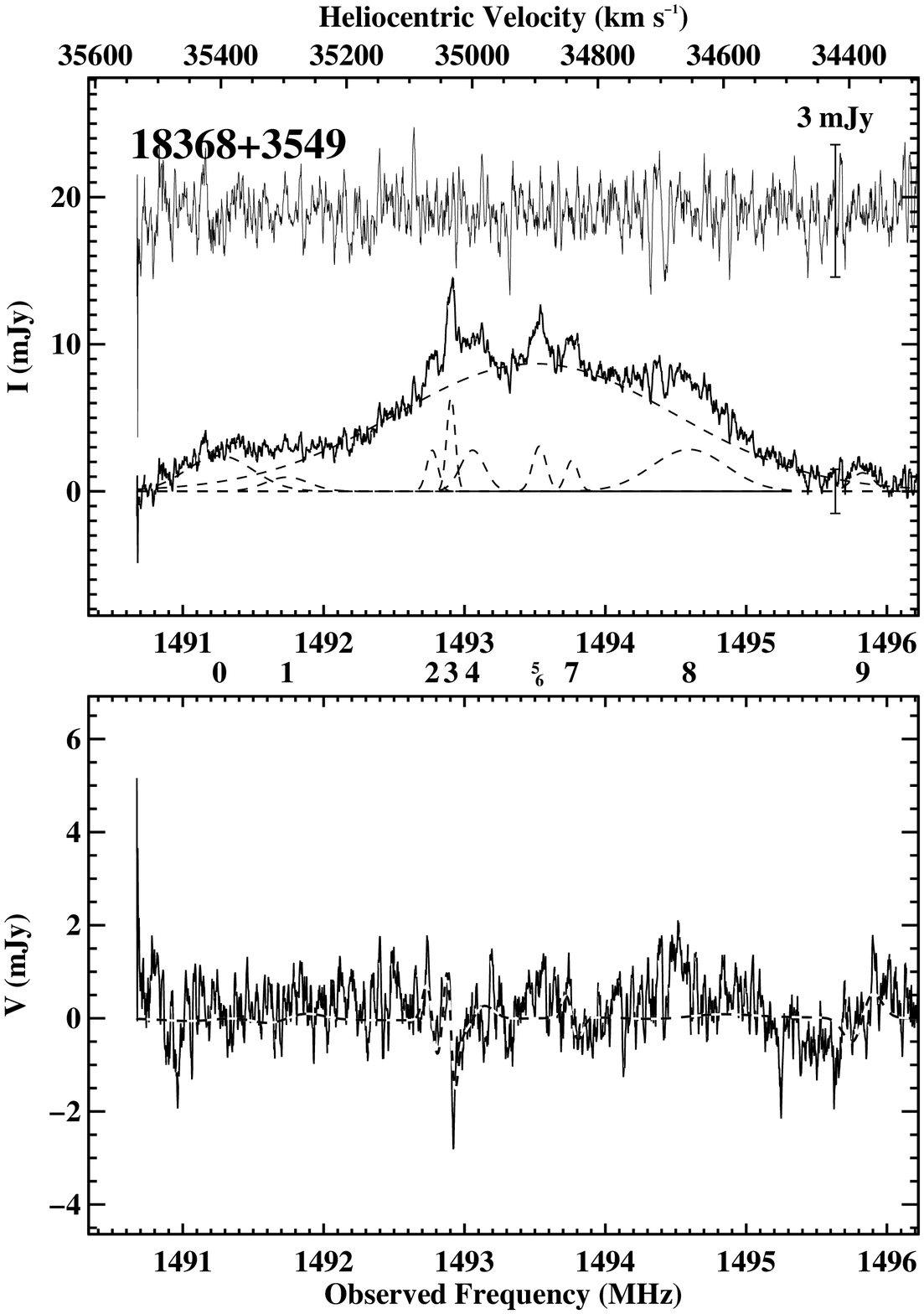}
    } \\
    \subfigure[IRAS F18588+3517: Zeeman splitting detection.]{
        \label{fig:18588}
        \includegraphics[width=3.4in] {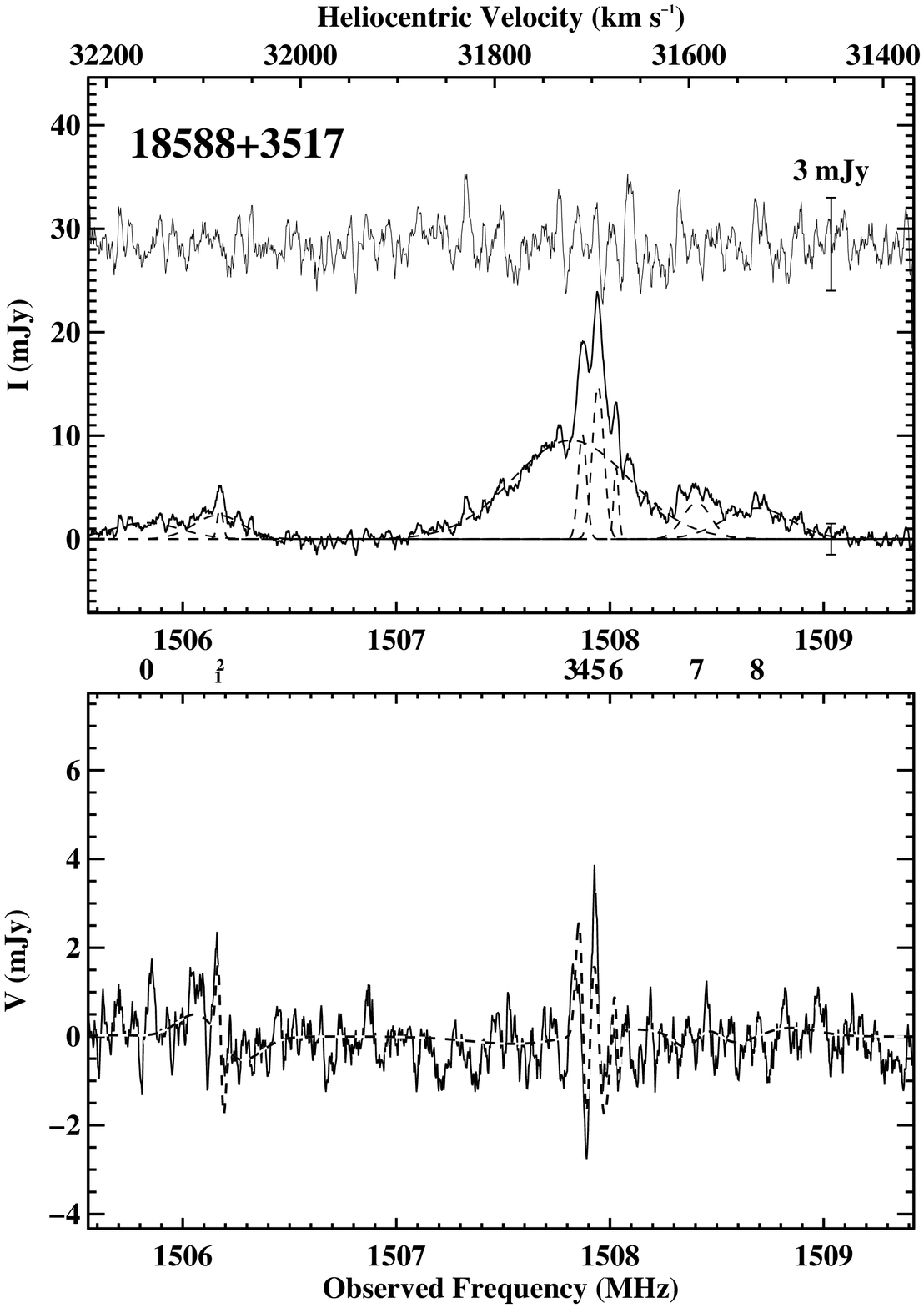}
    }
    \subfigure[IRAS F22134+0043: Zeeman splitting detection.]{
        \label{fig:F22134}
        \includegraphics[width=3.4in] {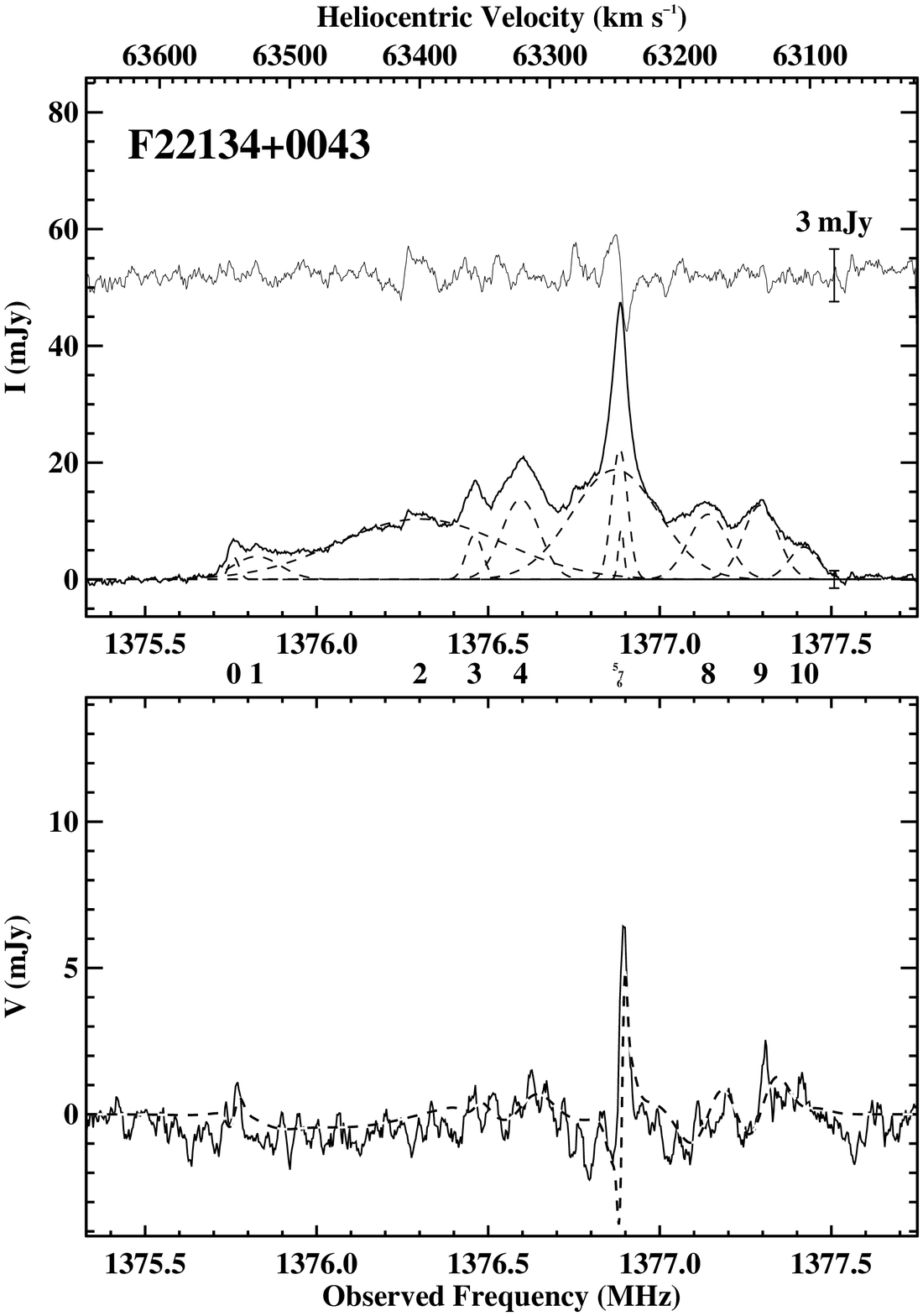}
    }
    \caption{(continued)}
    \label{fig:newiv4}
\end{center}
\end{figure*}

{\bf IRAS F00509+1225}: 
This OHM was first reported by 
\citet{Bottinelli1990}, with $\lloh = 1.78$, but has no published spectrum. 
Given the reported redshift $z_\odot = 0.060875$, the likely luminosity
distance they used is 243~Mpc. For the \citep{Hinshaw2009} cosmology, the 
luminosity distance at this redshift is 271~Mpc, 
implying an isotropic luminosity of $\lloh = 1.88$. At this redshift, 
maser emission should be expected near 1571 MHz. While the composite spectrum
has features near this frequency, those features completely disappear or
even appear as absorption in bootstrap resampling, suggesting they are 
instead due to interference from GPS L2 that is centered at 1575.42 MHz.
The rms error is 5~mJy, so we place an upper limit $\lloh \leq 1.82$ 
on the isotropic luminosity of OHM emission, weakly suggesting that 
this source is not actually an OHM. 

{\bf IRAS F03056+2034}: 
\citet{Bottinelli1990} reported this OHM with a an isotropic luminosity of 
$\lloh = 1.28$. Located at a redshift of $z = 0.027369$, the 
expected maser emission is in a frequency range dominated by
spiky interference likely from the {\em Iridium}
satellite, so much so that even the median composite 
spectrum had an rms of 15~mJy. From this, we place
an upper limit on the isotropic luminosity of the OHM emission that we
could detect of $\lloh \leq 1.60$. This is larger
than the previously reported value, so the
non-detection is unsurprising.

{\bf IRAS F11069+2711}: No spectrum of this OHM is available, nor 
are there any published properties. It was listed as a megamaser by 
\citet{Baan1998}. There is spiky interference in our spectra, so we took the 
median of shorter spectra to produce the total spectrum.
\citet{Baan1998} assumed a redshift for this source of 
of $z = 0.0703$, which places the expected frequency for OHM emission
at 1557.8 MHz. There is a roughly 5~mJy feature centered around 
1557.5 MHz, but this is likely due to terrestrial interference, as it 
appears intermittently in spectra both on and off source. 
A more recent redshift measure by \citet{Lawrence1999} puts the galaxy at
a redshift of $z = 0.072971$, shifting the expected frequency for OHM emission
to 1554 MHz. At this frequency, there is nothing resembling
emission. Outside of the interference at 1557.5 MHz, there is 
an rms error of 2.3~mJy, which we use to place an upper limit 
on the OHM luminosity of $\lloh < 1.65$. 

{\bf IRAS F13451+1232}: 
With a 5.4~Jy continuum flux at 1.4~GHz, this OHM is the strongest
continuum source in our sample. A tentative OH detection
was claimed by \citet{Dickey1990}, who reported 
a flux of 1.7~mJy and an isotropic
luminosity of $\lloh$ = 2.38. While most spectra in our sample have
rms errors of order 1~mJy, the combination of strong continuum emission and
interference in the spectrum for this source resulted in an rms error
of $\sim 20$~mJy, so we can neither confirm nor 
refute this tentative detection. 

{\bf IRAS F15233+0533}: \citet{Baan1998} listed this as a megamaser, but
no published spectrum or properties are available. At a redshift of 
$ z = 0.054064$, OHM emission is expected at 1581.8 MHz, which is 
near interference from GPS L2 located
at 1575.42 MHz. There are no features in the final spectrum that are 
clearly persistent. Bootstrap resampling of the 4-minute spectra shows that
features in the composite spectrum are not clearly persistent. 
With an rms error of 20~mJy, we place an upper limit
of $\lloh < 2.33$ for OHM emission from this galaxy. 

{\bf IRAS F17526+3253}: 
This source was first observed for
OH emission by \citet{Garwood1987}, who reported a non-detection with a
spectrum that had 1.0~mJy rms noise, which corresponds to an upper limit on
the isotropic luminosity of $\lloh < 0.60$ for the chosen cosmology. 
An OHM was reported two years later by
\citet{Martin1989b}, with an isotropic luminosity of $\lloh = 0.99$, but
no accompanying flux or spectrum was published.
In our spectrum, interference from the {\em Iridium} satellite is 
severe. After combining the median spectra, the rms error is  
$\sim$10~mJy, so we do not detect OHM emission, 
and cannot place any useful new limit on OHM emission from this galaxy. 

{\bf IRAS F20491+1846}: 
This OHM was first reported by \citet{Bottinelli1989}. There
is no published flux, and the published isotropic luminosity is $\lloh = 1.09$ 
at a redshift of 0.0290. Even looking at the median spectrum, 
spiky interference from the {\em Iridium} 
satellite dominates. With roughly 30~mJy rms error in the spectrum, the
upper limit on the isotropic luminosity is $\lloh = 1.96$, well above the
reported luminosity.

{\bf IRAS F23135+2517}: Interference from the {\em Iridium} satellite is
present in the spectrum for this source. Combining the median of the 
median spectra, the rms error in the spectrum is $\sim 20$~mJy.  
\citet{Mirabel1987} reported a flux of 2.4~mJy for this source, 
corresponding to 4.8~mJy for our definition, so the OHM is not detected. 

\subsubsection{Ambiguous Stokes $I$ detections}
{\bf IRAS F15250+3608}: This OHM was reported in an IAU Circular 
by \citet{Bottinelli1987} to have an isotropic luminosity of $\lloh = 2.58$.
At a redshift of $z = 0.0554$, OHM emission is expected at
roughly 1580 MHZ, near interference
from GPS L2 at 1575.42 MHz. There is at least one
persistent feature near 1581 MHz, and perhaps more than one, but
these are mixed among ripply interference that clearly fluctuates. 
It is therefore not possible to reliably estimate any of the OHM 
properties. 

{\bf IRAS F23050+0359}: 
This OHM has no published spectrum, or properties, but 
was listed as an OH megamaser by \citet{Baan1998}. 
It is located at a redshift of 0.0474, putting the OH lines at a frequency
near a strong source of interference, likely ringing from GPS L2, 
that contaminates the low frequency side of the spectrum. Even so, there
appears to be a detection of the 1667~MHz line, and possibly the 1665~MHz
line. The Stokes~$V$ spectrum has significant rippling, which prevents
any possible Zeeman splitting detection or placement of useful limits on
magnetic fields in this OHM.

{\bf IRAS F23365+3604}: \citet{Bottinelli1990} reported this OHM to have
an isotropic luminosity of $\lloh = 2.44$, but did not publish any other
properties of the source. Subsequently, 
\citet{Baan1992} cited the upper limit on the isotropic 
luminosity of the 1667~MHz line to be $\lloh < 1.93$,
assuming a 100 $\kms$ linewidth.
\citet{Darling2002a} considered the source suspect. 
The source is at a redshift of 0.064531, which puts the 1667~MHz OH line
at a frequency of 1566.3 MHz. The composite spectrum created by taking the
mean of individual 1-second spectra shows significant broadband interference.
Median combination of individual spectra to create a composite removes most
features, leaving what appears to be a reasonably flat spectrum with 
broad, low SNR emission from 1566.5--1568 MHz, plausibly close to 
the expected frequency of the OH line, and would correspond to
an isotropic luminosity of $\lloh = 2.56$.
Spectra of the 1612 MHz line of IRAS F01417+1651, another source observed
in this survey, spanned the frequency range 1566.1--1572.3 MHz. Examination
of these spectra showed features of comparable shape and intensity 
to those that were appeared in the
median composite spectrum for IRAS F23365+3604, strongly suggesting that
the apparent emission is actually interference.

\subsubsection{Stokes $V$ non-detections} \label{sec:non_v_dets}

For the sources described in this section, maser emission was detected, but due to low signal-to-noise in the Stokes $I$ and Stokes $V$ spectra, it was not possible to identify any features in the Stokes $V$ as being associated with features in the Stokes $I$ spectrum. For the majority of sources, we do not present the spectrum, instead providing a reference where a published spectrum can be found. Exceptions are made for three sources, IRAS~F08071+0509, IRAS~F10339+1548, and IRAS~F23028+0725. For IRAS~F08071+0509, no Stokes $I$ spectrum exists in the literature. In the case of IRAS~F10339+1548, our data provided an improvement over the existing Stokes $I$ data, allowing determination of the hyperfine ratio, 
which is defined as the ratio of the flux in the 1667~MHz main line to the flux in the 1665~MHz main line, or $R_H = F_{1667} / F_{1665}$. IRAS~F23028+0725 is included as an example of a source where our fitting routine reported marginally significant detection of Zeeman splitting that we do not find credible.  

Where possible, we estimate an upper limit to the magnetic field strength associated with the masing regions. Following \citet{Troland1982}, the upper limit is taken to be the fitted field strength plus three times the error associated with the fitted field. Most OHMs are fit with multiple Gaussian components; for sources where an upper limit is reported, the value is that of the Gaussian component with the lowest associated error that appeared to be real. This is based on the shape and strength of features in the Stokes $I$ spectrum, and the structure and amplitude of noise in the Stokes $V$ spectrum. For example, limits associated with Gaussians fit to wide features in spectra are not reported, as they are likely bandpass features. We consider only one digit in these upper limits to
be significant, but reported two digits for sources with upper limits between
10--20~mG. \\

{\bf IRAS F01562+2528}: This OHM, reported in DG02, has three broad, 
blended features, one of which is
at the expected frequency of the redshifted 1665~MHz line. 
Despite the well-detected
signal in Stokes $I$, there are no features in the Stokes $V$ spectrum that look
promising. Given the absence of any narrow structures, it is not possible
to constrain the magnetic field in any meaningful way. 

{\bf IRAS 03521+0028}: DG02 discovered this OHM, noting the excellent agreement
between the peak of the 1667~MHz emission and the published optical redshift,
as well as marginal detection
of the 1665~MHz line. Our spectrum has an additional narrow feature at
1448.3 MHz not seen in the DG02 observations.
Examination of multiple spectra for different objects that 
have observations in this frequency range show that this new, narrow feature 
is interference. No features in the Stokes $V$ spectrum can be identified as
associated with the astrophysical Stokes $I$ emission, and the lack of bright,
narrow emission prevents placing a useful limit on the magnetic field in this
OHM. 

{\bf IRAS F03566+1647}: This OHM was reported by DG02, and features broad, low amplitude emission. 
The Stokes $V$ spectrum has no apparent features, and the broad
Stokes $I$ features do not provide useful limits on the magnetic fields in
this source. 

{\bf IRAS F04121+0223}: The spectrum for this OHM discovered by DG01 
contains one narrow component centered at 1485.20 MHz and one broad feature
centered at 1486.04 MHz. 
The Stokes $V$ spectrum does not show any features associated with the Stokes
$I$ emission, and we place an upper limit of 40~mG on the magnitude of 
magnetic fields associated with the narrow emission in this OHM. 

{\bf IRAS 06487+2208}: DG00 reported this OHM, featuring two main 1667~MHz
peaks centered at 1457.7 MHz and 1458.3 MHz, 
and weak, corresponding 1665~MHz emission. 
There is a great deal of structure in the Stokes $I$
spectrum, requiring a total of nine Gaussians to produce a good fit. 
The Stokes $V$ spectrum features 1~mJy ripples, but no apparent Zeeman
splitting signature. We place an upper limit of 10~mG on the magnetic field 
associated with the narrow emission at 1457.7 MHz as well as the broader
emission centered at 1458.3 MHz. 

{\bf IRAS 07163+0817}: This OHM discovered by DG01 features three distinct
1667~MHz peaks, one of which is centered at 1501.3 MHz and is about twice as
bright as the other two peaks. The Stokes $V$ spectrum does not have any
notable features, and we place an upper limit of 20~mG on the magnitude of a
magnetic field associated with the narrow line at 1501.3 MHz.

{\bf IRAS 07572+0533}: Only one broad feature can be distinguished for this
OHM discovered by DG01. They noted an apparent second OH line 
offset blueward from the main feature by 400 $\kms$, but we do not confidently
detect this. 
Without any narrower lines, it is not possible to place useful limits on the
magnetic fields in this source. 

{\bf IRAS F07556+2859}: This OHM was discovered by \citet{Willett2012}, and
will be discussed in more detail in a future work by Willett, Darling, Kent, 
\& Braatz. The Stokes $I$ spectrum features broad, blended 1667~MHz emission
and weak 1665~MHz emission. Without any distinct narrow emission, we do not
place any limits on magnetic fields in this OHM. 

{\bf IRAS F08071+0509}:
\citet{Bottinelli1989} reported this OHM with $\lloh = 2.25$, but did
not publish a spectrum or the flux of the 1667~MHz line. \citet{Baan1998}
classified it as a composite AGN/starburst in a survey to optically classify
megamaser galaxies. 
We clearly detect the maser, but there are no circular polarization features,
as can be seen in Figure \ref{fig:08071}.
The flux of the 1667~MHz line is 35~mJy, and the isotropic luminosity is
$\lloh = 2.12$. For their likely choice of cosmology, this would correspond
to $\lloh = 2.03$, which is roughly 70\% less luminous than they reported.
It is not possible to identify Zeeman splitting in the 
Stokes $V$ spectrum. There is a large dip in the Stokes $V$ 
spectrum at the location of emission, but similar dips appear elsewhere in
the spectrum. The upper limit on the magnetic field associated with Gaussian 4,
which is a bright, relatively narrow, feature
located near the peak of emission, is 20~mG. The rest of the parameters used
to fit emission are shown in Table \ref{tab:08071}.

\begin{deluxetable*}{lccccc}
\tablecolumns{6}
\tablewidth{0pt}
\tablecaption{IRAS F08071$+$0509 Gaussian Fit Parameters \label{tab:08071}}
\tabletypesize{\footnotesize}
\tablehead{
& \colhead{$S$} & \colhead{$\nu$} & \colhead{$\Delta\nu$} & \colhead{$v_{\odot}$} & \colhead{$B_{\parallel}$} \\
\colhead{Gaussian} & \colhead{(mJy)} & \colhead{(MHz)} & \colhead{(MHz)} & \colhead{(km s$^{-1}$)} & \colhead{(mG)} \\
\colhead{(1)} & \colhead{(2)} & \colhead{(3)} & \colhead{(4)} & \colhead{(5)} & \colhead{(6)}
}
\startdata
     0 \dotfill & $\phantom{8}    2.74 \pm   0.20$ & $   1582.7638 \pm  0.0143$ & $   0.4223 \pm   0.0377$ & $  16023.2$ & $                         -89.27 \pm  29.37           $ \\
     1 \dotfill & $\phantom{8}    3.03 \pm   0.21$ & $   1583.8481 \pm  0.0130$ & $   0.3814 \pm   0.0338$ & $  15807.0$ & $                         -38.07 \pm  25.31           $ \\
     2 \dotfill & $\phantom{8}    2.90 \pm   0.91$ & $   1584.5591 \pm  0.0023$ & $   0.0902 \pm   0.0082$ & $  15665.4$ & $           \phantom{ }    -8.82 \pm  13.27           $ \\
     3 \dotfill & $              27.84 \pm   0.80$ & $   1584.6262 \pm  0.0067$ & $   0.4756 \pm   0.0243$ & $  15652.1$ & $\phantom{-}\phantom{8}     6.26 \pm   3.23\phantom{8}$ \\
     4 \dotfill & $              10.49 \pm   0.80$ & $   1584.7340 \pm  0.0024$ & $   0.1168 \pm   0.0082$ & $  15630.6$ & $           \phantom{ }    -9.77 \pm   4.19\phantom{8}$ \\
     5 \dotfill & $\phantom{8}    4.10 \pm   0.32$ & $   1585.2492 \pm  0.0058$ & $   0.1532 \pm   0.0142$ & $  15528.1$ & $\phantom{-}               23.32 \pm  11.83           $
\enddata
\end{deluxetable*}

{\bf IRAS F08201+2801}: DG01 discovered this OHM, which features broad 1667
MHz emission, and low flux emission redward of the main 1667~MHz features,
which DG01 suggested could be either 1665~MHz emission or high velocity 1667~MHz
emission. Overall, it has rich
enough structure that ten Gaussian components are needed
to produce a reasonable fit to the full 1665/1667~MHz emission. 
The Stokes $V$ spectrum has little in the way of structured noise, and shows
no compelling features. We place an upper limit of 20~mG on fields in this
OHM.  

{\bf IRAS F08279+0956}: This OHM was reported by DG01. It features
two broad, blended 1667~MHz features in the
Stokes $I$ spectrum, but at such low signal-to-noise that
even with minimal structured noise in the Stokes $V$ spectrum, 
it is not possible to place any
useful constraint on magnetic fields associated with this OHM. 

{\bf IRAS F08449+2332}: The maser emission of this OHM discovered by DG01 
overlaps with the same 1448.3 MHz interference that appeared in the spectrum of 
IRAS F03521+0028. As such, no magnetic field is detected, nor is it possible
to place any useful limit on magnetic field properties.  

{\bf IRAS F08474+1813}: DG01 discovered this OHM, which has a roughly 4~mJy
peak flux density, and three broad emission features that are distinguishable. 
The low flux density and broad lines means the Stokes $V$ observations
do not provide any useful limits on magnetic fields in this OHM. 

{\bf IRAS F09531+1430}: This OHM was discovered by DG01, and the spectrum has
two moderately narrow, well separated peaks atop a broad region of emission. 
Though no field is detected for either of the narrow features, we may place
an upper limit of 30~mG on magnetic fields associated with either peak. 

{\bf IRAS F09539+0857}: This OHM, discovered by DG01, 
has broad, blended 1665/1667~MHz emission and  
rich structure. We use nine Gaussian components to the total
1665/1667~MHz emission, including multiple narrower features, and achieve
an adequate fit. The Stokes
V spectrum has reasonably well behaved noise, but there are no suggestions
of Zeeman splitting. The upper limit on the magnetic field associated with
multiple narrow Gaussians fit to peaks in the emission is 20~mG.

{\bf IRAS F10035+2740}: There are two broad, low signal-to-noise features in the
spectrum of this OHM, which was discovered by DG02. 
These broad features do not provide a useful limit on the magnetic fields
in the masing region of this galaxy.

{\bf IRAS F10339+1548}: The OHM, discovered by DG01, 
is clearly detected at 1667~MHz. 
Weak 1665~MHz emission is also visible in our spectrum, shown in Figure
\ref{fig:10339}, while standing waves frustrated the 
detection of the 1665~MHz emission in the DG01 observations. We fit one
Gaussian to the 1665~MHz emission, and one broad and one narrow Gaussian to
the 1667~MHz emission, which are shown in Table \ref{tab:10339}. From the Gaussian, we compute a hyperfine ratio $R_H = 21$. 
The peak of emission is fit by Gaussian 2, 
and with well behaved noise in the Stokes $V$ spectrum, we can
place an upper limit of 20~mG on magnetic fields in this source.

\begin{deluxetable*}{lccccc}
\tablecolumns{6}
\tablewidth{0pt}
\tablecaption{IRAS F10339$+$1548 Gaussian Fit Parameters \label{tab:10339}}
\tabletypesize{\footnotesize}
\tablehead{
& \colhead{$S$} & \colhead{$\nu$} & \colhead{$\Delta\nu$} & \colhead{$v_{\odot}$} & \colhead{$B_{\parallel}$} \\
\colhead{Gaussian} & \colhead{(mJy)} & \colhead{(MHz)} & \colhead{(MHz)} & \colhead{(km s$^{-1}$)} & \colhead{(mG)} \\
\colhead{(1)} & \colhead{(2)} & \colhead{(3)} & \colhead{(4)} & \colhead{(5)} & \colhead{(6)}
}
\startdata
     0 \dotfill & $\phantom{8}    2.39 \pm   0.14$ & $   1391.4459 \pm  0.0198$ & $   0.7723 \pm   0.0550$ & $  59446.6$ & $                        -136.05 \pm  49.28           $ \\
     1 \dotfill & $\phantom{8}    6.69 \pm   0.41$ & $   1393.1475 \pm  0.0101$ & $   0.5373 \pm   0.0203$ & $  59007.8$ & $\phantom{ }              -50.12 \pm  15.57           $ \\
     2 \dotfill & $              11.26 \pm   0.45$ & $   1393.2540 \pm  0.0025$ & $   0.1841 \pm   0.0079$ & $  58980.4$ & $\phantom{ }\phantom{ }    -2.35 \pm   5.41\phantom{8}$
\enddata
\end{deluxetable*}

{\bf IRAS F11028+3130}: The spectrum for this OHM, reported by DG01, 
has two broad features, one corresponding to the 1667~MHz line and one to the
1665~MHz line, and a low amplitude 1667~MHz feature on the redward side of the
broad 1667~MHz emission. Given the low snr, it is not possible to place any
useful limit on the magnetic field strength for this source. 

{\bf IRAS F11180+1623}: DG02 discovered this OHM, which has only
one easily identified emission component with a peak flux density of roughly 
4~mJy. With such a weak signal, no useful limit may be placed
on the presence of magnetic fields in this galaxy.  

{\bf IRAS F11524+1058}: This OHM was discovered by DG01. It has broad emission, 
which we fit with four
reasonably broad Gaussians. The lack of distinguishable narrow spectral
features prevents placing useful limits on the magnetic field in this galaxy.

{\bf IRAS F12005+0009}: Though this OHM, discovered by DG02, has a peak flux
density of only 9~mJy, it features multiple distinguishable emission components,
including one relatively
narrow peak centered on a broad hump. The fits to the Stokes $V$ spectrum
yields a 3$\sigma$ claimed detection for a 21~mG field 
associated with the narrow peak. Unfortunately, the Stokes $V$
spectrum for this source includes many narrow ripples of the same 
characteristic size as the narrow feature, and shifting the Stokes $I$ and $V$ 
spectra relative to each other reveals many shifts that produce a 
claimed magnetic field detection. We place an upper limit of 40~mG
on magnetic fields associated with masing clouds in this galaxy.

{\bf IRAS F12018+1941}: Only one broad feature may be distinguished in
the spectrum of this OHM discovered by \citet{Martin1988}. 
It is not possible to place a useful limit on the magnetic field in this OHM. 

{\bf IRAS F12162+1047}: DG02 reported this OHM, which has broad, low flux 
density emission. The Stokes $I$ spectrum  
may be adequately fit with two broad Gaussians, but does not allow any
useful limit to be placed on magnetic fields in this galaxy. 

{\bf IRAS F12243-0036}: \citet{Martin1988} discovered this OH kilomaser, 
which has apparent absorption
bracketing the maser emission. The overall profile is complicated, and may
only be adequately fit. Nevertheless, there are a few narrow emission
features in the spectrum, and one that is aligned with a feature in the
Stokes $V$ spectrum. The fit to Stokes $V$ suggests a field detection, but
shifts between Stokes $I$ and $V$ also produce claimed fields. We place an
upper limit of 20~mG on the magnetic field associated with this maser. 

{\bf IRAS F12549+2403}: DG02 discovered this OHM, which has 
two broad, 1667~MHz emission components and marginally detected 1665~MHz
emission in the Stokes $I$ spectrum. No useful limit may be placed
on the magnetic field in this galaxy. 

{\bf IRAS F13126+2453}:  
This source is classified as an OH absorber and kilomaser.
\citet{Darling2002a} reported that no published
spectrum exists, but a published spectrum is available in 
\citet{Schmelz1986} under the designation IC 860, 
so the spectrum is not provided again here. 
The absorption in both the 1665 and 1667~MHz lines is clearly detected, as
well as emission at the wings of the absorption features. For the
absorption alone, the hyperfine ratio is $R_H = 1.27 \pm 0.18$,  
while for the emission, we find
a hyperfine ratio of $R_H = 1.3 \pm 0.36$, where the errors are based on
reported errors to Gaussian fits to the spectra. Both are consistent with
OH in local thermodynamic equilibrium. 

The Stokes $V$
spectrum features bandpass structure on scales comparable 
in width to features in the Stokes $I$ spectrum, leading to claimed fields 
with magnitudes of $\sim$30--40 mG associated with the absorption features. 
We do not consider these claims to be reliable. The noise in the vicinity of 
the emission components has less structure than in the regions of absorption, 
but there is no sign of
Zeeman splitting. In the emission regions, we place an upper limit of 30~mG
on the magnitude of magnetic fields.

{\bf IRAS F13218+0552}: Among broad emission in the spectrum of this OHM
discovered by DG02, 
there is one narrow feature at 1384.8 MHz. The Stokes $V$ spectrum has
well-behaved noise in the vicinity of the narrow emission seen in Stokes $I$,
but there is no sign of Zeeman splitting. We place an upper limit of 50~mG
on the magnetic field associated with the masing region in this source. 

{\bf IRAS F14043+0624}: This OHM, discovered by DG02, has an 
asymmetric 1667~MHz emission 
profile best fit by one narrow and one moderately wide Gaussian, along
with one broad, weak 1665~MHz line. 
No magnetic field is detected, and we place an upper limit of 
40~mG on magnetic fields in this galaxy. 

{\bf IRAS F14059+2000}: DG02 discovered this OHM, which has bright 
1667~MHz emission that is well
fit by one very broad Gaussian and one moderately broad Gaussian, and 1665~MHz
emission that blends with the red wing of 1667~MHz emission. No Zeeman
splitting is apparent in the Stokes $V$ spectrum, and we place an upper limit
of 20~mG on the magnitude of magnetic fields associated with this OHM. 

{\bf IRAS F14553+1245}: There is one relatively narrow, low SNR feature
in the spectrum of this OHM. It was discovered by DG02, and like them, 
we do not see a 
clear 1665~MHz component in the spectrum. The Stokes $V$ spectrum is nearly
featureless, and we place an upper limit of 60~mG on
magnetic fields associated with this OHM. 

{\bf IRAS F14586+1432}: This OHM, discovered by DG02, 
has a rich, blended spectrum, spread over more than 5~MHz.
Despite this richness, there are few
identifiable narrow features in the spectrum. Fits to Stokes $V$ provide 
claimed magnetic fields, but so too do fits when the Stokes $V$ is shifted
relative to Stokes $I$, thanks to a number of ripples in the Stokes $V$
spectrum. This is regarded as a non-detection, and it is difficult
to provide a meaningful upper limit on magnetic fields in this galaxy. 

{\bf IRAS F15107+0724}: \citet{Bottinelli1986} discovered this OHM, and a 
spectrum is available in \citet{Baan1987} and \citet{Martin1988a}. The 1667
MHz emission has a dual-peaked structure, which is closely matched by the
1665~MHz emission. A total of six Gaussian components are needed to fit the
Stokes $I$ spectrum, four of which are narrow. While no Zeeman splitting is
seen in the Stokes $V$ spectrum, each of the narrow components is consistent
with an upper limit of 12--15~mG to the magnitude of magnetic fields in this
OHM. 

{\bf IRAS F16100+2527}: DG01 discovered this OHM, which has emission 
made up of a broad 1667~MHz line with a marginally distinct peak, 
and a similarly broad and well separated 1665~MHz line. 
The 1667~MHz emission can be reasonably fit with one 
wide and one narrow Gaussian. The nearly 6~mJy narrow feature provides a 
40~mG upper limit on the magnitude of magnetic fields in this OHM.

{\bf IRAS F16300+1558}: This OHM was discovered by DG00, but RFI plagued their
observations. Our spectra combined by taking the median do not have serious 
narrow RFI, but do feature significant bandpass structure. Like DG00, we do not 
clearly detect the 1665~MHz line. Structure of unclear origin in both
the Stokes $I$ and $V$ spectra prevents placing any useful limits on magnetic
fields in this OHM. 

{\bf IRAS F17161+2006}: The 1667~MHz emission from this source, discovered by
DG02, has a peak flux density of less than 10~mJy. Four
Gaussians are required to reasonably fit the profile, three for the 1667~MHz
emission and one for the broad 1665~MHz emission. One of the 1667~MHz 
components is reasonably narrow, and provides an upper limit of 60~mG
on the magnitude of the magnetic field associated with this OHM. 

{\bf IRAS F17207-0014}:  \citet{Bottinelli1985} reported this OHM, which has a spectrum with
strong interference from GLONAS.
Despite the interference, the maser is still clearly detected, as
interference is negligible at frequencies lower
than about 1601 MHz, based on inspection
of individual spectra used to create the composite spectrum. Comparison
of previously published spectra for the OHM in IRAS F17207-0014 
\citep{Martin1989a, Momjian2006}, as well as with spectra of other sourcing
overlapping in frequency, confirms that the features below
1600 MHz are maser emission rather than interference. 
The Stokes~$I$ profile is bright, but there are no distinct peaks in the 
emission. Moreover, the Stokes~$V$ spectrum has significant broadband
structure. While interference below 1601 MHz is mild compared to that
above 1601 MHz, it still makes it difficult to detect Zeeman splitting in
this source or place useful limits. 

{\bf IRAS F17539+2935}: This OHM, discovered by DG00, 
is only marginally detected, with a peak flux density of just over 1~mJy. 
The SNR is insufficient to detect magnetic fields
or place any useful upper limits. 

{\bf IRAS 20248+1734}: While this OHM, discovered by DG00, 
is clearly detected, the spectrum for this OHM features standing waves
that complicate fitting and identification of real emission, and it is not
possible to detect Zeeman splitting or place useful limits on 
magnetic fields in this galaxy. 

{\bf IRAS 20286+1846}: DG00 discovered this OHM, which has 
extremely broad, blended 1665/1667~MHz masing lines. To fit the total 
profile adequately requires nine components, three of which are relatively
narrow. No Zeeman splitting is detected for any of the three narrow
components, but we place an upper limit of 
20~mG on magnetic fields in this OHM. 

{\bf IRAS 20450+2140}: This OHM was reported in DG00. 
The Stokes $I$ spectrum shows two blended 1667~MHz emission components,
adequately fit by two Gaussians. The
width of the fitted Gaussians is too large to place any useful limits on
magnetic fields in this galaxy. 

{\bf IRAS 21077+3358}: DG00 reported this OHM, which has broad 1667~MHz
that possibly blends with its 1665~MHz component.
It is not possible to clearly distinguish any narrow features in the spectrum, 
or place any useful limits on magnetic fields in this galaxy. 

{\bf IRAS 21272+2514}: The spectrum of this OHM, discovered by DG00, 
is rich in structure, and very broad.
It is not possible to clearly distinguish the 1665~MHz and 1667~MHz lines.
The full spectrum is reasonably well-fit by ten Gaussians, six of which are
moderately narrow. 
No magnetic fields are detected, and we place an upper limit of
20~mG on magnetic fields in the masing region in this galaxy. 

{\bf IRAS 22055+3024}: The spectrum of this OHM, reported in DG01,
has moderately broad 1667~MHz emission and a distinct 1665~MHz line. The 
1667~MHz features are well fit with a wide base and two narrow
Gaussians. There is some rippling structure in the Stokes $V$ in the 
vicinity of the narrow lines, but the noise is well enough behaved to
provide a 30~mG upper limit on magnetic fields in this OHM.  

{\bf IRAS F22116+0437}: This OHM was detected by DG00. Though it is 
re-detected here, it is insufficiently bright to place any
useful limit on magnetic fields. 

{\bf IRAS F23019+3405}: Emission from this OHM, reported in DG01, 
is quite narrow, with a FWHM of roughly $60\kms$. It nevertheless appears to
have a bi-peaked structure, though higher SNR spectra would be required to
confirm this. No Zeeman splitting is apparent in the Stokes $V$ spectrum,
but we are able to place an upper limit of 30~mG on 
magnetic fields in this OHM.

{\bf IRAS F23028+0725}: DG01 reported this OHM, which has a 
broad 1667~MHz emission that blends with a clear 1665~MHz component. 
Five Gaussians are needed to adequately fit the combined 1665/1667~MHz 
emission, shown in Figure \ref{fig:23028}. Gaussian components 1 and 2 are 
moderately narrow, and produce fits with reported $\sim 3\sigma$ 
magnetic fields having magnitudes of $\sim40$~mG. 
The spectrum is not convincing to the eye, however, and shifting the Stokes
$I$ and $V$ relative to each other confirms this, as non-zero shifts produce
stronger fits to the Stokes~$V$ spectrum, as seen 
in Figure \ref{fig:compare_shifts}.
For this reason, it is difficult to place useful
limits on magnetic fields in this galaxy.

{\bf IRAS F23129+2548}: Emission in this OHM, reported in DG01,
is broad, and the 1665~MHz and 1667~MHz lines are blended. 
The total spectrum is well fit with six Gaussians, though only one 
is narrow. Zeeman splitting is not observed, but noise in the Stokes $V$
spectrum is well enough behaved to provide an upper limit of 50~mG
on magnetic fields in this OHM. 

{\bf IRAS F23199+0123}: This OHM was reported by DG01. The 1667~MHz emission
is broad, and has a peak flux density of only 3~mJy. It is not clear if
1665~MHz emission is detected. Given the low flux density, no 
meaningful limit may be placed on magnetic fields in this OHM.

{\bf IRAS F23234+0946}: DG01 discovered this OHM. It has broad 1667~MHz
emission with an apparent blue tail and 1665~MHz line. Four
total Gaussians adequately fit the spectrum, one of which is narrow and near
the peak of 1667~MHz emission. No Zeeman splitting is detected, but the
narrow feature provides an upper limit of 50~mG on the magnitude of magnetic
fields in this OHM. 

\subsubsection{Marginal Stokes $V$ detections} \label{sec:marg_v_dets}
The three sources presented in this section, IRAS~F15224+1033, 
IRAS~F15587+1609, and IRAS~F20550+1655, all have features in their Stokes $V$ 
spectra consistent with Zeeman splitting. They also have Stokes $V$ 
structure that does not appear to be associated with any maser emission. 
For this reason, they are considered only ``marginal'' detections. In Table
\ref{tab:15224}, Table \ref{tab:15587}, and Table \ref{tab:20550}, the
fitted features that are consistent with detections of Zeeman splitting
are highlighted with bold text. 

{\bf IRAS F15224+1033}: 
This OHM, discovered by DG01, has a rich, wide Stokes $I$ spectrum, and a 
fairly noisy Stokes $V$ spectrum, shown in Figure \ref{fig:15224}. 
Any 1665~MHz emission that might be present
is blended with the broad 1667~MHz emission. The features of the OHM
have not changed significantly since DG01. The complexity requires
nine Gaussians to adequately fit, and even then, the residuals show structure
that coincides with the two brightest features, which are fit by Gaussian
components 3, 4, and 6. The full fits are shown in Table \ref{tab:15224}.
At the same frequency as Gaussian 6, there is a weak feature in the 
Stokes $V$ spectrum, with a peak-to-trough
amplitude only marginally larger than that of others
in the spectrum. The fits regard this
as a 3.9$\sigma$ detection of a -8.3~mG magnetic field. There is also a 
marginally significant fit associated with Gaussian 4, with a reported field
of -26.7$\pm$8.5~mG, but the fitted structure in Stokes~$V$ does not appear
markedly different to eye than other features in the spectrum. 
Altogether, the structure in the Stokes~$V$ spectrum that is apparently
unassociated with features in the Stokes~$I$ spectrum, and the marginal
appearance of the fitted features, limits our confidence in
the two reported fields.

\begin{deluxetable*}{lccccc}
\tablecolumns{6}
\tablewidth{0pt}
\tablecaption{IRAS F15224$+$1033 Gaussian Fit Parameters \label{tab:15224}}
\tabletypesize{\footnotesize}
\tablehead{
& \colhead{$S$} & \colhead{$\nu$} & \colhead{$\Delta\nu$} & \colhead{$v_{\odot}$} & \colhead{$B_{\parallel}$} \\
\colhead{Gaussian} & \colhead{(mJy)} & \colhead{(MHz)} & \colhead{(MHz)} & \colhead{(km s$^{-1}$)} & \colhead{(mG)} \\
\colhead{(1)} & \colhead{(2)} & \colhead{(3)} & \colhead{(4)} & \colhead{(5)} & \colhead{(6)}
}
\startdata
     0 \dotfill & $\phantom{8}    1.35 \pm   0.18$ & $   1467.4444 \pm  0.0212$ & $   0.3575 \pm   0.0593$ & $  40841.7$ & $                                   -179.44 \pm  43.83           $ \\
     1 \dotfill & $\phantom{8}    1.11 \pm   0.23$ & $   1468.0327 \pm  0.0197$ & $   0.2063 \pm   0.0520$ & $  40705.2$ & $\phantom{-}\phantom{8}               52.26 \pm  40.79           $ \\
     2 \dotfill & $\phantom{8}    3.25 \pm   0.19$ & $   1468.7777 \pm  0.0272$ & $   0.4532 \pm   0.0595$ & $  40532.5$ & $           \phantom{ }              -40.08 \pm  23.32           $ \\
     3 \dotfill & $\phantom{8}    5.36 \pm   0.69$ & $   1469.1573 \pm  0.0293$ & $   0.2430 \pm   0.0619$ & $  40444.5$ & $           \phantom{ }              -14.76 \pm  10.59           $ \\
     4 \dotfill & $\phantom{8}    4.75 \pm   1.78$ & $   1469.3149 \pm  0.0118$ & $   0.1446 \pm   0.0255$ & $  40408.0$ & $           \phantom{ }              {\bf -26.62 \pm   8.45}\phantom{8}$ \\
     5 \dotfill & $\phantom{8}    8.30 \pm   0.32$ & $   1469.7945 \pm  0.0010$ & $   0.8947 \pm   0.0029$ & $  40297.0$ & $           \phantom{ }\phantom{ }    -0.09 \pm  12.39           $ \\
     6 \dotfill & $              15.76 \pm   0.22$ & $   1469.7968 \pm  0.0094$ & $   0.1116 \pm   0.0338$ & $  40296.5$ & $           \phantom{ }\phantom{ }   {\bf -8.27 \pm   2.13}\phantom{8}$ \\
     7 \dotfill & $\phantom{8}    4.34 \pm   0.29$ & $   1470.5376 \pm  0.0686$ & $   6.3929 \pm   0.4799$ & $  40125.2$ & $\phantom{-}\phantom{8}               86.82 \pm  66.19           $ \\
     8 \dotfill & $\phantom{8}    6.07 \pm   0.20$ & $   1470.8441 \pm  0.0059$ & $   0.3314 \pm   0.0156$ & $  40054.3$ & $\phantom{-}\phantom{8}               14.55 \pm   9.88\phantom{8}$ \\
     9 \dotfill & $\phantom{8}    2.09 \pm   0.19$ & $   1471.3402 \pm  0.0175$ & $   0.3661 \pm   0.0478$ & $  39939.7$ & $           \phantom{ }\phantom{ }    -8.39 \pm  30.00           $
\enddata
\end{deluxetable*}

{\bf IRAS F15587+1609}: The spectrum of this OHM, reported in DG01, 
has a great deal of structure, with three distinct peaks, 
and many narrow features. 
Figure \ref{fig:15587} shows our fits, and Table \ref{tab:15587} lists the
fits, which use eleven total Gaussians 
for the 1667~MHz emission, and two more for the 1665~MHz emission. 
In fitting the Stokes $V$ spectrum, there are three reported magnetic
fields, associated with Gaussians 6, 7, and 8, and each is 
just above 3$\sigma$ claimed detections. These are three of the four brightest
Gaussian components used, have similar reported values of $6.1\pm2.0$,
$7.5\pm1.9$, and $12.2\pm3.5$, respectively, and are contained within 
a roughly 100$\kms$ range of velocities. 

We regard this claimed magnetic field detection as marginal primarily because 
the fitted features do not look particularly compelling to the eye. 
Though there are noticeable dips in the Stokes~$V$ spectrum near 
the same frequency as the brightest emission seen in the Stokes~$V$, there 
are no features strongly resembling Zeeman splitting.  
Given the complex structure of the spectrum, shifting the 
Stokes $I$ and $V$ spectra relative to each 
other does not provide significant insight.
The fits are of highest quality for no shift, but are of only 
slightly lower quality at a shift of 0.12~MHz, which is roughly
the average of the frequency separation between Gaussian 5, 6, 7, and 8. 
Shifting the Stokes $I$ and $V$ by roughly 0.5~MHz relative to each other,
corresponding to the separation between the central peak and the redward peak,
also produces claimed fields, though they are also of lower quality 
than when no shift is applied. Taken together, we are not confident in these
reported fields. 

\begin{deluxetable*}{lccccc}
\tablecolumns{6}
\tablewidth{0pt}
\tablecaption{IRAS F15587$+$1609 Gaussian Fit Parameters \label{tab:15587}}
\tabletypesize{\footnotesize}
\tablehead{
& \colhead{$S$} & \colhead{$\nu$} & \colhead{$\Delta\nu$} & \colhead{$v_{\odot}$} & \colhead{$B_{\parallel}$} \\
\colhead{Gaussian} & \colhead{(mJy)} & \colhead{(MHz)} & \colhead{(MHz)} & \colhead{(km s$^{-1}$)} & \colhead{(mG)} \\
\colhead{(1)} & \colhead{(2)} & \colhead{(3)} & \colhead{(4)} & \colhead{(5)} & \colhead{(6)}
}
\startdata
     0 \dotfill & $\phantom{8}    1.91 \pm   0.28$ & $   1464.5345 \pm  0.0130$ & $   0.2546 \pm   0.0426$ & $  41117.9$ & $\phantom{-}               24.22 \pm  16.49           $ \\
     1 \dotfill & $\phantom{8}    6.23 \pm   0.14$ & $   1465.0808 \pm  0.0129$ & $   0.8858 \pm   0.0418$ & $  40990.7$ & $                         -14.98 \pm   9.46\phantom{8}$ \\
     2 \dotfill & $\phantom{8}    7.08 \pm   0.53$ & $   1466.3288 \pm  0.0038$ & $   0.0686 \pm   0.0076$ & $  41100.8$ & $\phantom{-}\phantom{8}     0.22 \pm   4.21\phantom{8}$ \\
     3 \dotfill & $\phantom{8}    7.75 \pm   0.44$ & $   1466.4171 \pm  0.0039$ & $   0.0862 \pm   0.0100$ & $  41080.3$ & $\phantom{-}\phantom{8}     0.84 \pm   4.32\phantom{8}$ \\
     4 \dotfill & $              15.30 \pm   0.61$ & $   1466.5661 \pm  0.0337$ & $   0.7174 \pm   0.0456$ & $  41045.7$ & $           \phantom{ }    -2.04 \pm   6.34\phantom{8}$ \\
     5 \dotfill & $\phantom{8}    7.06 \pm   0.87$ & $   1466.6249 \pm  0.0045$ & $   0.0843 \pm   0.0122$ & $  41032.0$ & $\phantom{-}               12.01 \pm   4.90\phantom{8}$ \\
     6 \dotfill & $              18.78 \pm   1.10$ & $   1466.7254 \pm  0.0017$ & $   0.0847 \pm   0.0053$ & $  41008.7$ & $\phantom{-}\phantom{8}     {\bf 6.12 \pm   2.01}\phantom{8}$ \\
     7 \dotfill & $              25.75 \pm   3.70$ & $   1466.8546 \pm  0.0036$ & $   0.1567 \pm   0.0125$ & $  40978.6$ & $\phantom{-}\phantom{8}     {\bf 7.51 \pm   1.94} \phantom{8}$ \\
     8 \dotfill & $              20.80 \pm   1.71$ & $   1467.0327 \pm  0.0110$ & $   0.2574 \pm   0.0458$ & $  40937.3$ & $\phantom{-}               {\bf 12.23 \pm   3.48}\phantom{8}$ \\
     9 \dotfill & $\phantom{8}    4.19 \pm   0.64$ & $   1467.2375 \pm  0.0026$ & $   0.0482 \pm   0.0085$ & $  40889.7$ & $           \phantom{ }    -6.87 \pm   5.47\phantom{8}$ \\
    10 \dotfill & $              25.91 \pm   1.29$ & $   1467.3137 \pm  0.0076$ & $   0.2382 \pm   0.0130$ & $  40872.0$ & $\phantom{-}\phantom{8}     0.60 \pm   2.57\phantom{8}$ \\
    11 \dotfill & $\phantom{8}    6.45 \pm   0.17$ & $   1467.6886 \pm  0.0156$ & $   0.5179 \pm   0.0375$ & $  40785.0$ & $\phantom{-}               16.38 \pm  13.72           $ \\
    12 \dotfill & $\phantom{8}    2.97 \pm   0.13$ & $   1468.7666 \pm  0.0258$ & $   1.2354 \pm   0.0881$ & $  40535.0$ & $                         -65.71 \pm  40.91           $
\enddata
\end{deluxetable*}

{\bf IRAS F20550+1655}: 
This source was reported in an IAU Circular by \citet{Bottinelli1986},
and a published spectrum is available in \citet{Baan1989a}, under the
designation II Zw 96. 
\citet{Bottinelli1986} cite an isotropic
luminosity of $\lloh = 2.27$, and later sources use $\lloh = 2.11$, in 
agreement with the value of $\lloh = 2.13$ that we find. 
The spectrum, shown in Figure \ref{fig:20550},
suffers from interference in the 1602--1608.5 MHz part of the spectrum that 
is caused by the GLONAS satellite, including 
apparent absorption that just overlaps with redward side of the 1667~MHz 
line, redshifted to 1609 MHz. 
The interference is strongly linearly polarized, and also
produces features in the Stokes $V$ spectrum. 
It is not clear if this interference appears beyond 1608.5 MHz
in either the Stokes $I$ or $V$ spectra. 
Using eleven Gaussian components, provided in Table \ref{tab:20550}, 
to fit the emission blueward of 1608.5~MHz 
does an adequate job, and produces four magnetic field fits greater than
$3\sigma$. Two of these are to wide components, and are not considered
believable, but Gaussians 2 and 5 are narrow, and produce claimed fields of
$-18.5\pm2.4$~mG and $-17.4\pm2.9$~mG, respectively, and also appear plausible
examining the fits visually. The two Stokes $V$ features that are fit are the
largest amplitude, from peak-to-trough, of any blueward of 1608.5~MHz, but
there are two other dips that are not well accounted for. The interference
at 1608~MHz is also shown, and is of comparable amplitude to the fitted 
features. Though it has a different shape, it casts 
doubt on the validity of the 
detection. Examining the variance of each channel of the Stokes $V$ spectrum
shows a local maximum at 1608~MHz, corresponding to the interference, but 
has no apparent structure blueward of 1608.5~MHz, which suggests the fitted
features in the Stokes $V$ spectrum are real. Given the conflicting evidence,
we consider this a plausible magnetic field detection.

\begin{deluxetable*}{lccccc}[!h]
\tablecolumns{6}
\tablewidth{0pt}
\tablecaption{IRAS F20550$+$1655 Gaussian Fit Parameters \label{tab:20550}}
\tabletypesize{\footnotesize}
\tablehead{
& \colhead{$S$} & \colhead{$\nu$} & \colhead{$\Delta\nu$} & \colhead{$v_{\odot}$} & \colhead{$B_{\parallel}$} \\
\colhead{Gaussian} & \colhead{(mJy)} & \colhead{(MHz)} & \colhead{(MHz)} & \colhead{(km s$^{-1}$)} & \colhead{(mG)} \\
\colhead{(1)} & \colhead{(2)} & \colhead{(3)} & \colhead{(4)} & \colhead{(5)} & \colhead{(6)}
}
\startdata
     0 \dotfill & $\phantom{8}    7.17 \pm   1.14$ & $   1608.6408 \pm  0.0014$ & $   0.0180 \pm   0.0034$ & $  10942.9$ & $           \phantom{ }    -2.27 \pm   3.20\phantom{8}$ \\
     1 \dotfill & $              31.72 \pm   1.10$ & $   1608.9151 \pm  0.0050$ & $   0.2567 \pm   0.0085$ & $  10890.0$ & $                         -10.01 \pm   2.91\phantom{8}$ \\
     2 \dotfill & $              17.19 \pm   1.06$ & $   1608.9286 \pm  0.0019$ & $   0.0481 \pm   0.0040$ & $  10887.4$ & $                         {\bf -18.53 \pm   2.39}\phantom{8}$ \\
     3 \dotfill & $              15.47 \pm   1.50$ & $   1608.9974 \pm  0.0024$ & $   0.0693 \pm   0.0095$ & $  10874.1$ & $           \phantom{ }    -2.73 \pm   3.13\phantom{8}$ \\
     4 \dotfill & $              28.65 \pm   1.32$ & $   1609.0765 \pm  0.0016$ & $   1.1363 \pm   0.0044$ & $  10858.8$ & $                         -31.89 \pm   7.26\phantom{8}$ \\
     5 \dotfill & $              10.13 \pm   0.91$ & $   1609.1010 \pm  0.0049$ & $   0.0276 \pm   0.0139$ & $  10854.1$ & $                         {\bf -17.42 \pm   2.87}\phantom{8}$ \\
     6 \dotfill & $\phantom{8}    5.91 \pm   0.66$ & $   1609.1543 \pm  0.0064$ & $   0.0635 \pm   0.0216$ & $  10843.8$ & $                         -15.02 \pm   7.42\phantom{8}$ \\
     7 \dotfill & $\phantom{8}    9.12 \pm   0.83$ & $   1609.2766 \pm  0.0018$ & $   0.0493 \pm   0.0056$ & $  10820.2$ & $           \phantom{ }    -5.59 \pm   4.02\phantom{8}$ \\
     8 \dotfill & $              12.72 \pm   0.63$ & $   1609.4236 \pm  0.0017$ & $   0.0810 \pm   0.0054$ & $  10791.8$ & $           \phantom{ }    -7.27 \pm   3.69\phantom{8}$ \\
     9 \dotfill & $\phantom{8}    3.20 \pm   0.56$ & $   1609.6839 \pm  0.0067$ & $   0.0820 \pm   0.0180$ & $  10741.6$ & $\phantom{-}               48.31 \pm  14.76           $ \\
    10 \dotfill & $\phantom{8}    3.70 \pm   0.68$ & $   1609.8747 \pm  0.0046$ & $   0.0515 \pm   0.0115$ & $  10704.8$ & $           \phantom{ }    -4.71 \pm  10.14           $
\enddata
\end{deluxetable*}

\subsubsection{Stokes $V$ detections} \label{sec:v_dets}
The following sources all exhibit features in their Stokes $V$ spectra that 
are consistent with Zeeman splitting of OH maser lines. In the tables 
presented for sources with detections, the features that are considered
detections of Zeeman splitting are highlighted with bold text. 
In some cases, however, it was not possible to reliably estimate the 
magnetic field strengths associated with maser components. 
Of the new detections, none of them
have any existing VLBI observations, so it is not possible to provide 
detailed comments on the structure of the magnetic fields in the OHMs. 

{\bf IRAS F02524+2046}:  
DG02 discovered this very luminous OHM, which features three 
regions of narrow 1667~MHz emission, and two distinct areas of 1665~MHz 
emission that correspond to two of the 1667~MHz emission regions. Figure
\ref{fig:02524} shows a
Stokes $V$ spectrum that is similarly rich, relative to the other Stokes $V$ 
spectra of sources in this survey. There are three large amplitude 
features right next to one another, at frequencies corresponding to the
brightest emission in the Stokes $I$ spectrum. 

Fitting all of this structure was challenging. Given the relatively lower flux
of the 1665~MHz emission, five Gaussian components were sufficient to 
produce residuals absent of any structure. Most structure in the 1667~MHz
emission in the Stokes $I$ spectrum 
could be reasonably fit with nine or ten Gaussian components, but
produced extremely poor fits to the most obvious features in Stokes $V$, and 
left residuals around some of the narrower peaks. 
Ultimately, we used thirteen components for fitting 1667~MHz features, which 
are all listed in Table \ref{tab:02524}, and
produced a fit to Stokes $I$ that still has difficulty with the brightest
feature in the spectrum, and only partially captures the complex Stokes $V$
structure. In particular, a strong dip near 1412.1~MHz is completely missed
by the fits, and no combination of Gaussians we attempted was able to
simultaneously fit that feature and the other strong features in the
Stokes~$V$ spectrum of this source. 

Nevertheless, the fields and errors associated with these fits yield multiple
confident detections, as well as two fits that exceed 3$\sigma$ but in which
we have no confidence. 
The claimed, but ultimately dubious, fields are those associated
with Gaussians 2 and 10. Each of these two Gaussians is fitting broad emission
at a velocity of roughly 54,100--55,250~$\kms$, Gaussian 2 for 1665~MHz 
emission and Gaussian 10 for 1667~MHz emission. We generally place no 
confidence in such wide features, as discussed in 
Section \ref{sec:broad_rippling}. 

Aside from these reported fields, there are
four strong fields associated with Gaussians 8, 9, 12, and 13. All are narrow 
Gaussians fit to the bright, central 1667~MHz emission, and each gives a 
strong positive field, ranging from 12.3$\pm$1.8~mG for Gaussian 9 
to 23.9$\pm$5.8~mG for Gaussian 13. While this result is dependent upon the
input parameters of the fit, different combinations of Gaussians consistently
reported strong, positive fields associated with the masing regions that
produced the central 1667~MHz emission. A weak 1665~MHz feature, fit by 
Gaussian 3, provides further confirmation of a strong, positive field. On its
own, this weak component and $<3\sigma$ fitted field would not be considered
significant, but it has a similar velocity and width as some of the 
significant 1667~MHz features. The component is best fit
by a field of $15.5\pm 5.2$~mG, right in the same range of magnetic field
strengths measured by the 1667~MHz lines. VLBI observations of this OHM do not, 
unfortunately, exist, which limits further interpretation of the structure in 
the masing region.

\begin{deluxetable*}{lccccc}
\tablecolumns{6}
\tablewidth{0pt}
\tablecaption{IRAS F02524$+$2046 Gaussian Fit Parameters \label{tab:02524}}
\tabletypesize{\footnotesize}
\tablehead{
& \colhead{$S$} & \colhead{$\nu$} & \colhead{$\Delta\nu$} & \colhead{$v_{\odot}$} & \colhead{$B_{\parallel}$} \\
\colhead{Gaussian} & \colhead{(mJy)} & \colhead{(MHz)} & \colhead{(MHz)} & \colhead{(km s$^{-1}$)} & \colhead{(mG)} \\
\colhead{(1)} & \colhead{(2)} & \colhead{(3)} & \colhead{(4)} & \colhead{(5)} & \colhead{(6)}
}
\startdata
     0\footnote{Velocities and magnetic fields for Gaussians 0--4 correspond to values for the 1665~MHz line.}\dotfill & $              11.37 \pm   0.68$ & $   1410.0094 \pm  0.0014$ & $   0.0910 \pm   0.0053$ & $  54300.9$ & $\phantom{-}\phantom{8}     1.36 \pm   2.41\phantom{8}$ \\
     1 \dotfill & $\phantom{8}    6.87 \pm   0.67$ & $   1410.0398 \pm  0.0078$ & $   0.2969 \pm   0.0259$ & $  54293.2$ & $                         -13.10 \pm   7.58\phantom{8}$ \\
     2 \dotfill & $              11.07 \pm   0.36$ & $   1410.4608 \pm  0.0044$ & $   0.3152 \pm   0.0122$ & $  54187.5$ & $                         -17.83 \pm   4.49\phantom{8}$ \\
     3 \dotfill & $\phantom{8}    3.63 \pm   0.47$ & $   1410.6113 \pm  0.0030$ & $   0.0510 \pm   0.0083$ & $  54149.8$ & $\phantom{-}               {\bf 15.50 \pm   5.19}\phantom{8}$ \\
     4 \dotfill & $\phantom{8}    7.40 \pm   0.28$ & $   1410.8600 \pm  0.0276$ & $   1.3718 \pm   0.0806$ & $  54087.4$ & $\phantom{-}               21.66 \pm  13.77           $ \\
     5 \dotfill & $              17.43 \pm   0.74$ & $   1411.6473 \pm  0.0056$ & $   0.1795 \pm   0.0080$ & $  54305.7$ & $\phantom{-}\phantom{8}     7.50 \pm   3.93\phantom{8}$ \\
     6 \dotfill & $\phantom{8}    7.10 \pm   4.52$ & $   1411.8708 \pm  0.0438$ & $   0.2179 \pm   0.1105$ & $  54249.6$ & $           \phantom{ }    -1.15 \pm  12.19           $ \\
     7 \dotfill & $\phantom{8}    9.53 \pm   3.46$ & $   1412.0160 \pm  0.0029$ & $   0.0562 \pm   0.0098$ & $  54213.2$ & $           \phantom{ }    -6.73 \pm   3.86\phantom{8}$ \\
     8 \dotfill & $              38.02 \pm   7.69$ & $   1412.0955 \pm  0.0057$ & $   0.1146 \pm   0.0185$ & $  54193.3$ & $\phantom{-}               {\bf 16.22 \pm   1.61}\phantom{8}$ \\
     9 \dotfill & $              32.04 \pm   8.95$ & $   1412.2010 \pm  0.0089$ & $   0.1133 \pm   0.0155$ & $  54166.8$ & $\phantom{-}               {\bf 12.27 \pm   1.83}\phantom{8}$ \\
    10 \dotfill & $              35.20 \pm   5.23$ & $   1412.2861 \pm  0.0513$ & $   0.4675 \pm   0.0484$ & $  54145.5$ & $                         -12.07 \pm   3.18\phantom{8}$ \\
    11 \dotfill & $              11.97 \pm   1.02$ & $   1412.3222 \pm  0.0007$ & $   0.0230 \pm   0.0022$ & $  54136.4$ & $           \phantom{ }    -0.78 \pm   1.85\phantom{8}$ \\
    12 \dotfill & $              41.25 \pm   2.55$ & $   1412.3425 \pm  0.0012$ & $   0.0813 \pm   0.0027$ & $  54131.4$ & $\phantom{-}              {\bf 13.65 \pm   1.07}\phantom{8}$ \\
    13 \dotfill & $\phantom{8}    5.56 \pm   0.52$ & $   1412.5320 \pm  0.0021$ & $   0.0538 \pm   0.0061$ & $  54083.9$ & $\phantom{-}               {\bf 23.88 \pm   5.82} \phantom{8}$ \\
    14 \dotfill & $\phantom{8}    4.28 \pm   0.50$ & $   1412.7663 \pm  0.0023$ & $   0.0411 \pm   0.0060$ & $  54025.2$ & $\phantom{-}               10.94 \pm   6.63\phantom{8}$ \\
    15 \dotfill & $              12.57 \pm   0.48$ & $   1412.9512 \pm  0.0012$ & $   0.0832 \pm   0.0038$ & $  53978.9$ & $\phantom{-}\phantom{8}     1.75 \pm   3.33\phantom{8}$ \\
    16 \dotfill & $              11.75 \pm   0.38$ & $   1412.9866 \pm  0.0055$ & $   0.3496 \pm   0.0161$ & $  53970.0$ & $           \phantom{ }    -9.07 \pm   7.83\phantom{8}$ \\
    17 \dotfill & $\phantom{8}    3.60 \pm   0.22$ & $   1413.5314 \pm  0.0269$ & $   0.6639 \pm   0.0795$ & $  53833.7$ & $\phantom{-}               75.98 \pm  33.51           $
\enddata
\end{deluxetable*}
{\bf IRAS F04332+0209}: 
This OH kilomaser was first reported by \citet{Martin1989b} 
with $\lloh = 0.44$, but
no spectrum or other properties of the maser were published. 
We clearly detect this source, shown in Figure \ref{fig:04332}. It has
a 1667~MHz peak flux density of 16~mJy, and
an isotropic luminosity $\lloh = 0.29$, assuming a distance of 51~Mpc. 
This is slightly lower than the value reported by \citet{Martin1989b}. 
If they assumed a distance of 48~Mpc, our measured flux suggests
possibility that the luminosity of the source has changed of order 40--50\%.
Given the unknown uncertainty in their luminosity estimate, uncertainty in
their assumed distance, and 
roughly 10\% uncertainty in ours, this evidence for variability is merely
suggestive. 

As a kilomaser host, this galaxy is unlike the majority of the galaxies
observed in this survey. It has an infrared luminosity of 2.9 $\times 10^{10}
L_\odot$, well short of meeting the definition of a LIRG, which requires
$\llfir > 11$. Nevertheless,
\citet{Baan1998} characterized this galaxy as having a starburst nucleus in
their optical classification of maser hosts. 

The maser is also unusual in that the Stokes $V$ spectrum for 
this maser is unlike any other in this survey. There
are three distinct features in the spectrum at the same frequency as
emission in the Stokes $I$ spectrum, and the features in Stokes $V$ 
have the same width and magnitude of the amplitude as those in the Stokes $I$
spectrum. This suggests that the narrow splitting assumption is
invalid for this source; instead, the Zeeman splitting exceeds the
linewidth, and the $LHCP$ and $RHCP$ components are distinct. In the absence of
VLBI observations for this source, this makes deriving estimates of
magnetic fields difficult. For this reason, we do not present a table showing
fits, as we do for other sources. 

Nevertheless, we offer a plausible scenario, while acknowledging that this
single dish spectrum provides incomplete evidence. 
The two reddest peaks in the Stokes $I$ spectrum, centered at 1647.43~MHz
and 1647.53~MHz (labelled on the axis of the spectrum with a 0), 
have comparable widths, and amplitudes that differ by a 
factor of roughly two. The 1647.43~MHz peak is the weaker of the two, and
almost entirely $RHCP$, while the 1647.53~MHz peak is $LHCP$. If the two
are paired components, then the difference between centers of the
Gaussian components, $\Delta \nu$, gives the total magnetic field
\begin{equation}
	B = \frac{\Delta \nu}{b}, \label{eq:wide_split}
\end{equation} 
where $b$ is the same splitting coefficient described before. 
The roughly 100~kHz separation would correspond to a 46.8~mG
field. The Stokes $V$ spectrum contains another large dip at the same frequency
as the bluest feature in the Stokes $I$ spectrum. This component is nearly
all in the $LHCP$ spectrum. If the component is associated with a field that
is also strong, and positive, the corresponding $RHCP$ emission would fall
near the central, brightest emission from this maser. The Stokes $V$ spectrum
is relatively flat in the region of the brightest emission. This could result
from overlap of the $RHCP$ emission associated with the reddest feature 
and $LHCP$ emission associated with the peak, 
along with asymmetric amplification of the
$LHCP$ and $RHCP$ components in the source. Even if this interpretation is
incorrect, we regard this as a clear detection of Stokes $V$ features. 
Follow-up VLBI observations would be very interesting in better understanding
this source.

{\bf IRAS F09039+0503}: DG01 reported this OHM, identifying weak 1665~MHz 
emission as well as a broad blueshifted 1667~MHz component at 37,300~$\kms$ 
in addition to the main features between 37,500--37,900~$\kms$. The
1665~MHz feature is apparently absent in this more recent spectrum, shown
in Figure \ref{fig:09039}, and the
blueshifted 1667~MHz component is marginally detected. The central part of
the spectrum features multiple narrow peaks, ideal for Zeeman detections. The
Stokes $V$ spectrum shows two clear features, each associated with the 
bright, narrow peaks, and fit by Gaussians 0 and 2. The fitted magnetic fields
are each considered significant by the fits, and none of the checks we tried
suggested that the features were not real. The field reported for Gaussian 0
is -16.1$\pm$2.7~mG, while Gaussian 2 has an associated field of 
-27.6$\pm$5.5~mG. The Stokes~$V$ feature fit by Gaussian 2 is clearly
asymmetric, possibly as a result of the relatively wide splitting relative to
the line width. None of the other fitted components, shown in Table 
\ref{tab:09039}, are considered significant. 
Of all sources in the survey with magnetic fields detections in which we 
are confident, the fields reported here for IRAS~F09039+0503 are the strongest.

\begin{deluxetable*}{lccccc}
\tablecolumns{6}
\tablewidth{0pt}
\tablecaption{IRAS F09039$+$0503 Gaussian Fit Parameters \label{tab:09039}}
\tabletypesize{\footnotesize}
\tablehead{
& \colhead{$S$} & \colhead{$\nu$} & \colhead{$\Delta\nu$} & \colhead{$v_{\odot}$} & \colhead{$B_{\parallel}$} \\
\colhead{Gaussian} & \colhead{(mJy)} & \colhead{(MHz)} & \colhead{(MHz)} & \colhead{(km s$^{-1}$)} & \colhead{(mG)} \\
\colhead{(1)} & \colhead{(2)} & \colhead{(3)} & \colhead{(4)} & \colhead{(5)} & \colhead{(6)}
}
\startdata
     0 \dotfill & $    8.98 \pm   0.63$ & $   1480.2093 \pm  0.0013$ & $   0.0375 \pm   0.0031$ & $  37904.2$ & $              {\bf -16.05 \pm   2.70}\phantom{8}$ \\
     1 \dotfill & $    5.25 \pm   0.25$ & $   1480.4762 \pm  0.0097$ & $   0.2791 \pm   0.0264$ & $  37843.3$ & $\phantom{ }    -3.39 \pm  13.57           $ \\
     2 \dotfill & $    6.75 \pm   1.73$ & $   1480.6593 \pm  0.0052$ & $   0.0754 \pm   0.0129$ & $  37801.5$ & $              {\bf -27.55 \pm   5.46}\phantom{8}$ \\
     3 \dotfill & $    3.69 \pm   0.57$ & $   1480.7525 \pm  0.0260$ & $   0.1400 \pm   0.0520$ & $  37780.3$ & $\phantom{ }    -5.00 \pm  14.36           $ \\
     4 \dotfill & $    8.66 \pm   0.36$ & $   1481.1259 \pm  0.0175$ & $   0.4110 \pm   0.0400$ & $  37695.2$ & $              -17.81 \pm  10.73           $ \\
     5 \dotfill & $    4.63 \pm   0.35$ & $   1481.5707 \pm  0.0368$ & $   0.4333 \pm   0.0699$ & $  37593.9$ & $              -46.32 \pm  20.02           $ \\
     6 \dotfill & $    6.40 \pm   0.43$ & $   1481.9562 \pm  0.0029$ & $   0.0932 \pm   0.0077$ & $  37506.1$ & $              -12.54 \pm   5.97\phantom{8}$
\enddata
\end{deluxetable*}

{\bf IRAS F10378+1108}: 
This OHM was discovered by \citet{Kazes1991}, and re-observed by DG02. The
spectrum, shown in Figure \ref{fig:10378}, 
features extremely broad emission, with two distinct narrow features overlaid.
The broad spectrum requires 12~Gaussians for a reasonable fit, listed in
Table \ref{tab:10378}, though
the residuals show that the fits near the two narrow features could still
be improved. 
The Stokes $V$ spectrum contains two suggestive features. One of these is well
aligned with the peak in Stokes $I$ fit by Gaussian 3, and the field
associated with this feature is -13.9 $\pm$ 3.3 mG. The other apparent 
feature is just redward of the peak fit by Gaussian 7, and is fit in part
by Gaussian 6, and a fitted field of -28.0 $\pm$ 7.0 mG associated with it.
Given the width of Gaussian 6, we do not regard this reported field as 
being meaningful.

\begin{deluxetable*}{lccccc}
\tablecolumns{6}
\tablewidth{0pt}
\tablecaption{IRAS F10378$+$1108 Gaussian Fit Parameters \label{tab:10378}}
\tabletypesize{\footnotesize}
\tablehead{
& \colhead{$S$} & \colhead{$\nu$} & \colhead{$\Delta\nu$} & \colhead{$v_{\odot}$} & \colhead{$B_{\parallel}$} \\
\colhead{Gaussian} & \colhead{(mJy)} & \colhead{(MHz)} & \colhead{(MHz)} & \colhead{(km s$^{-1}$)} & \colhead{(mG)} \\
\colhead{(1)} & \colhead{(2)} & \colhead{(3)} & \colhead{(4)} & \colhead{(5)} & \colhead{(6)}
}
\startdata
     0 \dotfill & $\phantom{8}    1.75 \pm   0.20$ & $   1465.5685 \pm  0.0197$ & $   0.4041 \pm   0.0575$ & $  41277.7$ & $                         -99.68 \pm  43.72           $ \\
     1 \dotfill & $\phantom{8}    3.56 \pm   0.37$ & $   1466.3600 \pm  0.0043$ & $   0.0868 \pm   0.0110$ & $  41093.6$ & $\phantom{-}\phantom{8}     0.56 \pm  10.05           $ \\
     2 \dotfill & $\phantom{8}    2.43 \pm   0.54$ & $   1466.5915 \pm  0.0043$ & $   0.0401 \pm   0.0106$ & $  41039.8$ & $           \phantom{ }    -3.31 \pm  10.24           $ \\
     3 \dotfill & $              10.43 \pm   0.40$ & $   1467.1249 \pm  0.0015$ & $   0.0804 \pm   0.0039$ & $  40915.9$ & $                         {\bf -13.87 \pm   3.31}\phantom{8}$ \\
     4 \dotfill & $              17.46 \pm   0.37$ & $   1467.3636 \pm  0.0310$ & $   3.6798 \pm   0.1391$ & $  40860.4$ & $                         -10.08 \pm  17.20           $ \\
     5 \dotfill & $\phantom{8}    4.13 \pm   0.55$ & $   1467.4195 \pm  0.0066$ & $   0.1265 \pm   0.0201$ & $  40847.5$ & $\phantom{-}               20.81 \pm  10.92           $ \\
     6 \dotfill & $              16.14 \pm   0.76$ & $   1467.5052 \pm  0.0120$ & $   0.7169 \pm   0.0195$ & $  40827.6$ & $                         -28.03 \pm   6.95\phantom{8}$ \\
     7 \dotfill & $              14.99 \pm   0.74$ & $   1467.6153 \pm  0.0011$ & $   0.0730 \pm   0.0039$ & $  40802.0$ & $\phantom{-}\phantom{8}     4.20 \pm   2.21\phantom{8}$ \\
     8 \dotfill & $\phantom{8}    9.86 \pm   0.84$ & $   1467.7385 \pm  0.0090$ & $   0.2996 \pm   0.0237$ & $  40773.4$ & $\phantom{-}               20.95 \pm   7.09\phantom{8}$ \\
     9 \dotfill & $\phantom{8}    2.70 \pm   0.28$ & $   1468.3149 \pm  0.0082$ & $   0.1569 \pm   0.0210$ & $  40639.7$ & $\phantom{-}\phantom{8}     9.23 \pm  17.79           $ \\
    10 \dotfill & $\phantom{8}    2.10 \pm   0.28$ & $   1469.5167 \pm  0.0100$ & $   0.1596 \pm   0.0255$ & $  40361.3$ & $\phantom{-}               22.77 \pm  22.98           $ \\
    11 \dotfill & $\phantom{8}    5.19 \pm   0.30$ & $   1471.2592 \pm  0.1248$ & $   2.9057 \pm   0.4441$ & $  39958.5$ & $\phantom{-}               61.95 \pm  70.23           $
\enddata
\end{deluxetable*}

{\bf IRAS F16255+2801}:  
This OHM was discovered and characterized by DG01. Like DG01,
we only detect 1667~MHz emission. The Stokes $I$ spectrum, shown in Figure
\ref{fig:16255} (with fits listed in Table \ref{tab:16255}), features four 
main overlapping peaks, and there is significant structure in the Stokes $V$
spectrum at the same frequency. The Stokes $I$ could be fit well with five
Gaussian components, 
but the addition of one more low amplitude, narrow 
component on the red side of the spectrum significantly improved the fit
to the Stokes $V$ spectrum, and even then the Stokes $V$ fit 
is still poor. 
There is clear, strong structure, and the fields associated with
Gaussian 1, 2, and 3 are all significant: 7.0 $\pm$ 1.9~mG, 9.6$\pm$2.0~mG, 
and 26.8$\pm$4.0~mG. Given the large field claimed for Gaussian 3, and the
relative narrowness of the lines, the narrow line assumption we use in
fitting fields is perhaps pushed past its limit, as was the case for
IRAS~F04332+0209. We experimented with
separately fitting the RCP and LCP, using only five Gaussian components, and
deriving magnetic field estimates using the frequency separation.
The results are actually fairly similar, though the reported fields for each 
component are 20--30\% smaller. While we regard this as a clear detection of 
Zeeman splitting, we are less confident in the derived magnetic field
strengths.

\begin{deluxetable*}{lccccc}
\tablecolumns{6}
\tablewidth{0pt}
\tablecaption{IRAS F16255$+$2801 Gaussian Fit Parameters \label{tab:16255}}
\tabletypesize{\footnotesize}
\tablehead{
& \colhead{$S$} & \colhead{$\nu$} & \colhead{$\Delta\nu$} & \colhead{$v_{\odot}$} & \colhead{$B_{\parallel}$} \\
\colhead{Gaussian} & \colhead{(mJy)} & \colhead{(MHz)} & \colhead{(MHz)} & \colhead{(km s$^{-1}$)} & \colhead{(mG)} \\
\colhead{(1)} & \colhead{(2)} & \colhead{(3)} & \colhead{(4)} & \colhead{(5)} & \colhead{(6)}
}
\startdata
     0 \dotfill & $\phantom{8}    8.46 \pm   2.45$ & $   1470.5518 \pm  0.0079$ & $   0.0528 \pm   0.0096$ & $  40121.9$ & $           \phantom{ }    -3.45 \pm   3.87\phantom{8}$ \\
     1 \dotfill & $              21.22 \pm   0.82$ & $   1470.6097 \pm  0.0030$ & $   0.0657 \pm   0.0115$ & $  40108.5$ & $\phantom{-}\phantom{8}     {\bf 7.01 \pm   1.91}\phantom{8}$ \\
     2 \dotfill & $              16.63 \pm   1.20$ & $   1470.6763 \pm  0.0020$ & $   0.0482 \pm   0.0044$ & $  40093.1$ & $\phantom{-}\phantom{8}     {\bf 9.61 \pm   2.02}\phantom{8}$ \\
     3 \dotfill & $\phantom{8}    9.58 \pm   0.65$ & $   1470.7556 \pm  0.0023$ & $   0.0771 \pm   0.0073$ & $  40074.8$ & $\phantom{-}               {\bf 26.78 \pm   4.02}\phantom{8}$ \\
     4 \dotfill & $\phantom{8}    6.60 \pm   0.60$ & $   1470.8160 \pm  0.0091$ & $   0.5459 \pm   0.0264$ & $  40060.8$ & $\phantom{-}               64.13 \pm  14.37           $ \\
     5 \dotfill & $\phantom{8}    5.99 \pm   0.52$ & $   1470.9155 \pm  0.0025$ & $   0.0824 \pm   0.0088$ & $  40037.9$ & $\phantom{-}\phantom{8}    10.00 \pm   6.07\phantom{8}$
\enddata
\end{deluxetable*}

{\bf IRAS 18368+3549}: The emission from this OHM discovered by DG01 
is extremely broad, causing any 1665~MHz to be blended with the 
1667~MHz emission, as Figure \ref{fig:18368} shows.  
The brightest flux density feature in the Stokes $I$ spectrum is nicely aligned
with the only suggestive feature in the Stokes $V$ spectrum. The fits
agree with what can be seen by eye, and suggest a field of 21.9 $\pm$ 5.9 mG 
associated with Gaussian 3. None of the other components fitted to the
overall profile, listed in Table \ref{tab:18368}, 
produce a fit regarded as significant. The feature in 
Stokes $V$ being fitted is fairly asymmetric; given the large reported field,
and relatively narrow line, the narrow splitting assumption may be breaking 
down here, and the LCP is amplified more strongly than the RCP.

\begin{deluxetable*}{lccccc}
\tablecolumns{6}
\tablewidth{0pt}
\tablecaption{IRAS 18368$+$3549 Gaussian Fit Parameters \label{tab:18368}}
\tabletypesize{\footnotesize}
\tablehead{
& \colhead{$S$} & \colhead{$\nu$} & \colhead{$\Delta\nu$} & \colhead{$v_{\odot}$} & \colhead{$B_{\parallel}$} \\
\colhead{Gaussian} & \colhead{(mJy)} & \colhead{(MHz)} & \colhead{(MHz)} & \colhead{(km s$^{-1}$)} & \colhead{(mG)} \\
\colhead{(1)} & \colhead{(2)} & \colhead{(3)} & \colhead{(4)} & \colhead{(5)} & \colhead{(6)}
}
\startdata
     0 \dotfill & $    2.43 \pm   0.21$ & $   1491.2619 \pm  0.0483$ & $   0.5514 \pm   0.1134$ & $  35401.3$ & $           \phantom{ }              -18.02 \pm  45.18           $ \\
     1 \dotfill & $    0.96 \pm   0.39$ & $   1491.7403 \pm  0.0825$ & $   0.3640 \pm   0.1574$ & $  35293.8$ & $           \phantom{ }              -72.24 \pm  92.19           $ \\
     2 \dotfill & $    2.79 \pm   0.41$ & $   1492.7714 \pm  0.0069$ & $   0.0945 \pm   0.0180$ & $  35062.3$ & $\phantom{-}\phantom{8}               40.39 \pm  15.10           $ \\
     3 \dotfill & $    6.38 \pm   0.57$ & $   1492.9042 \pm  0.0029$ & $   0.0714 \pm   0.0082$ & $  35032.5$ & $\phantom{-}\phantom{8}               {\bf 21.86 \pm   5.88}\phantom{8}$ \\
     4 \dotfill & $    2.80 \pm   0.30$ & $   1493.0578 \pm  0.0146$ & $   0.2039 \pm   0.0400$ & $  34998.1$ & $           \phantom{ }              -33.69 \pm  22.34           $ \\
     5 \dotfill & $    8.68 \pm   0.28$ & $   1493.5034 \pm  0.0304$ & $   2.3252 \pm   0.1040$ & $  34898.2$ & $           \phantom{ }              -17.35 \pm  23.97           $ \\
     6 \dotfill & $    3.07 \pm   0.38$ & $   1493.5322 \pm  0.0066$ & $   0.1154 \pm   0.0178$ & $  34891.8$ & $\phantom{-}\phantom{8}\phantom{8}     0.03 \pm  14.88           $ \\
     7 \dotfill & $    2.11 \pm   0.40$ & $   1493.7634 \pm  0.0089$ & $   0.1037 \pm   0.0237$ & $  34840.0$ & $\phantom{-}\phantom{8}               37.01 \pm  20.54           $ \\
     8 \dotfill & $    2.85 \pm   0.30$ & $   1494.5953 \pm  0.0191$ & $   0.6221 \pm   0.0705$ & $  34653.7$ & $           \phantom{ }              -17.80 \pm  37.16           $ \\
     9 \dotfill & $    1.26 \pm   0.29$ & $   1495.8268 \pm  0.0203$ & $   0.1868 \pm   0.0530$ & $  34378.4$ & $                                   -123.47 \pm  46.07           $
\enddata
\end{deluxetable*}

{\bf IRAS F18588+3517}: DG01 discovered this OHM,  
which has multiple narrow components in the 1667~MHz line, as well as a 
weak, but clearly distinguished, 1665~MHz component.  

Figure \ref{fig:18588} shows the Stokes~$I$ and $V$ features
for this source, and Table \ref{tab:18588} lists the fits. 
To fit the 1667~MHz features, 
we used six components, including one broad component and three narrow
components where the emission is strongest. We also fit three
components to the 1665~MHz line, which at first glance seems a large number
given the strength of the 1665~MHz features. 
Gaussian 2, in particular, is fit to a feature that is of smaller amplitude
and width than we would ordinarily fit, 
but it is motivated by the fact that the velocity of the slight peak
to which Gaussian 2 is fitted matches extremely well with the velocity of
Gaussian 5. The widths are moderately different, but close enough to
plausibly be the same feature.
The reported magnetic fields are also in good agreement, with
Gaussian 2 giving 16.3 $\pm$ 4.2 mG, and Gaussian 5 giving 12.8 $\pm$ 2.1 mG. 
Combining the two gives 13.9 $\pm$ 1.9 mG for the region that produces the 
matched 1665/1667 lines. The field fit to Gaussian 4 is also significant, 
with a field of 18.3 $\pm$ 2.6 mG. A large field is also reported for 
Gaussian 1, but the Stokes $V$ spectrum has noisy features of comparable
width, and the reported field is not regarded as meaningful.

\begin{deluxetable*}{lccccc}
\tablecolumns{6}
\tablewidth{0pt}
\tablecaption{IRAS F18588$+$3517 Gaussian Fit Parameters \label{tab:18588}}
\tabletypesize{\footnotesize}
\tablehead{
& \colhead{$S$} & \colhead{$\nu$} & \colhead{$\Delta\nu$} & \colhead{$v_{\odot}$} & \colhead{$B_{\parallel}$} \\
\colhead{Gaussian} & \colhead{(mJy)} & \colhead{(MHz)} & \colhead{(MHz)} & \colhead{(km s$^{-1}$)} & \colhead{(mG)} \\
\colhead{(1)} & \colhead{(2)} & \colhead{(3)} & \colhead{(4)} & \colhead{(5)} & \colhead{(6)}
}
\startdata
     0\footnote{Velocities and magnetic fields for Gaussians 0, 1, and 2 correspond to values for the 1665~MHz line.} \dotfill & $\phantom{8}    1.58 \pm   0.21$ & $   1505.8294 \pm  0.0662$ & $   0.3804 \pm   0.1439$ & $  31768.9$ & $\phantom{-}\phantom{8}     6.72 \pm  30.63           $ \\
     1 \dotfill & $\phantom{8}    2.32 \pm   0.48$ & $   1506.1660 \pm  0.0328$ & $   0.2691 \pm   0.0600$ & $  31694.8$ & $\phantom{-}               57.68 \pm  17.10           $ \\
     2 \dotfill & $\phantom{8}    2.98 \pm   0.65$ & $   1506.1751 \pm  0.0032$ & $   0.0318 \pm   0.0086$ & $  31692.9$ & $\phantom{-}               {\bf 16.29 \pm   4.21}\phantom{8}$ \\
     3 \dotfill & $\phantom{8}    9.53 \pm   0.20$ & $   1507.8172 \pm  0.0057$ & $   0.6455 \pm   0.0149$ & $  31721.0$ & $                         -17.45 \pm   9.92\phantom{8}$ \\
     4 \dotfill & $              10.10 \pm   0.53$ & $   1507.8706 \pm  0.0016$ & $   0.0463 \pm   0.0036$ & $  31709.2$ & $\phantom{-}               {\bf 18.31 \pm   2.60}\phantom{8}$ \\
     5 \dotfill & $              14.82 \pm   0.48$ & $   1507.9445 \pm  0.0012$ & $   0.0633 \pm   0.0033$ & $  31693.0$ & $\phantom{-}               {\bf 12.78 \pm   2.07}\phantom{8}$ \\
     6 \dotfill & $\phantom{8}    6.90 \pm   0.67$ & $   1508.0288 \pm  0.0013$ & $   0.0270 \pm   0.0032$ & $  31674.5$ & $\phantom{-}\phantom{8}     6.05 \pm   2.79\phantom{8}$ \\
     7 \dotfill & $\phantom{8}    3.50 \pm   0.41$ & $   1508.4066 \pm  0.0077$ & $   0.1553 \pm   0.0207$ & $  31591.4$ & $                         -17.25 \pm  13.58           $ \\
     8 \dotfill & $\phantom{8}    2.99 \pm   0.19$ & $   1508.6885 \pm  0.0200$ & $   0.3726 \pm   0.0489$ & $  31529.5$ & $                         -38.22 \pm  24.61           $
\enddata
\end{deluxetable*}

{\bf IRAS F22134+0043}: This OHM was discovered with the GBT by \citet{Willett2012}. The heliocentric
redshift is 0.212 \citep{Stanford2000}, but the peak 1667~MHz emission 
occurs at a redshift of 0.211. The Stokes $I$ spectrum features broad emission
mostly between 10--20~mJy, and one narrow peak that reaches 50~mJy. There is
possible 1665~MHz emission corresponding to the 1667~MHz emission redward of
the peak, but the low amplitude and lack of any 1665~MHz emission that
corresponds to the 1667~MHz peak suggest that the broad feature is bandpass
structure.

Our fit to the emission from this source is shown in Figure \ref{fig:F22134},
with parameters listed in Table \ref{tab:F22134}. 
To adequately capture the broad emission and complex peaked structure in the 
Stokes $I$ spectrum required a total of eleven Gaussian components. Fitting 
the peak was particularly difficult. The primary Stokes $V$ feature is an 
asymmetric ``S'' at the same frequency as the Stokes $I$ peak, and producing
what was ultimately only a decent fit to both spectra at that frequency 
required three Gaussian components all centered there, with three very
different widths. These three Gaussians, labeled 5, 6, and 7, each report
very similar magnetic fields: -12.2 $\pm$ 4.7, -7.1 $\pm$ 1.7, and 
-13.5 $\pm$ 2.6. While the field reported with Gaussian 5 is not, on its own,
considered significant given the associated error and width of the line,
it is encouraging that all three lines agree
so well, given that they likely are produced by the same masing clouds. 

In addition to these reported fields, there were also 3$\sigma$ fields
fitted to Gaussians 2, 8, and 9. As Gaussian 2 is wide, it is not considered
significant. Gaussians 8 and 9 are moderately narrow, and the quality of the
fits are good, with fields of -26.8 $\pm$ 5.3~mG and -22.1 $\pm$ 4.6~mG, 
respectively. The features being fit are small amplitude 
relative to those associated with the peak emission, but are larger than 
Stokes~$V$ structures unassociated with Stokes~$I$ emission. 
Unfortunately, shifting the Stokes $I$ and $V$ are relative to each other does 
not provide any added support, as the brighter features dominate. So, despite
the high quality of the fit, we consider the fields associated with Gaussians
8 and 9 to be strongly suggestive, but not confident detections. 

\begin{deluxetable*}{lccccc}
\tablecolumns{6}
\tablewidth{0pt}
\tablecaption{IRAS F22134+0034 Gaussian Fit Parameters \label{tab:F22134}}
\tabletypesize{\footnotesize}
\tablehead{
& \colhead{$S$} & \colhead{$\nu$} & \colhead{$\Delta\nu$} & \colhead{$v_{\odot}$} & \colhead{$B_{\parallel}$} \\
\colhead{Gaussian} & \colhead{(mJy)} & \colhead{(MHz)} & \colhead{(MHz)} & \colhead{(km s$^{-1}$)} & \colhead{(mG)} \\
\colhead{(1)} & \colhead{(2)} & \colhead{(3)} & \colhead{(4)} & \colhead{(5)} & \colhead{(6)}
}
\startdata
     0 \dotfill & $\phantom{8}    3.83 \pm   0.62$ & $   1375.7583 \pm  0.0025$ & $   0.0351 \pm   0.0072$ & $  63542.9$ & $           \phantom{ }    -8.27 \pm   6.69\phantom{8}$ \\
     1 \dotfill & $\phantom{8}    3.92 \pm   0.35$ & $   1375.8254 \pm  0.0079$ & $   0.1614 \pm   0.0179$ & $  63525.2$ & $\phantom{-}               14.41 \pm  14.07           $ \\
     2 \dotfill & $              10.34 \pm   0.18$ & $   1376.3018 \pm  0.0136$ & $   0.5674 \pm   0.0392$ & $  63399.4$ & $                         -42.05 \pm  10.64           $ \\
     3 \dotfill & $\phantom{8}    7.61 \pm   0.54$ & $   1376.4600 \pm  0.0017$ & $   0.0564 \pm   0.0048$ & $  63357.7$ & $           \phantom{ }    -2.88 \pm   4.34\phantom{8}$ \\
     4 \dotfill & $              13.73 \pm   0.70$ & $   1376.5947 \pm  0.0020$ & $   0.1293 \pm   0.0060$ & $  63322.2$ & $           \phantom{ }    -9.37 \pm   3.82\phantom{8}$ \\
     5 \dotfill & $              18.78 \pm   0.65$ & $   1376.8707 \pm  0.0032$ & $   0.2974 \pm   0.0156$ & $  63249.4$ & $                         -12.15 \pm   4.74\phantom{8}$ \\
     6 \dotfill & $              22.40 \pm   1.68$ & $   1376.8832 \pm  0.0007$ & $   0.0544 \pm   0.0030$ & $  63246.1$ & $           \phantom{ }    {\bf -7.13 \pm   1.67}\phantom{8}$ \\
     7 \dotfill & $\phantom{8}    8.60 \pm   1.80$ & $   1376.8874 \pm  0.0009$ & $   0.0191 \pm   0.0036$ & $  63245.0$ & $                         {\bf -13.54 \pm   2.59}\phantom{8}$ \\
     8 \dotfill & $              11.16 \pm   0.49$ & $   1377.1415 \pm  0.0040$ & $   0.1348 \pm   0.0113$ & $  63178.0$ & $                         {\bf -26.75 \pm   5.36}\phantom{8}$ \\
     9 \dotfill & $              12.71 \pm   0.36$ & $   1377.2937 \pm  0.0032$ & $   0.1187 \pm   0.0124$ & $  63137.9$ & $                         {\bf -22.10 \pm   4.64}\phantom{8}$ \\
    10 \dotfill & $\phantom{8}    5.55 \pm   0.49$ & $   1377.4203 \pm  0.0078$ & $   0.1020 \pm   0.0137$ & $  63104.5$ & $           \phantom{ }    -5.46 \pm   8.85\phantom{8}$
\enddata
\end{deluxetable*}

\subsection{Linear Polarization}
As in R08, we derived the linear polarized intensity for each source from the
Stokes $Q$ and $U$ spectra. Unfortunately, we did not find believable linear 
polarization in any of the newly observed OHMs. For this reason, we do not 
provide discussion of the process, instead referring the reader to R08. 
\section{Discussion}
This work confirms the most basic result of R08, showing that OH megamasers are viable targets for detection of extragalactic Zeeman splitting, and direct measurement of line-of-sight magnetic fields. Six of the eight sources observed by R08 were re-observed here, and the Stokes $V$ features and derived magnetic fields were quite consistent for five of the six sources. The exception, IRAS~F10173+0829, featured weak interference in the Stokes $V$ spectrum that coincided with the peak of Stokes $I$ emission, at a level barely distinguishable from noise in the Stokes $I$ spectrum for the source, and masqueraded as Zeeman splitting. 

\begin{figure}
\includegraphics[width=3.5in]{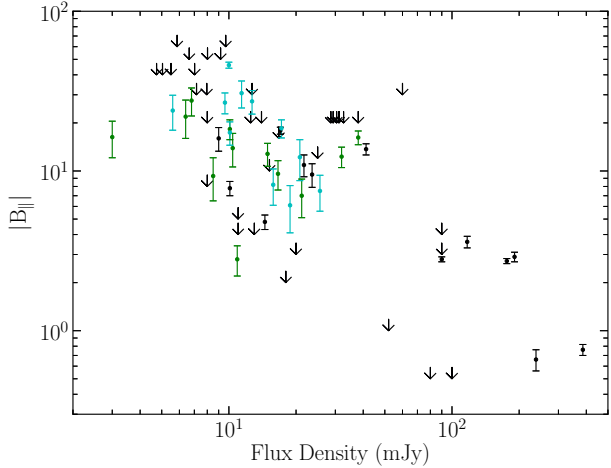}
\caption{Magnetic field detections and upper limits are shown relative to the flux density of the 1667~MHz line. For OHMs with detected Zeeman splitting, the flux density used is that of the Gaussian component with which the Zeeman splitting was associated. For upper limits, the plotted flux density is the peak flux density of the line. The most confident detected fields are black, moderate quality detections are green, and the marginal detections are cyan.} \label{fig:bmag_flux}
\end{figure}

The success rate in detecting Zeeman splitting was significantly lower than in R08. Not counting IRAS~F10173+0829, four of the eight sources R08 observed displayed Zeeman splitting, compared to eleven of the seventy-one newly observed sources. This is not surprising, as R08 selected the highest peak flux density OHMs for observations. The peak-to-trough Stokes~$V$ amplitude of Zeeman splitting features that are less than the line width will generally be much smaller than the Stokes~$I$ feature to which it corresponds. High flux density lines are necessary then for detecting weak Zeeman splitting. For this reason, all of the newly detected magnetic fields presented here are above the median magnetic field amplitude in R08, which was $\sim$3~mG. Of newly detected fields alone, the median magnitude of magnetic fields was $\sim$16~mG, and the median of the sample, counting re-detections and new detections, is $\sim$~12 mG. The magnetic field fits and errors, as a function of the Stokes $I$ flux density, is shown in Figure \ref{fig:bmag_flux}. 

The detections from R08 are clustered in the lower right hand corner, while the new detections mostly fall along a trend that reflects that the weaker the flux density of lines, the stronger field required for a Zeeman splitting detection. The lowest flux density point, at 3~mJy, corresponds to the lone 1665~MHz Zeeman splitting detection, which nicely agreed with a 1667~MHz detection at the same velocity. Many low flux density sources are omitted from Figure \ref{fig:bmag_flux}, as they did not provide meaningful upper limits on the magnitude of magnetic fields. 

It is possible that in addition to missing weak magnetic fields, this survey also misses very strong magnetic fields present in OHMs. A few of the detected fields were strong enough that the assumption of narrow line widths was only marginally applicable. This typically occurred in sources with very complex emission, in which it is likely that circularly polarized emission from physically unassociated masing clouds blended together. In IRAS~04332+0209, the Stokes~$I$ and $V$ spectra suggest the possibility that the right and left circularly polarized components are completely separated, corresponding to a 47~mG magnetic field detection. While confirmation of this interpretation is not possible without VLBI observations, in a source with more complex emission and multiple blended components, it is not even possible to identify that such a strong field may exist. The absence of fields stronger than $\sim$30~mG inferred using our standard analysis procedure may reflect a limitation in our approach, rather than strongly disavowing the possibility of such strong magnetic fields in OHMs. 

\subsection{Comparison with Galactic OH masers}
R08 compared their measured magnetic field strengths with results from OH masers in the Milky Way. Combining results from \citet{Reid1990} and \citet{Fish2003}, R08 noted that in Galactic OH maser Zeeman pairs, the mean was 0~mG, within the 3.3~mG error in the distribution. In a similar survey, \citet{Fish2005} detected a total of 184 Zeeman pairs among all OH lines. A companion paper presenting detailed analysis of these data \citep{Fish2006} shows their distribution of magnetic field magnitudes peaked at 4~mG, and fell off quickly thereafter. The two strongest observed fields, in W51 e2, had magnitudes of 19.8~mG and 21~mG. The results from R08 were broadly similar, as the median field detected had a magnitude of $\sim$3~mG, and from that R08 concluded that when examined on small scales, star formation in (U)LIRGs proceeds similarly to star formation in the Milky Way. The results from this larger survey suggest that strong fields may be more common in OHMs than in Galactic OH masers.

\begin{figure}
\includegraphics[width=3.5in]{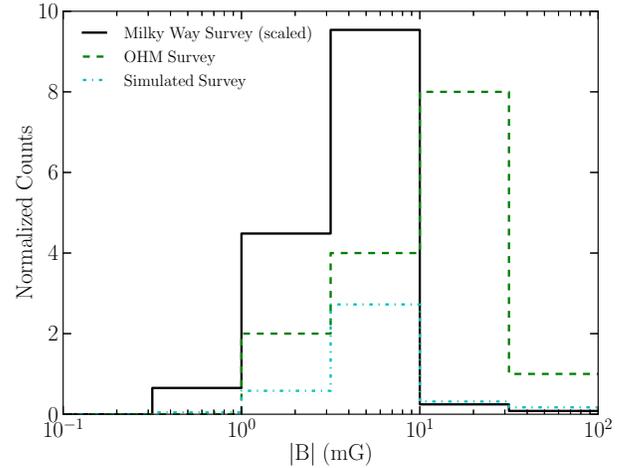}
\caption{Three distributions of magnetic field strengths are compared. The black solid line shows the results from the survey of Galactic OH masers by \citet{Fish2005}. The dashed green line shows the distribution of magnetic fields we found in OHMs. The dash-dot cyan line represents the expected distribution we would have observed in OHMs if magnetic fields in OHMs had comparable strength to magnetic fields in Galactic OH masers.} \label{fig:sampled_dist}
\end{figure}

Direct comparison of the two distributions of detected field strengths is misleading, however, as the \citet{Fish2005} magnetic field limits were far lower than the limits of this OHM survey. To better compare the two distributions, we simulated an experiment where magnetic field strengths were drawn from the observed distribution of magnetic fields observed in Galactic OH masers. This was made up of the \citet{Fish2005} distribution, plus a $\sim$40~mG field discovered in W75N \citep{Slysh2006}. The limits for the simulated survey were drawn from our own results. 

The sample of limits was created in the following way. For sources with detections, we counted each detected field associated with a Gaussian component, and used four times the error associated with that component as the limit, since our reported upper limits were found by taking the reported field strength and adding three times the reported error. For sources with no detection, we used the upper limit for magnetic fields in the source reported above. Then, we randomly drew a magnetic field strength from the Galactic OH maser magnetic field distribution and a limit from our distribution, and recorded fields larger than the limits as detections. For a single run, this process was repeated $n$ times, where $n$ was the number of values in the upper limit distribution. 

We then simulated $10^4$ surveys in which the magnetic field strengths in the OHM sample were distributed according to the distribution observed in the Milky Way. The results of this simulated survey are shown in Figure \ref{fig:sampled_dist}, alongside the distribution of fields in Milky Way OH masers reported in \citet{Fish2005}, and the distribution we actually observed. The three distributions are clearly distinct. If the magnetic fields in OHMs were drawn from the same distribution as those in Galactic OH masers, we would have detected many fewer fields overall, and, most notably, many fewer strong fields. Fields stronger than 10~mG are apparently much more common in OHMs than in the Milky Way. 

The results of this simulated survey were consistently different than what we observed. The median of the median magnitude magnetic field detected in the simulated surveys was 5.0~mG, and 95\% of the medians fell between 4.0--5.5~mG. The 12~mG median of OHM magnetic fields in the actual survey is more than a factor of two larger than the typical median in the simulated samples. Altogether, these results argue that the magnetic fields in OHMs are stronger than those in Milky Way OH masers.

\subsection{Strong Fields and Time Variability}
R08 reported a factor of $\sim$2 variation in one component of the emission
from IRAS~F12032+1707, and noted that the magnetic field associated with that component was the strongest observed in their observations. R08 noted that a connection between strong fields and time variability had been observed in Galactic OH masers, as \citet{Slysh2006} and \citet{Fish2007} both reported 40~mG fields in W75N that are associated with maser spots in the process of flaring. 

Our more recent observations of magnetic fields in OHMs do not provide any further evidence of this connection, though the observations do not provide strong contrary evidence either. There is weak variability, at the 1--10\% level, observed in four of the six OHMs re-observed in this survey. The strongest variability observed here occurs in IRAS~F12112+0305, in which the upper limit on magnetic field magnitude is $\sim$3~mG, and in III~Zw~35, which has magnetic fields with magnitudes of $\sim$3~mG in the region where variability occurs. In IRAS~F12032+1707, any further variability since the observations of R08 is limited. In comparing other sources with strong fields detected in our survey with existing spectra in the literature, we encounter no instances similar to IRAS~F12032+1707, where a component with a strong magnetic field has clearly varied relative to its previously published strength. 

\subsection{Dynamical importance}
The small number of OHMs with VLBI observations show reasonable diversity, with compact, parsec-scale emission providing the majority of emission in some sources (Arp~220, III~Zw~35, IRAS~F17208--0014), while compact emission is weak or absent in others (Mrk~231, IRAS~F14070+0525), and in one case, emission was compact on the $\sim$100~pc resolution of the observations (IRAS~F12032+1707). Of the strong Zeeman splitting signals observed in Arp~220 and III~Zw~35, most could be associated with these compact, parsec-scale features. If all strong magnetic fields detections presented here, including those in sources without VLBI observations, occur in parsec-scale regions, it suggests that magnetic fields may be dynamically important in masing clouds in OHMs. 

In the modeling of OHM emission in III~Zw~35 by \citet{Parra2005a}, they assumed a maximum density of $\nsubhtwo = 10^5 \cmc$ in their OHM clouds, a cloud size of roughly $R=1$~pc, and noted that typical internal velocity dispersions are $\Delta V = 20$~$\kms$. We take these values as being reasonably representative of OHMs at large, and explore the implications of the observed magnetic field strengths. While this is surely imperfect, the small sample of VLBI measurements indicates that III~Zw~35 is not unusual in its small scale properties.

Following the \citet{Stahler2005} application of the virial theorem to a spherical cloud, and assuming its mass is dominated by molecular hydrogen, the ratio of the magnetic energy density, $\mathcal{M}$, to the self gravitational energy, $\mathcal{W}$, is 
\begin{equation}
    \frac{\mathcal{M}}{\left|\mathcal{W}\right|} \simeq 0.5 \left(\frac{B}{3 {\rm \;mG}}\right)^2 \left(\frac{R}{1 {\rm \; pc}}\right)^{-2} \left(\frac{\nsubhtwo^2}{10^5 \cmc}\right)^{-2}.
\end{equation}

R08 found $\sim$3~mG line-of-sight magnetic fields in III~Zw~35. For a variety of plausible magnetic field probability distribution functions, the mean of the line-of-sight magnetic fields is roughly half the mean of the true magnetic field strength \citep{Heiles2005}. This suggests that the typical magnetic fields observed in OHMs are dynamically important, and perhaps even dominant.

\citet{Parra2005a} note, however, that the clouds in III~Zw~35 have typical velocity dispersions of $\sim$20~$\kms$ and are not likely gravitationally confined, and then go on to suggest that magnetic fields could play a role in magnetically confining the clouds.
The ratio of the magnetic energy density and the turbulent energy density, $\mathcal{T}$, are equal for 
\begin{equation}
    \frac{\mathcal{M}}{\left|\mathcal{T}\right|} \simeq 0.2 \left(\frac{B}{3 {\rm \;mG}}\right)^2 \left(\frac{\Delta V}{20 {\rm \; \kms}}\right)^{-2} \left(\frac{\nsubhtwo^2}{10^5 \cmc}\right)^{-1}.
\end{equation}
This expression shows that for total magnetic field strengths a factor of two stronger than the observed line-of-sight fields, and values for the velocity dispersion and cloud size used in \citet{Parra2005a}, the turbulent pressure and magnetic pressure are of roughly equal strength. Altogether, these results argue that magnetic fields are important in determining the structure of OH masing clouds in at least some fraction of OHMs. 

Densities in Galactic OH maser regions are of order $10^6$--$10^7 \cmc$ \citep{Reid1987}, as measured by NH$_3$. If extragalactic OH megamasers mase at similar densities, the weaker ($\sim$3~mG) magnetic fields detected in this survey would not be dynamically important, unless clouds are a factor of a few smaller than 1~pc. However, for sources with $\sim$20--30~mG fields, magnetic fields would still be dynamically relevant in a $\sim$1~pc cloud, even at densities of $10^6 \cmc$. More VLBI observations of OHMs with magnetic fields, particularly the strongest magnetic fields, will be necessary to developing a better understanding of the role magnetic fields play in the central regions of OHM hosts.

\section{Conclusions}
We performed Full-Stokes observations of seventy-seven OH megamasers, representing the entire known sample of Arecibo accessible OHMs known at the time of observations. Six of these sources had recently been observed in Full-Stokes mode by \citet{Robishaw2008}, and four of those six sources had features in their Stokes~$V$ spectra consistent with Zeeman splitting. We confirm the Stokes~$V$ detections in three of the four sources, the non-detections in the two sources without Stokes~$V$ features, and find that the apparent Zeeman splitting in one of the four sources was actually interference. Of the remaining seventy-one sources without previous Stokes~$V$ observations, eleven have features consistent with Zeeman splitting in the Stokes~$V$ spectra. 

For all sources with Stokes~$V$ features, we derived magnetic field strengths that could produce the observed Zeeman splitting signal. The median magnetic field magnitude associated with masing components in the fourteen OHMs with Zeeman detections is 12~mG. We show that the magnetic field strengths observed in OHM masing clouds are roughly a factor of 2--3 larger than those observed in Galactic OH masers. The strongest derived magnetic field in which we are confident has a magnitude of 27.6~mG. One source, IRAS~F04332+0209, appears to have Zeeman splitting that exceeds the linewidth. This interpretation cannot be confirmed without VLBI observations, but if it is correct, the required magnetic field strength would be $\sim$47~mG. In two other sources, IRAS~02524+2046 and IRAS~F18588+3517, we observed for the first time Zeeman splitting in the 1665~MHz OH line of an OHM. In each case, the reported magnetic field strength is consistent with that derived for the 1667~MHz line at the same velocity. For reasonable assumptions about conditions in OHMs, the magnetic field measurements presented here argue for magnetic fields playing a dynamically important role in the masing clouds. \\ \\

We thank Tim Robishaw for sharing code he developed for analyzing and reducing Full-Stokes Arecibo data. We also thank Eliot Quataert and Moshe Elitzur for many helpful discussions. We thank the anonymous referee for constructive comments that improved this paper. This research was supported in part by NSF grant AST-0908572. J. M. received support from a NSF Graduate Research Fellowship. This research used NASA's Astrophysics Data System Bibliographic Services, the SIMBAD database, operated at CDS, Strasbourg, France, and the NASA/IPAC Extragalactic Database (NED), which is operated by the Jet Propulsion Laboratory, California Institute of Technology, under contract with the National Aeronautics and Space Administration. Cosmological calculations used CosmoloPy (${\rm http://roban.github.com/CosmoloPy/}$), a cosmology package for Python.

\bibliographystyle{apj}
\bibliography{ms,extra}

\end{document}